\documentclass[sort,compress,11pt]{elsarticle}
\usepackage{orcidlink}

\usepackage{amssymb}
\usepackage{amsthm}
\usepackage{amsmath}
\usepackage{mathtools}
\usepackage{mathrsfs}
\usepackage{algorithm}
\usepackage[algo2e]{algorithm2e}
\usepackage{algpseudocode}
\usepackage{array}
\usepackage{multirow}
\usepackage{listings}
\usepackage{tabu}
\usepackage{booktabs}
\usepackage{enumerate}
\usepackage{fullpage}
\usepackage{float}
\usepackage{xcolor}
\usepackage[colorinlistoftodos]{todonotes}
\usepackage{bbm}
\usepackage{bm}
\usepackage{url}
\usepackage{textcomp}
\usepackage{gensymb}
\usepackage{soul}
\usepackage{lineno}
\usepackage{graphicx}
\usepackage{subcaption}
\usepackage{float}
\usepackage{caption}
\usepackage{tikz}
\usetikzlibrary{shapes.misc}
\tikzset{cross/.style={cross out, draw=black, minimum size=2*(#1-\pgflinewidth), inner sep=0pt, outer sep=0pt},
cross/.default={1pt}}

\theoremstyle{definition}

\theoremstyle{remark}

\biboptions{numbers,comma,round,square}

\graphicspath{ {./figs/} }

\journal{Elsevier}

\begin{document}
\begin{frontmatter}

\title{Conditional flow matching for generative modeling of near-wall turbulence with quantified uncertainty}

\author[cornell,ndAME]{Meet Hemant Parikh\orcidlink{0009-0008-1284-6368}}
\author[cornell,ndAME]{Xiantao Fan\orcidlink{0000-0002-0977-0330}}

\author[cornell,ndAME]{Jian-Xun Wang\orcidlink{0000-0002-9030-1733}\corref{corr}}

\address[cornell]{Sibley School of Mechanical and Aerospace Engineering, Cornell University, Ithaca, NY, USA}
\address[ndAME]{Department of Aerospace and Mechanical Engineering, University of Notre Dame, Notre Dame, IN, USA}

\cortext[corr]{Corresponding author: Jian-Xun Wang}
\ead{jw2837@cornell.edu}

\begin{abstract}

Reconstructing near-wall turbulence from wall-based measurements is a critical yet inherently ill-posed problem in wall-bounded flows, where limited sensing and spatially heterogeneous flow-wall coupling challenge deterministic estimation strategies. To address this, we introduce a novel generative modeling framework based on conditional flow matching for synthesizing instantaneous velocity fluctuation fields from wall observations, with explicit quantification of predictive uncertainty. Our method integrates continuous-time flow matching with a probabilistic forward operator trained using stochastic weight averaging Gaussian (SWAG), enabling zero-shot conditional generation without model retraining. We demonstrate that the proposed approach not only recovers physically realistic, statistically consistent turbulence structures across the near-wall region but also effectively adapts to various sensor configurations, including sparse, incomplete, and low-resolution wall measurements. The model achieves robust uncertainty-aware reconstruction, preserving flow intermittency and structure even under significantly degraded observability. Compared to classical linear stochastic estimation (LSE) and deterministic convolutional neural networks (CNN) methods, our stochastic generative learning framework exhibits superior generalization and resilience under measurement sparsity with quantified uncertainty. This work establishes a robust semi-supervised generative modeling paradigm for data-consistent flow reconstruction and lays the foundation for uncertainty-aware, sensor-driven modeling of wall-bounded turbulence.

\end{abstract}

\end{frontmatter}

\section{Introduction}

Wall-bounded turbulence is fundamental to a wide range of engineering and natural systems, governing critical processes such as skin friction drag, heat and mass transfer, and the onset of flow instabilities. The near-wall region, including the viscous sublayer, buffer layer, and lower logarithmic layer, plays a particularly dominant role in the dynamics of wall-bounded flows, as this is where turbulent kinetic energy is both generated and dissipated at the highest rates~\cite{townsend1961equilibrium,pope2001turbulent}. Coherent structures in these regions, such as streamwise streaks and quasi-streamwise vortices, are known to mediate the transfer of momentum and energy across scales, and their spatiotemporal organization governs the emergence of larger-scale turbulent motions farther from the wall~\cite{adrian2007hairpin,hwang2016inner}. Therefore, near-wall turbulence has long been the focus of flow control strategies aimed at reducing drag, delaying transition, or enhancing mixing~\cite{kim2007active}. However, implementing such strategies in practice demands high-fidelity instantaneous flow field information, which remains challenging due to small spatial scales, high temporal variability, and limited accessibility of measurements in this region. 

Wall-mounted sensors, such as those measuring pressure or wall shear stress, are relatively easy to deploy, offering time-resolved, scalable, and non-disruptive access to the surface signatures of near-wall turbulence~\cite{lofdahl1999mems,choi1994active}. This practical advantage has naturally led to the question of whether it is possible to reconstruct the full, off-wall velocity field using only wall-based measurements. If successful, such reconstructions would enable closed-loop control, real-time flow monitoring, and improved wall models for large-eddy simulations (LES) without requiring full-field sensing. The feasibility of this task is supported by the bidirectional coupling between near-wall and outer-layer structures~\cite{zaki2024turbulence}. Large-scale motions in the logarithmic and outer regions of wall-bounded flows modulate the near-wall turbulence, leaving observable imprints on wall quantities~\cite{abe2004very,mathis2009large,hwang2016inner}. Conversely, energetic near-wall events can propagate their influence outward and modulate the coherence of large-scale motions in the outer layer~\cite{adrian2007hairpin,lozano2014time}. These bidirectional interactions have inspired extensive efforts to infer inner velocity fields from wall measurements. Physics-based strategies often leverage data assimilation techniques to integrate wall observations into high-fidelity simulations such as direct numerical simulations (DNS) or LES, thereby reconstructing the full flow state~\cite{colburn2011state,wang2025variational,suzuki2017estimation}. While these methods benefit from governing equations that enforce physical consistency, they remain computationally prohibitive for most practical settings. Alternatively, resolvent analysis provides a linearized, reduced-order mapping between wall forcing and flow response~\cite{amaral2021resolvent}. While insightful, these models rely on simplifying assumptions (e.g., linearization around a mean flow) and cannot fully capture the nonlinear and intermittent nature of near-wall turbulence.

On the other hand, data-driven methods offer an attractive alternative for flow estimation and reconstruction, particularly in scenarios where repeated measurements are available, and the underlying physics are too complex to be modeled directly. Early work using linear stochastic estimation (LSE)~\cite{adrian1988stochastic,marusic2010predictive,baars2016spectral,encinar2019logarithmic} and proper orthogonal decomposition (POD)~\cite{towne2018spectral} laid the foundation by extracting statistically dominant flow features and their correlations with wall signals. These techniques, however, are constrained by their linearity, and often fail to resolve the complex, multi-scale turbulence structures. With the advent of deep learning, convolutional neural networks (CNNs) and their variants have emerged as powerful tools for nonlinear flow estimation from wall data~\cite{guemes2019sensing,guastoni2021convolutional,balasubramanian2023predicting,cuellar2024three,hora2024physics,yousif2023deep}. These models have demonstrated promising results in reconstructing velocity fields at various wall-normal locations using wall shear stress or pressure as input. Despite these advances, most of the existing approaches remain fundamentally deterministic and do not account for the inherent uncertainty of the reconstruction task, which is due to spatially heterogeneous correlations and wall measurement quality.  In regions where wall-flow coherence is strong (e.g., within the viscous sublayer), deterministic models may perform well. However, as the correlation decays with wall-normal distance, especially across the buffer and logarithmic layers~\cite{wang2022observable,arranz2024informative}, these models often fail to recover energetic but weakly observable flow features. Furthermore, practical limitations such as sensor sparsity, measurement noise, and resolution constraints exacerbate this challenge, making the reconstruction problem increasingly ill-posed~\cite{guemes2021coarse,cuellar2024some}. As a result, deterministic models tend to produce overly smooth or biased estimates that suppress physically realistic variability. These limitations highlight the need for probabilistic modeling frameworks that not only generate physically consistent velocity features that are weakly correlated to wall quantities, but also quantify predictive uncertainty, enabling robust inference under imperfect or incomplete wall data.

Generative modeling, particularly when combined with Bayesian learning, offers a promising solution to these challenges. Rather than producing a single estimate, generative models learn the conditional distribution of turbulent velocity fields given wall measurements, enabling the synthesis of multiple physically plausible flow realizations that reflect both the informative and non-informative components of the flow with respect to wall data. Recent advances in generative AI, particularly diffusion-based models, have demonstrated impressive capabilities in synthesizing high-fidelity flow realizations with accurate statistics~\cite{ruhling2023dyffusion,shu2023physics,kohl2023benchmarking,li2023multi,gao2024bayesian,gao2024generative,shehataimproved,gao2025generative,dong2024data,molinaro2024generative,zhuang2025spatially,du2024conditional,fan2025neural}. Specifically, Wang and co-workers have developed a conditional neural field-based latent diffusion model, which has been successfully demonstrated in generating spatiotemporal wall-bounded turbulence in 3D that is both inhomogeneous and anisotropic, enabling applications such as synthetic inflow turbulence generation and zero-shot spatiotemporal flow reconstruction~\cite{du2024conditional,liu2024confild}. However, diffusion-based sampling remains computationally expensive due to its iterative nature and is sensitive to posterior conditioning. Recently, flow matching has emerged as a scalable alternative to diffusion, offering fast and stable sampling by learning continuous-time transport maps between simple base distributions and complex data distributions through direct supervision~\cite{lipman2022flow}.   

In this work, we propose a novel generative learning framework for reconstructing near-wall turbulent velocity fields from wall-based measurements with quantified uncertainty. Specifically, we develop a conditional generative model that integrates conditional flow matching (CFM)~\cite{lipman2022flow} with Bayesian neural operators trained using stochastic weight averaging gaussian (SWAG)~\cite{maddox2019simple}. The CFM-based model enables efficient generation of diverse, physically consistent instantaneous velocity fluctuation fields across multiple wall-normal locations, while the SWAG-based operator serves as a probabilistic forward model that maps velocity fields to wall measurements with quantified epistemic uncertainty. At inference time, our framework performs zero-shot conditional generation by iteratively refining sampled velocity fields to satisfy wall measurements -- whether sparse, noisy, or partial -- without requiring retraining, transfer learning, or fine-tuning. 

To the best of our knowledge, this work presents the first attempt to leverage flow matching for reconstructing 3D inhomogeneous and anisotropic wall-bounded turbulence. Moreover, it is the first to demonstrate zero-shot conditional flow generation with uncertainty quantification in this context. By explicitly modeling both flow variability and predictive uncertainty, our framework addresses the ill-posed nature of near-wall turbulence reconstruction and enables robust, data-consistent inference under realistic sensing conditions. The remainder of this paper is organized as follows: Section~\ref{sec:method} details the methodology of the proposed generative modeling framework. Section~\ref{sec:results} presents reconstruction results under varying sensor conditions, demonstrating the effectiveness of the proposed methods. Section~\ref{sec:discuss} compares the performance of the proposed model against the state-of-the-art baseline methods and discusses each individual component of the proposed framework. Finally, Section~\ref{sec:conclude} summarizes the key findings and outlines directions for future research.

\section{Methodology}
\label{sec:method}
\subsection{Problem formulation and analysis}
\label{subsec:prob-form}

This study tackles the challenge of reconstructing instantaneous velocity fluctuations at various wall-normal locations using only measurements acquired at the wall. The complexity of this task varies significantly across the near-wall region due to spatially heterogeneous correlations between wall measurements (e.g., wall shear stress) and turbulent structures in wall-bounded flows. Rather than focusing on a specific off-wall plane, our goal is to develop a general-purpose generative learning framework capable of synthesizing physically realistic velocity fields throughout the entire near-wall region. Critically, the proposed framework explicitly quantifies predictive uncertainty, especially in regions where the coherence between wall signals and turbulent fluctuations is weak, or where wall measurements are sparse and contaminated by noise.

\begin{figure}[t!]
    \centering
    \includegraphics[width=1.\textwidth]{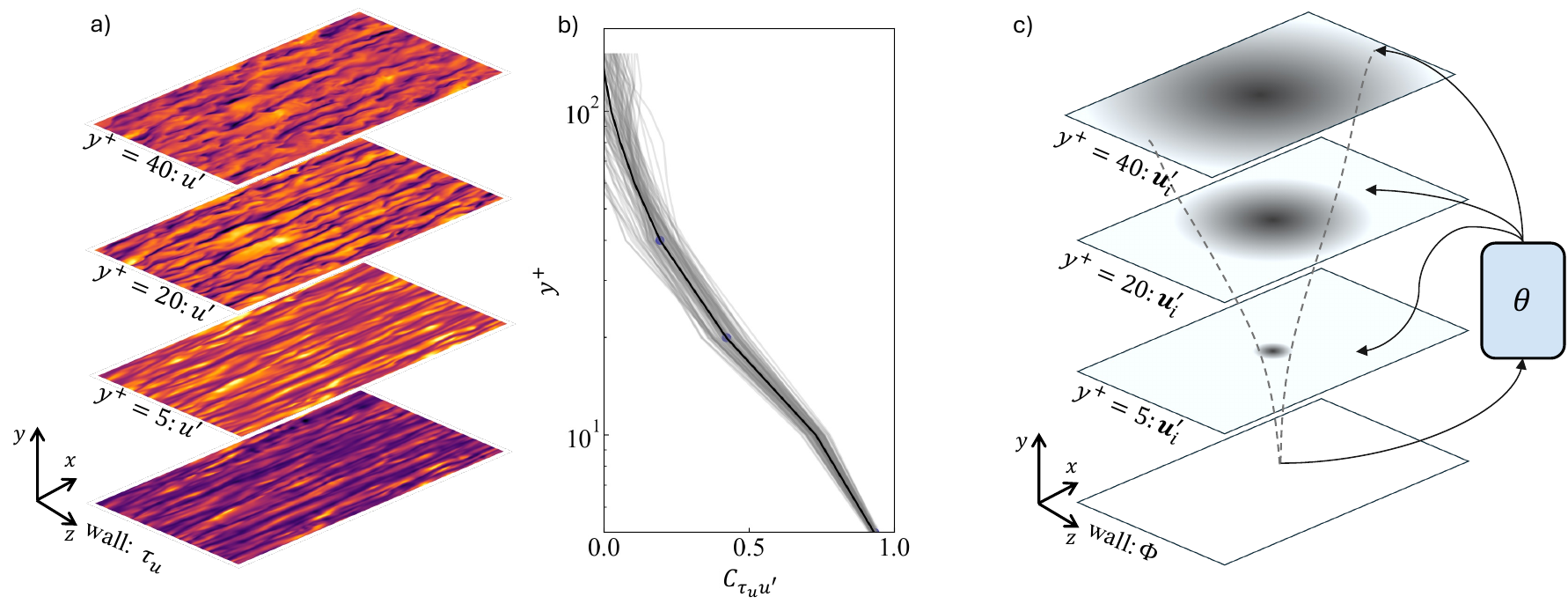}
    \caption{(a) Instantaneous streamwise velocity fluctuations $\bm{u}'$ of {the turbulent channel flow ($Re_\tau = 180$)} at three wall-normal locations ($y^+ = 5, 20, 40$) alongside the corresponding wall-shear stress field $\tau_{u}$; (b) Correlation coefficients $C_{\tau_u, {u}'}$ between wall shear stress and velocity fluctuations as a function of $y^+$; (c) Reconstruction concept: given wall input ${\bm\Phi}_{\mathrm{wall}}$, the model (parameterized by $\bm\theta$) predicts $\bm u'$ at different off-wall planes, with increasing uncertainty illustrated by blurred regions at higher $y^+$.}
    \label{fig:intro-and-motivation-schema}
\end{figure}

To better understand the inherent challenges associated with this reconstruction task, we first analyze the correlation between wall shear stress and velocity fluctuations across different wall-normal positions. As illustrated in Figure~\ref{fig:intro-and-motivation-schema}a and \ref{fig:intro-and-motivation-schema}b, the correlation strength between wall shear stress and velocity fluctuations decreases sharply as the wall-normal distance ($y^+$) increases. This reduction significantly limits the feasibility of accurately reconstructing instantaneous turbulent structures in outer regions using solely wall-based measurements~\cite{suzuki2017estimation, wang2022observable}. Nonetheless, large-scale flow structures in the outer region leave subtle yet detectable imprints on the wall through amplitude and wavelength modulation effects,  indirectly encoding useful information within wall shear stress fluctuations~\cite{arranz2024informative, mathis2009large}. Recognizing this indirect yet informative coupling motivates our generative model, which aims to reconstruct instantaneous velocity fluctuations at various off-wall locations, incorporating uncertainty quantification that reflects the progressive weakening of coherence, measurement sparsity, and observational noise.

Specifically, within the viscous sublayer, velocity fluctuations exhibit a nearly linear and direct relationship with wall measurements, characterized by a high correlation coefficient ($C_{\tau_{u} u'} \approx 1$)~\cite{wang2022observable}. Due to this strong coupling, velocity fields in this region can be reconstructed accurately with minimal uncertainty, as illustrated conceptually in Figure~\ref{fig:intro-and-motivation-schema}c. 
Moving outward into the buffer layer, however, this direct coherence rapidly decays~\cite{wang2022observable}, even though this region is critical for turbulent kinetic energy production and hosts energetic coherent structures, such as streamwise streaks and hairpin vortices~\cite{adrian2007hairpin,bae2021life}. These dynamically important structures are only indirectly captured by wall signals due to modulation effects, significantly complicating their instantaneous reconstruction. Thus, deterministic models struggle to adequately capture the variability and complexity inherent in buffer-layer fluctuations. This limitation underscores the need for probabilistic modeling approaches that explicitly quantify the uncertainty arising from weakened correlations and indirect observability.
Further away from the wall (logarithmic layer and beyond), large-scale turbulent motions, particularly low-speed streaks, persist and can extend significant into outer layer, sometimes reaching boundary-layer edge~\cite{adrian2007hairpin, encinar2019logarithmic}. Although their instantaneous correlation with wall signals is generally low, these large-scale structures indirectly influence near-wall dynamics through modulation effects, leaving distinguishable but indirect signatures on wall quantities~\cite{arranz2024informative, mathis2009large}. Despite being less dominant energetically compared to smaller-scale structures nearer to the wall, accurately capturing their statistical features remains essential to realistically reconstruct turbulence characteristics. Deterministic methods relying solely on wall measurements often significantly underestimate fluctuations in this region, as they only recover weakly correlated, low-energy large-scale motions~\cite{guastoni2021convolutional}. Recognizing this limitation, our framework not only predicts informative components directly from wall data, but also leverages generative learning to synthesize non-informative yet energetically crucial turbulence structures, thereby ensuring accurate statistical reconstructions.

\subsection{Overview of proposed generative modeling framework for near-wall turbulence}
\begin{figure}[t!]
    \centering
    \includegraphics[width=1.\textwidth]{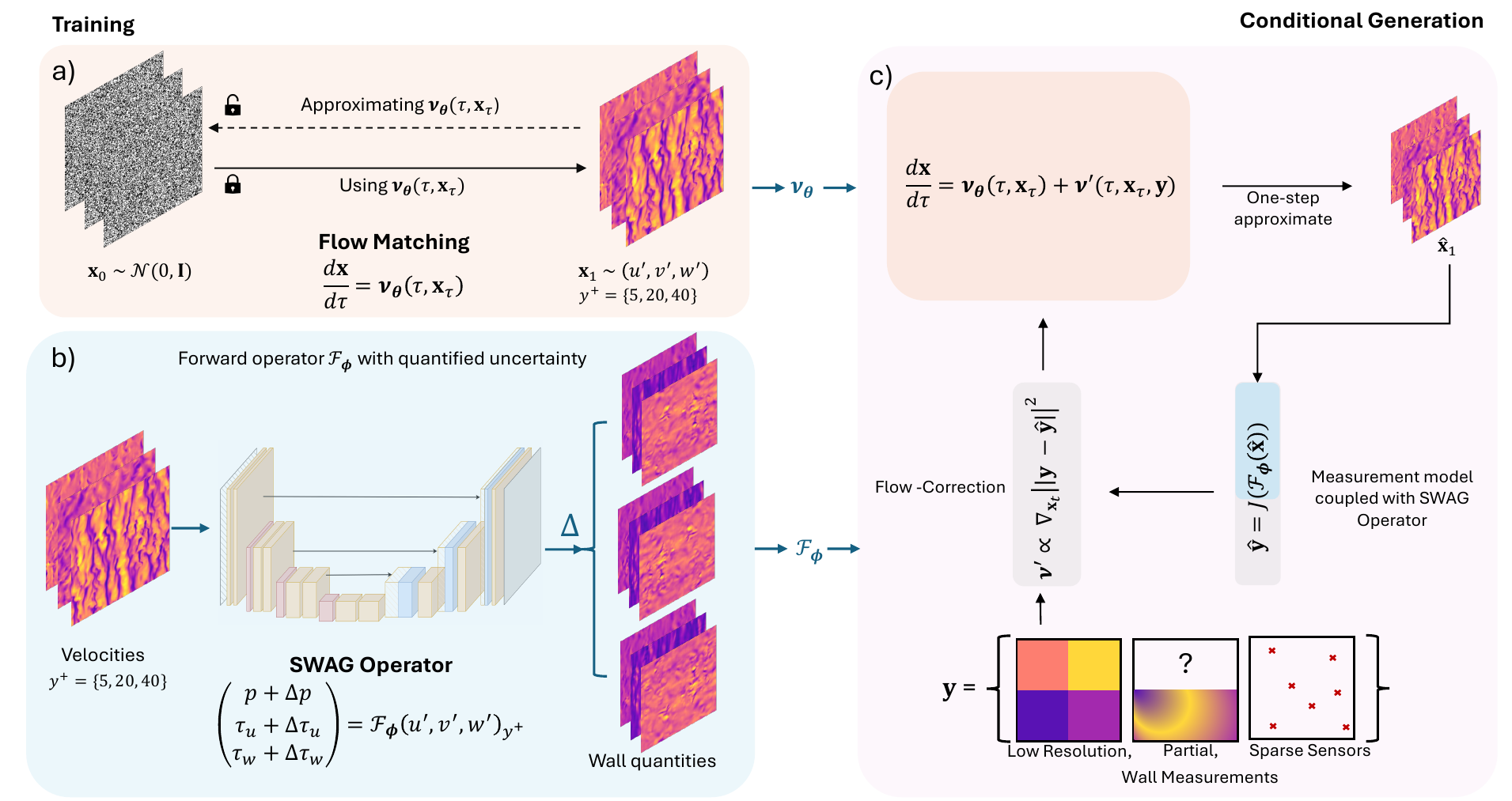}
    \caption{(a) Flow Matching based generative model for synthesizing novel instances of velocity fluctuations. (b) Forward operator with Stochastic Weight Averaging (SWAG) to quantify the epistemic uncertainty between velocity fluctuations and wall quantities. (c) Training-free conditional generation based on the predictor-corrector FM inference algorithm.}
    \label{fig:complete-methodology}
\end{figure}

Building upon our analysis of spatially heterogeneous correlations between wall measurements and turbulent structures, we propose a generative modeling framework explicitly designed to reconstruct instantaneous velocity fluctuations across multiple wall-normal locations. The core idea of the framework leverages generative learning combined with probabilistic neural operators within a Bayesian framework to effectively handle the inherent uncertainties and indirect observability in wall-bounded turbulence. Figure~\ref{fig:complete-methodology} illustrates the three key components of our framework.

In the training phase (Figure~\ref{fig:complete-methodology}a), a generative model based on flow matching~\cite{lipman2022flow} is built to generate instantaneous full-field velocity fluctuation samples at predefined wall-normal positions ($y^+ = 5, 20, 40$) by sampling from a standard Gaussian distribution, $\mathcal{N}(\bm{0}, \mathbf{I})$. Using unsupervised learning, the trained model is able to randomly synthesize novel, physically plausible instantaneous velocity fluctuation fields consistent with the statistical features observed in turbulent channel flows. The details of the flow matching training are further elaborated in Section~\ref{sub-sec:CFM}. Concurrently, as depicted in Figure~\ref{fig:complete-methodology}b, a probabilistic neural operator is designed to predict wall quantities (e.g., wall-shear stresses and pressures) from the velocity fluctuations with quantified uncertainties. This U-net based neural operator is trained probabilistically using SWAG~\cite{maddox2019simple}, a Bayesian learning technique enabling the quantification of epistemic uncertainty inherent in neural network predictions. The incorporation of the SWAG operator is critical as it allows our framework to explicitly learn the uncertainty arising from weak correlations, measurement sparsity, and noise. 

In the inference stage, depicted in Figure~\ref{fig:complete-methodology}c, our framework is able to perform conditional generation in a zero-shot manner (i.e., without re-training) using a predictor-corrector sampling approach. Specifically, given instantaneous wall measurements, the pretrained generative model synthesizes instantaneous velocity fluctuation fields that are consistent with these sensor data. This is accomplished through iterative refinement, where an initial prediction from the generative model is progressively updated using gradient-based corrections derived from discrepancies between predicted and actual wall data. The gradients are computed via the differentiable forward measurement model integrated with the probabilistic SWAG operator, thereby ensuring that the synthesized velocity fluctuations realistically reflect measurement uncertainties and the physical constraints imposed by available wall information. Further details are provided in Section~\ref{sub-sec:infer}.

In summary, our proposed generative modeling framework uniquely integrates generative learning and scalable uncertainty quantification techniques, providing robust and physically consistent reconstructions of turbulent velocity fluctuations from limited wall-based measurements.

\subsection{Flow matching for generative modeling of instantaneous velocity fluctuations}
\label{sub-sec:CFM}

To model the complex, high-dimensional distribution of instantaneous velocity fluctuations in near-wall turbulence, we adopt the recently proposed flow matching (FM) framework~\cite{lipman2022flow}, which enables tractable and scalable training of continuous normalizing flows (CNFs) by directly learning the velocity fields that transport probability mass from a simple base distribution to the target data distribution. Unlike conventional CNFs, which require density estimation via the change-of-variables formula and involve solving ordinary differential equations (ODEs) during training, FM bypasses these limitations by reframing the generative modeling as a regression task over velocity fields that define transport paths between distributions. This makes FM particularly suitable for large-scale, high-dimensional turbulent flow data. In this work, we present the first use of FM for generating inhomogeneous, anisotropic 3D turbulence.

Let $\mathbf{x}_1 \sim P_1(\mathbf{x})$ denote a sample drawn from the target data distribution, where each sample represents a snapshot of the instantaneous three-component velocity fluctuation field at a specified wall-normal location $y^+$. Specifically, $\mathbf{x}_1 = [u'(x, z, t), v'(x, z, t), w'(x, z, t)]$ are defined on a discrete $(x, z)$ grid, where $x, z$ are wall-parallel coordinates and $t$ denotes physical time. We define the base distribution $P_0(\mathbf{x}) = \mathcal{N}(\mathbf{0}, \mathbf{I})$ as a standard multivariate Gaussian in the same space as the target data. Our goal is to learn a transport velocity field $\bm{\nu}_{\bm\theta}(\tau, \mathbf{x}_\tau)$, parameterized by neural network weights $\bm\theta$, that smoothly transforms samples from $P_0$ into samples from $P_1$ along a continuous path indexed by a fictitious time variable $\tau \in [0, 1]$. The generative process is defined by the flow ODE,
\begin{equation}
\frac{d\mathbf{x}_\tau}{d\tau} = \bm{\nu}_{\bm\theta}(\tau, \mathbf{x}_\tau),
\end{equation}
where $\mathbf{x}_\tau$ denotes the sample at intermediate time $\tau$ along the transport trajectory from $P_0$ to $P_1$.

To ensure probability conservation along the flow, the transport velocity field $\bm{\nu}(\tau, \mathbf{x}_\tau)$ must satisfy the continuity equation, 
\begin{equation}
\frac{\partial P_\tau}{\partial \tau} = - \nabla \cdot (P_\tau \bm{\nu}_\tau),
\end{equation}
where $P_\tau(\mathbf{x})$ denotes the intermediate distribution at time $\tau$.
If the true velocity field $\mathbf{\nu}(\tau, \bm{x}_\tau)$, is known, we can directly minimize the following loss
\begin{equation} 
\mathcal{L}(\bm\theta) := \mathbb{E}_{\tau\sim\mathrm{U}[0,1], \mathbf{x}_\tau \sim P_\tau(\mathbf{x}_\tau)}\Big\|\bm{\nu}_{\bm\theta}(\tau, \mathbf{x}_\tau) - \bm{\nu}(\tau, \mathbf{x}_\tau)\Big\|^2.
\label{eq:fm-loss} 
\end{equation}
In practice, however, directly evaluating the true velocity $\bm{\nu}(\tau, \mathbf{x}_\tau)$ and sampling from $P_\tau(\mathbf{x}_\tau)$ is intractable. To address this, FM introduces a conditional formulation by introducing a latent variable $\mathbf{z} \sim q(\mathbf{z})$ that leads to conditional intermediate distributions $P_\tau(\mathbf{x}|\mathbf{z})$ and velocities $\bm{\nu}(\mathbf{x}_\tau|\mathbf{z})$ such that,
\begin{subequations}\label{eq:cfm}
\begin{align}
P_\tau(\mathbf{x}_\tau) &= \int P_\tau(\mathbf{x}_\tau|\mathbf{z})q(\mathbf{z})d\mathbf{z}, \\
\bm{\nu}_\tau(\tau,\mathbf{x}_\tau) &= \int \frac{\bm{\nu}_\tau(\mathbf{x}_\tau|\mathbf{z}) P_\tau(\mathbf{x}_\tau|\mathbf{z})}{P_\tau(\mathbf{x}_\tau)}q(\mathbf{z})d\mathbf{z} 
         = \mathbb{E}_{q(\mathbf{z})}\bigg[ \frac{\bm{\nu}_\tau(\mathbf{x}_\tau|\mathbf{z}) P_\tau(\mathbf{x}_\tau|\mathbf{z})}{P_t(\mathbf{x}_\tau)}\bigg].
\end{align}
\end{subequations}
Substituting Eq.~\ref{eq:fm-loss} into the FM loss and swapping the gradient and expectation under suitable regularity conditions leads to the modified loss,
\begin{equation} 
\tilde{\mathcal{L}}(\bm\theta) = \mathbb{E}_{\tau \sim \mathcal{U}[0,1], \mathbf{z} \sim q(\mathbf{z}), \mathbf{x}_\tau \sim P_\tau(\mathbf{x} | \mathbf{z})} \Big\| \bm{\nu}_{\bm\theta}(\tau, \mathbf{x}_\tau) - \bm{\nu}(\tau, \mathbf{x}_\tau | \mathbf{z}) \Big\|^2. \label{eq:cfm-loss} 
\end{equation}
Under certain assumptions on $q(\mathbf{z})$ and $P_\tau(\mathbf{x}_\tau|\mathbf{z})$~\cite{lipman2022flow}, it can be proved that the gradient of the modified loss yields the same optimization objective as the original FM loss, i.e.,
\begin{equation}
    \nabla_{\bm\theta} \mathcal{L}(\bm\theta) = \nabla_{\bm\theta} \tilde{\mathcal{L}}(\bm\theta).
\end{equation}
Therefore, the neural velocity field can be trained based on the conditional formulation of the loss function, where both the conditional intermediate distributions and velocities are available.  In practice, we use empirical samples from the data distribution as conditioning variables. Specifically, we let $q(\mathbf{z}) = \frac{1}{N} \sum_{i=1}^{N} \delta(\mathbf{z} - \mathbf{x}_1^{(i)})$, where $\{\mathbf{x}_1^{(i)}\}_{i=1}^N$ are training velocity fluctuation data. Then we have the following tractable definitions for the conditional intermediate distribution and velocity: 
\begin{subequations} 
\begin{align} 
P_\tau(\mathbf{x}_\tau | \mathbf{x}_1) &= \mathcal{N}\Big(\tau \mathbf{x}_1, \big[1 - (1 - \sigma_{\min})\tau\big]^2\mathbf{I}\Big), \\
\bm{\nu}(\tau, \mathbf{x}_\tau | \mathbf{x}_1) &= \frac{\mathbf{x}_1 - (1 - \sigma_{\min})\mathbf{x}_\tau}{1 - (1 - \sigma_{\min})\tau}, 
\end{align} 
\end{subequations} 
which corresponds to an isotropic Gaussian interpolation with linearly evolving mean and covariance from $\mathcal{N}(0, \mathbf{I})$ to $\mathcal{N}(\mathbf{x}_1, \sigma_{\min}^2 \mathbf{I})$.
Here, the mean interpolates linearly from $\mathbf{0}$ to $\mathbf{x}_1$, while the covariance linearly decays from $\mathbf{I}$ to $\sigma_{\min}^2 \mathbf{I}$. This results in a simple and tractable sampling scheme, where both the intermediate samples $\mathbf{x}_\tau$ and their associated transport velocities $\bm{\nu}(\tau, \mathbf{x}_\tau | \mathbf{x}_1)$ are available in closed form for supervised training.
Additionally, the neural velocity field $\bm{\nu}_{\bm \theta}(\tau, \mathbf{x}_\tau)$ is trained in a class-conditional way for the different wall-normal distances of $y^+=\{5,\space 20,\space 40\}$ considered in this study. The model learns a trainable neural embedding corresponding to each wall-normal distance. The details of network architectures are provided in~\ref{sec:appendix-training}.

\subsection{Training-free conditional generation guided by wall measurements}
\label{sub-sec:infer}

The generative model described in Section~\ref{sub-sec:CFM} enables efficient sampling of instantaneous velocity fluctuation fields from the learned distribution at specified wall-normal locations. While this unconditional generation captures the statistical structure of turbulent fluctuations, it does not incorporate any instance-specific observations. In practical settings, however, partial information about a specific turbulent flow realization is often available, either in the form of direct but sparse velocity measurements within the domain, or more commonly, through indirect wall-based observations such as wall shear stress or pressure signals. These measurements $\mathbf{y}$ encode valuable information about the underlying flow state and can be leveraged during inference. This motivates the need for conditional generative inference that can synthesize velocity fields consistent with available observations, while retaining uncertainty quantification over unobserved regions.

A common strategy to condition generative models is to incorporate the observation vector $\mathbf{y}$, whether wall measurements, sparse velocity probes, or both, into the sampling process. This is typically achieved by modifying the training objective to explicitly encode $\mathbf{y}$ as input, for example through input concatenation, modulation via feature-wise affine transformations, or injection via a hypernetwork $F_\phi(\mathbf{y})$~\cite{dhariwal2021diffusion,zhuang2025spatially, jacobsen2025cocogen,fu2024unveil}. However, these approaches require retraining the generative model for each new type of conditioning input and often scale poorly when the structure, modality, or spatial extent of $\mathbf{y}$ varies across instances, as is the case in sensor-limited turbulence setups.

To overcome these limitations, we introduce a \emph{training-free} conditional inference strategy that augments the unconditional flow matching model with a correction term derived from the conditioning measurements. Inspired by training-free conditional methods in diffusion models such as diffusion posterior sampling~\cite{chung2022diffusion,du2024conditional} and Bayesian classifier guidance~\cite{dhariwal2021diffusion,gao2024bayesian}, our approach adapts these ideas to the flow-matching paradigm. Specifically, we define a guided sampling procedure in which the learned transport velocity $\bm{\nu}_{\bm\theta}(\tau, \mathbf{x}_\tau)$ is augmented during inference by a correction term $\bm{\nu}'(\tau, \mathbf{x}_\tau, \mathbf{y})$ that nudges the generative trajectory toward compatibility with the measurements,
\begin{equation}
    \frac{d\mathbf{x}}{d\tau} = \bm{\nu}_{\bm\theta}(\tau, \mathbf{x}_\tau) + \bm{\nu}'(\tau, \mathbf{x}_\tau, \mathbf{y}).
\end{equation}

To derive this correction term, we first assess the discrepancy between the observations $\mathbf{y}$ and the predicted measurements $\hat{\mathbf{y}}$ derived from the current state $\mathbf{x}_\tau$, and then propagate this discrepancy back to $\mathbf{x}_\tau$. This process involves three components: (i) an efficient approximation of the terminal state $\hat{\mathbf{x}}_1$ from intermediate states $\mathbf{x}_\tau$, (ii) a forward state-to-observable operator that maps the generated state to measurement space, and (iii) a differentiable loss to quantify mismatch between predicted and observed measurements. 
For the first step, we adopt a one-step linear approximation of the terminal state:
\begin{equation}
\label{eqn:one-step-approx}
    \hat{\mathbf{x}}_{1|\tau} = \mathbf{x}_\tau + (1-\tau)\bm{\nu}_{\bm\theta}(\tau, \mathbf{x}_\tau),
\end{equation}
which estimates the endpoint of the flow by extrapolating along the learned transport velocity. This is computationally efficient and justified by the nearly linear nature of the flow trajectory learned through conditional flow matching (Section~\ref{sub-sec:CFM}).
Next, we define the forward operator $\mathcal{F}$ that maps the predicted state $\hat{\mathbf{x}}_{1|\tau}$ to the observable domain. Depending on the context, $\mathcal{F}$ may represent a fixed linear mapping (e.g., extracting velocity values at sensor locations) or a learned neural operator that maps the off-wall flow states to indirect measurements (e.g., wall measurements). Considering measurement noises $\bm\epsilon$, this operator is defined as,
\begin{align}
    \hat{\mathbf{y}} &= \mathcal{F}(\hat{\mathbf{x}}_{1|\tau}) + \bm{\epsilon}; \qquad \bm{\epsilon} \sim \mathcal{N}(0, \Sigma_e)
\end{align}
where $\Sigma_e = \sigma_e^2 \mathbf{I}$ captures the assumed level of aleatoric uncertainty.

The discrepancy $\mathscr{D}(\hat{\mathbf{y}}, \mathbf{y})$ between predicted and observed measurements is quantified via a loss function, typically the mean squared error. This loss is then differentiated with respect to $\mathbf{x}_\tau$, yielding a gradient that points in the direction of greater alignment between the generated sample and the observations. We define the correction term as: 
\begin{equation}
\label{eqn:condition-term}
    \bm{\nu}'(\tau, \mathbf{x}_\tau, \mathbf{y}) = -b ||\bm{\nu}_{\bm \theta}(\tau, \mathbf{x}_\tau)|| \frac{\nabla_{\mathbf{x}_\tau} \mathscr{D}(\hat{\mathbf{y}}, \mathbf{y})}{||\nabla_{\mathbf{x}_\tau} \mathscr{D}(\hat{\mathbf{y}}, \mathbf{y})||},
\end{equation}
where $b$ is a scalar that controls the guidance strength. The correction direction is aligned with the normalized loss gradient, while the magnitude is scaled to match the norm of the current transport velocity, ensuring that the guidance term and learned flow are balanced in strength.

Importantly, this conditional sampling strategy is agnostic to the nature and modality of the measurement vector $\mathbf{y}$. While in this paper we focus on wall-based observations, the framework can also accommodate sparse velocity probes, inpainting of partially known flow fields, or any other differentiable observation models. In fact, the proposed corrector can be viewed as a variational approximation to the gradient of the log-posterior distribution $\nabla_{\mathbf{x}} \log P(\mathbf{y} | \mathbf{x})$, and thus relates conceptually to likelihood-based guidance in diffusion posterior sampling (DPS)~\cite{chung2022diffusion,gao2024bayesian,du2024conditional}. We demonstrate the flexibility and effectiveness of this approach through reconstruction experiments with wall-based and in-domain conditional signals in Section~\ref{sec:results}, and further illustrate its applicability to sparse inpainting tasks in Section~\ref{sec:train-free-method-inpainting}.

\subsection{Learning forward operators from velocity fields to wall measurements with uncertainty}
\label{sub-sec:SWAG}

As established in Section~\ref{sub-sec:infer}, our conditional generation framework requires evaluating the discrepancy between synthesized velocity fields and observed flow measurements, which depends on a forward operator $\mathcal{F}$ that maps the generated velocity fluctuation field $\bm{u}'$ to predicted measurements $\hat{\bm{y}} = \mathcal{F}(\bm{u}')$. In general, $\bm{y}$ may consist of direct velocity samples at sparse spatial locations or more commonly, wall-based signals such as pressure or shear stress, which are nonlinearly related to the off-wall velocity field. While mapping from velocity to sparse velocity measurements can often be expressed as a linear masking operation, the velocity-to-wall mapping is inherently nonlinear and spatially nonlocal, reflecting the complex modulation mechanisms by which outer-layer motions influence near-wall quantities~\cite{mathis2009large,arranz2024informative}.

To enable training-free conditional inference using wall measurements, we need a forward observation operator, which should be differentiable and capable of quantifying predictive uncertainty, especially where the velocity-wall correlation is weak. This is particularly important because in regions such as the logarithmic layer, wall signals encode only indirect and noisy information about the flow. The forward operator must therefore propagate not only observations but also their epistemic confidence into the sampling process. To this end, we design a probabilistic forward operator $\mathcal{F}_{\bm\phi}$, built as a convolutional U-Net, that learns to predict wall measurements $\bm{\Phi}_\mathrm{wall} = [p, \tau_u, \tau_w]$ from input 3D velocity fluctuations $\bm{u}' = [u', v', w']$. Crucially, to enable robust uncertainty-aware conditioning during inference, we train $\mathcal{F}_{\bm\phi}$ using the SWAG framework~\cite{maddox2019simple}, which approximates the posterior distribution over the network parameters and propagates uncertainty through the prediction. Specifically, we treat the operator's parameters $\bm{\phi}$ as a distribution rather than a point estimate, modeling their posterior as a multivariate Gaussian with a diagonal plus low-rank covariance structure,
\begin{equation}
    \bm{\phi} \sim \mathcal{N}\left(\bm{\phi}_{\mathrm{SWA}}, \frac{1}{2}\Big(\Sigma_\mathrm{diag} + \Sigma_\mathrm{low-rank}\Big)\right),
\end{equation}
where $\bm{\phi}_{\mathrm{SWA}}$ is the running average of network weights collected over stochastic gradient decent (SGD) trajectories after a burn-in period. The covariance matrices are computed as: 
\begin{subequations} 
\begin{align} 
\Sigma_{\mathrm{diag}} = \mathrm{diag}(\overline{\bm\phi^2} - \bm{\phi}_{\mathrm{SWA}}^2), \\ 
\Sigma_{\mathrm{low-rank}} = \frac{1}{K - 1} \hat{H} \hat{H}^\top, 
\end{align}
\end{subequations}
where $\Sigma_\mathrm{low-rank}$ is a low-rank approximation of the covariance using the last $K$ epochs, with $\hat{H} = [\bm{\phi}_{N-K+1} - \bm{\phi}_{\mathrm{SWA}}, \dots, \bm{\phi}_N - \bm{\phi}_{\mathrm{SWA}}]$. 

During inference, the weights of the trained function $\mathcal{F}_{\bm{\phi}^*}$ are sampled from this posterior,
\begin{equation}
    \bm{\phi}^* = \bm{\phi}_\mathrm{SWA} + \frac{1}{\sqrt{2}}\Sigma_{\mathrm{diag}}^{\frac{1}{2}}\bm{z}_1 + \frac{1}{\sqrt{2(K-1)}}\hat{H}\bm{z}_2,
\end{equation}
where $\bm{z}_1$ and $\bm{z}_2$ are independently sampled from a multivariate standard normal distribution $\mathcal{N}(\bm{0}, \mathbf{I})$. With the weigh ensemble of size $m$, the mean and epistemic uncertainty of wall quantity predictions are computed as,
\begin{subequations} 
\begin{align}
    \label{eqn:forward_model_epistemic}
    \overline{\bm\Phi}_{\mathrm{wall}} &= \frac{1}{m}\sum_{i=1}^m \mathcal{F}_{\bm{\phi}^*_i}(\bm{u}'_i)\\
    \bm{\sigma}_{\bm{\Phi}_{\mathrm{wall}}} &= \sqrt{\frac{1}{m}\sum_{i=1}^m \left( \mathcal{F}_{\bm{\phi}^*_i}(\bm{u}'_i) - \overline{\bm\Phi}_{\mathrm{wall}}\right)^2}
\end{align}
\end{subequations}

Moreover, we adopt a patchwise training strategy: the operator is trained on small spatial subdomains of size $n_x \times n_z = 32$, much smaller than the full domain size $N_x \times N_z$. This localized learning strategy is inspired by the inherently local structure of near-wall turbulence, which increases the diversity of training examples, reduces memory usage, and improves generalization. Importantly, due to the fully convolutional nature of the U-Net, the trained model can be applied to full-domain velocity inputs during inference, leveraging translation invariance of convolution operations. 

In summary, our patch-trained, SWAG-based forward operator $\mathcal{F}_{\bm\phi}$ provides a scalable, interpretable, and uncertainty-aware mechanism to connect full-field turbulence predictions to surface measurements, and is essential to enabling flexible and rigorous conditional flow reconstruction in the proposed generative framework.

\section{Results}
\label{sec:results}

In this section, we evaluate the performance of the proposed FM framework on reconstructing instantaneous velocity fluctuation fields from wall-based measurements in wall-bounded turbulent flow. The primary goal is to assess how well the model can synthesize physically realistic velocity fluctuations under varying wall measurement conditions, and to analyze the effect of wall-flow coherence on reconstruction fidelity and uncertainty quantification. The framework is tested on canonical turbulent channel flow at a friction Reynolds number of $Re_\tau = 180$. During training, the generative model is exposed to unconditional samples at wall-normal positions $y^+ = {5, 20, 40}$, without access to any measurement-based conditioning information. Conditional inference is performed entirely at test time using the training-free guidance strategy described in Section~\ref{sub-sec:infer}, which allows flexible assimilation of wall-based measurements across different levels of sparsity and noise.

\subsection{DNS case setup, data generation, and evaluation metrics}

The dataset used for training and evaluation is generated via direct numerical simulation (DNS) of incompressible channel flow at $Re_\tau = 180$. The computational domain is defined as $[L_x, L_y, L_z] = [4\pi, 2, 2\pi]$ in the streamwise ($x$), wall-normal ($y$), and spanwise ($z$) directions, respectively. The simulation employs a uniform grid resolution of $[N_x, N_y, N_z] = [320, 400, 200]$ with periodic boundary conditions in $x$ and $z$, and no-slip boundary conditions at the walls. The flow is driven by a constant mean pressure gradient in the streamwise direction. Snapshots of both the velocity field and wall quantities are recorded at intervals of $\Delta T^+ = 0.4$ in viscous time units.

To construct the dataset, statistical symmetry about the channel centerline is exploited by reflecting samples from the upper and lower halves of the domain, yielding a total of $43.8k$ snapshots. These snapshots span approximately 60 flow-through times. Each data sample comprises the velocity fluctuations $\bm{u}'_i = [u', v', w']$ at a given wall-normal location $y^+$ and the corresponding wall quantities $\bm{\Phi}_{\mathrm{wall}} = [p, \tau_u, \tau_w]$. The dataset is denoted as $\mathcal{D} = { (\bm{\Phi}_{\mathrm{wall}}, \bm{u}'_i)^j }_{j=1}^{N}$, where $j$ indexes the wall-normal planes and $N$ is the total sample size.

For training the generative flow matching model, we use the full $\mathcal{D}$ but only the velocity fluctuation fields $\bm{u}'_i$, since the training is entirely unsupervised. The forward operator $\mathcal{F}_{\bm\phi}$, which maps $\bm{u}'_i$ to $\bm{\Phi}_{\mathrm{wall}}$ with uncertainty quantification, is trained using approximately $9,000$ velocity-wall data pairs randomly drawn from $\mathcal{D}$. For evaluation, an entirely separate set of $500$ uncorrelated samples is generated with a larger sampling interval to ensure statistical independence from the training data. This evaluation set corresponds to approximately $7$ flow-through times. Such strict  separation allows us to benchmark generalization performance in fully unseen flow states. A summary of the simulation setup and dataset statistics is provided in Table~\ref{tab:dataset for 3D channel}.
\begin{table}[H]
    \centering
    \small
    \caption{Dataset of channel flow for training and testing the model}
    \begin{tabular}{c c c c c c c}
    \toprule
        Domain($L_x \times L_y \times L_z$)  & Cell($N_x \times N_y \times N_z$) &  ($\Delta T^+, \space N$)$_{\bm \theta}$ &  ($\Delta T^+, \space N$)$_{\bm \phi}$  &  $(\Delta T^+, \space N)_\mathrm{test}$\\ \midrule
        $4\pi \times 2 \times 2 \pi$ & $320 \times 400 \times 200$ & $0.4, \space 43.8k$ & $2.0, \space 9k$ & $4.0, \space 500$\\ 
    \bottomrule
    \end{tabular}
    \label{tab:dataset for 3D channel}
\end{table}

To demonstrate the effectiveness of the proposed framework, we evaluate its conditional generation capabilities under various wall measurement scenarios, including fully observed, partially masked, and spatially downsampled sensor configurations. Both qualitative and quantitative analyses are conducted to assess how well the generated velocity fluctuation fields $\bm{u}'_i = [u', v', w']$ agree with the ground truth flow and available wall-based observations. As qualitative assessment, we begin by visualizing representative conditional samples generated under different wall measurement settings, allowing us to inspect the model's ability to synthesize flow realizations that preserve the physical structure and intermittency of near-wall turbulence while adapting to available sensor data. For quantitative evaluation, we compute the ensemble mean (EM) and ensemble spread (ES) across multiple conditional samples. For example, the mean and standard deviation of $u'$ are defined as:
\begin{equation}                 
\label{eqn:Ensemble_Mean}
    \begin{aligned}
        EM(u') &= \frac{1}{N_\mathrm{ens}}\sum^{N_\mathrm{ens}}_{i=1} {u'}_{gen,i},\\
        ES(u') &= \sqrt\frac{{\sum^{N_\mathrm{ens}}_{i=1}|{u'}_{gen,i}  -EM(u')|}}{N_\mathrm{ens}}
    \end{aligned}
\end{equation}
where $N_\mathrm{ens}$ denotes the number of generated samples. Analogous definitions are used for $v'$ and $w'$ components. We further assess the accuracy of the ensemble mean by comparing it to the ground truth DNS field using the Pearson correlation coefficient $r$. After flattening the $x$–$z$ plane into a 1D vector, the correlation for $u'$ is calculated as follows:
\begin{equation}
    r = \frac{\sum_{i}^{N_x \times N_z} \left({u'}_{gt, i} - \overline{{u'}_{gt}}\right) \left(EM({u'}_{gen, i}) - \overline{EM({u'}_{gen})}\right)}{\sqrt{\sum_{i}^{N_x \times N_z} \left({u'}_{gt, i} - \overline{{u'}_{gt}}\right)^2 \sum_{i}^{N_x \times N_z} \left(EM({u'}_{gen, i}) - \overline{EM({u'}_{gen})}\right)^2}},
\end{equation}
where $\overline{(\cdot)}$ denotes spatial mean over the $x$–$z$ plane. To quantify how well each generated sample match the given wall measurements, we use the normalized pointwise $L_2$ error $\Delta_{\mathbf{y}}(x,z)$ as proposed in~\cite{li2023multi,du2024conditional}, 
\begin{equation} \label{eqn:Delta_y}
    \Delta_{\mathbf{y}}(x, z) = \frac{1}{E_{\mathbf{y}}}\left|\mathbf{y}_{gen}(x,z) - \mathbf{y}_{gt}(x,z)\right|^2,
\end{equation}
where the normalization factor $E_{\mathbf{y}}$ accounts for the spatial variability in both the generated and ground-truth wall measurements, and is given by $E_{\mathbf{y}} = \sigma^{gen}_{\mathbf{y}} \times \sigma^{gt}_{\mathbf{y}}$, where $\sigma^{gen}_{\mathbf{y}}$ denotes the ensemble-averaged spatial standard deviation of the generated wall measurements:
\begin{equation}
\sigma^{gen}_{\mathbf{y}} = \left\langle \frac{1}{A} \int |\mathbf{y}_{gen}|^2(x, z)\space dxdz\right\rangle^{\frac{1}{2}}.
\end{equation}
and $\sigma^{gt}_{\mathbf{y}}$ is defined analogously using the ground truth measurements $\mathbf{y}_{\mathrm{gt}}$. 

Finally, to further assess the physical realism of the generated samples, we perform a statistical turbulence analysis on an ensemble of 500 conditional realizations. Specifically, we evaluate the pre-multiplied two-dimensional energy spectra, $E_{u_i'u_i'}$. This quantity is a standard diagnostic for characterizing the energetic and structural fidelity of turbulent flows and are compared directly against the corresponding statistics computed from the DNS reference dataset. The pre-multiplied energy spectra are defined as, 
\begin{equation}
    E_{u_i'u_i'}(k_x, k_z, y) = k_x k_z \Big\langle  \hat{u}_i (k_x, k_z, y) \hat{u}_i^* (kx, kz, y) \Big\rangle
\label{eq:two_dim_spectra}
\end{equation}
where $k_x$ and $k_z$ are the streamwise and spanwise wavenumbers, respectively; $i = 1, 2, 3$ corresponds to the streamwise, wall-normal, and spanwise components of the velocity fluctuations; $\hat{u}_i$ denotes the Fourier transform of $u_i'$ in the horizontal plane, and $(\cdot)^*$ is the complex conjugate. The angle brackets $\langle \cdot \rangle$ indicate averaging over both spatial directions and ensemble members.

For conciseness, this section presents results primarily at the buffer-layer height of $y^+ = 20$, where the reconstruction task is most challenging due to reduced wall-flow coherence. Additional results at $y^+ = 5$ and $y^+ = 40$ are included in~\ref{sec:appendix-extra-res}.

\subsection{Off-wall velocity generation conditioned on fully observed wall measurements}
We begin our evaluation with an idealized scenario in which all wall quantities, i.e., streamwise and spanwise shear stress ($\tau_u$, $\tau_w$) and pressure ($p$), are available across the entire surface. Although full-resolution wall measurements are rarely available in practice, this setting serves as a controlled benchmark to assess our generative model's capacity to reconstruct physically realistic velocity fluctuations when provided with complete boundary information. It also allows us to examine how well the learned conditional flow matching framework captures the nonlinear, multiscale mapping from wall data to off-wall turbulent structures.

\begin{figure}[t!]
    \centering
    \includegraphics[width=1.0\textwidth]{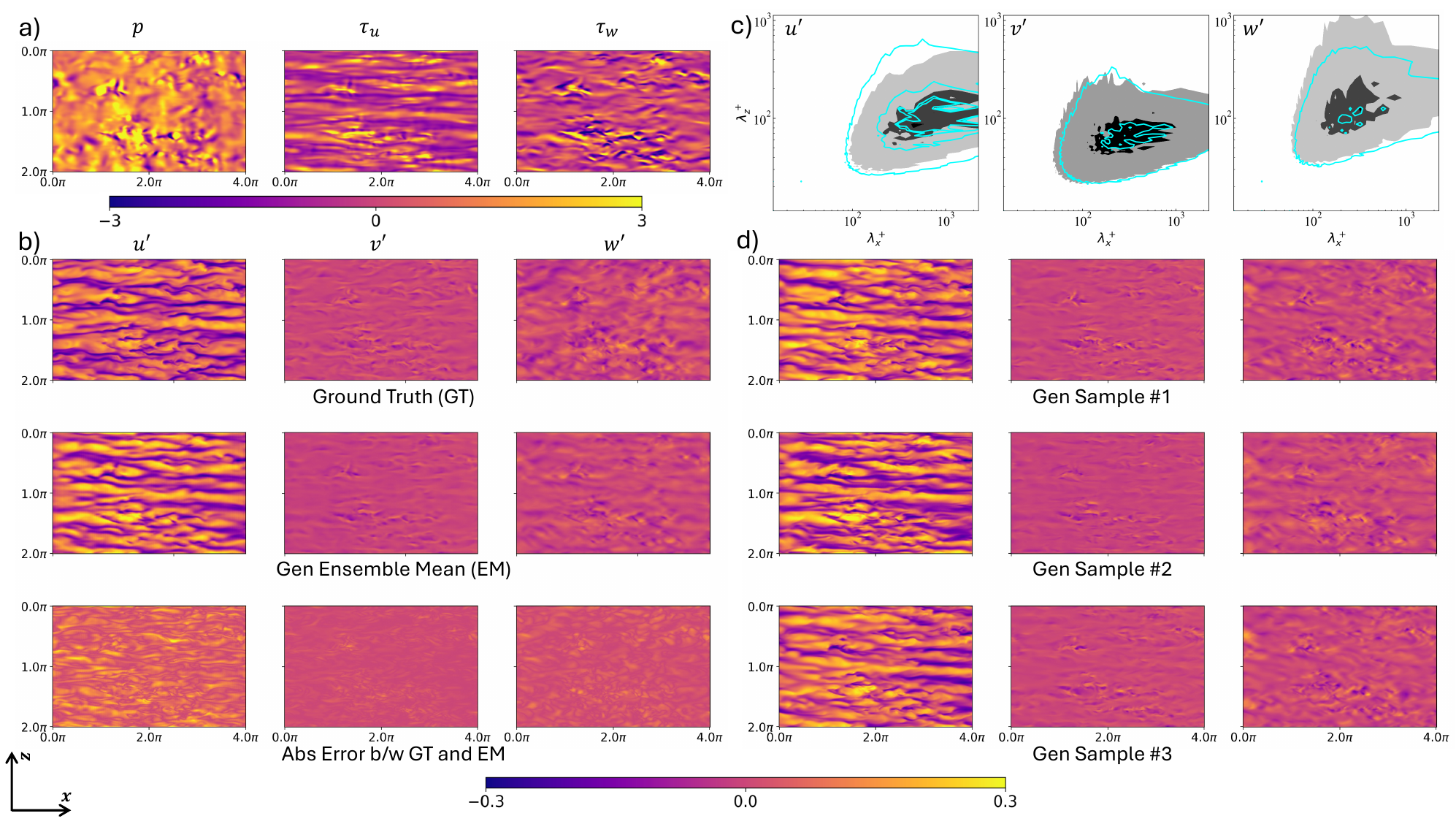}
    \caption{(a) An example of fully observed wall measurements $\bm{\Phi}_{\mathrm{wall}} = [p, \tau_u, \tau_w]$ used as the condition for generating corresponding velocity fluctuations $\bm{u}'_i$. (b) Comparison between the ground truth velocity fluctuations (top row), the ensemble mean of $50$ conditionally generated samples (middle row), and the absolute error between the ensemble mean and ground truth (bottom row), for all three velocity components. (c) Pre-multiplied two-dimensional energy spectra of the generated samples (\protect\tikz[baseline=-0.5ex]\protect\draw [cyan, thick] (0,0) -- (0.5,0); lines) versus ground truth (\protect\tikz \protect\fill[gray!40] (0,0) rectangle (0.5,0.25); contours), computed from $500$ different test cases. Streamwise and spanwise wavelengths $\lambda_x^+$ and $\lambda_z^+$ are normalized by wall units; contours indicate $10\%$, $50\%$, and $90\%$ of the maximum ground-truth energy. (d) Three representative samples from the ensemble, illustrating the diversity of generated flow realizations consistent with the wall measurements in (a).}
    \label{fig:full_contour_and_stats_y20}
\end{figure}

Figure~\ref{fig:full_contour_and_stats_y20} presents the results under this setting, where panel~(a) displays one example of instantaneous wall measurements (i.e., wall shear stress and wall pressure fields) from one of the $500$ test cases. Given this conditioning input, our model generates an ensemble of $N_{\mathrm{ens}} = 50$ velocity fluctuation fields at $y^+ = 20$. The ensemble mean of these samples, $EM(\bm{u}'_i)$, is shown in the middle row of panel~(b), while the top row presents the ground truth DNS field. Qualitatively, the ensemble mean closely resembles the ground truth across all velocity components. For $u'$, the large-scale streamwise streaks are clearly recovered with accurate orientation and spatial coherence. Similarly, high-amplitude fluctuations in $v'$ and $w'$ are faithfully reconstructed in both location and intensity. The bottom row in panel~(b) displays the pointwise absolute error between the ensemble mean and the ground truth. Errors are concentrated in fine-scale regions and interstitial zones where flow structures are weakly correlated with wall data, consistent with expected limitations of wall-based observability.
To further illustrate the diversity and realism of individual predictions, panel~(d) displays three representative samples from the ensemble conditioned on the wall input in panel~(a). These realizations exhibit certain spatial variability and small-scale structures while remaining consistent with the wall constraints. Compared to the ensemble mean, which smooths out fine-scale variations, these samples reflect the stochastic nature of the conditional velocity distribution, highlighting the model's ability to capture multimodal uncertainty and preserve the intermittency of turbulent structures. This is a key strength of the proposed framework, enabling generation of multiple plausible flow states rather than a single deterministic estimate.

To validate the statistical accuracy of the generated samples, we compute the pre-multiplied two-dimensional energy spectra $E_{u_i'u_i'}(\lambda_x^+, \lambda_z^+)$ for all 500 test cases and compare them with DNS reference statistics in Panel~(c). Across all velocity components, the spectra of generated samples closely match the true spectra in both shape and magnitude. In particular, the dominant energetic scales in $u'$ and the anisotropic energy distributions in $v'$ and $w'$ are well preserved. This strong agreement confirms that the model not only reconstructs individual samples with realistic structure but also maintains consistency with the underlying energy distribution of wall-bounded turbulence.

\subsection{Off-wall velocity generation conditioned on sparse wall measurements}
\begin{figure}[t!]
    \centering
    \includegraphics[width=1.0\textwidth]{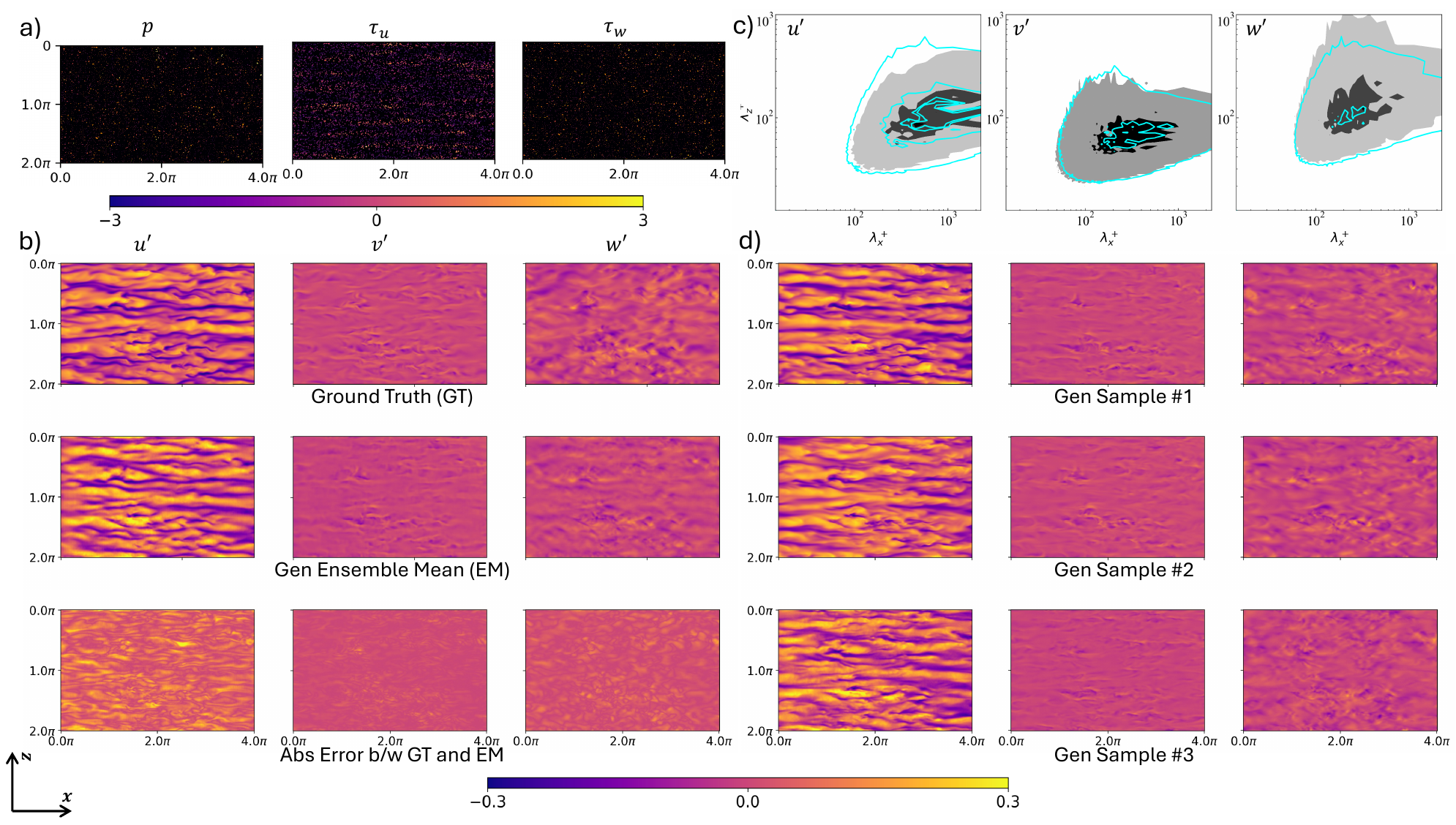}
    \caption{(a) An example of sparse wall measurements ($10\%$ data availability) ${\bm\Phi}_\mathrm{wall}$ used as the condition for generating corresponding velocity fluctuations $\bm{u}'_i$. (b) Comparison between the ground truth velocity fluctuations (top row), ensemble mean of $50$ velocity fluctuations samples generated using proposed method (middle row), and the absolute error between the ground truth and ensemble mean velocity fluctuations (bottom row). (c) Pre-multiplied two-dimensional energy spectra of the generated samples (\protect\tikz[baseline=-0.5ex]\protect\draw [cyan, thick] (0,0) -- (0.5,0); lines) versus ground truth (\protect\tikz \protect\fill[gray!40] (0,0) rectangle (0.5,0.25); contours), computed from $500$ different test cases. Streamwise and spanwise wavelengths $\lambda_x^+$ and $\lambda_z^+$ are normalized by wall units; contours indicate $10\%$, $50\%$, and $90\%$ of the maximum ground-truth energy. (d) Three representative samples from the ensemble, illustrating the diversity of generated flow realizations consistent with the sparse wall measurements in (a).}
    \label{fig:sparse10_contour_and_stats_y20}
\end{figure}

We now consider a more practical and challenging scenario where the model is conditioned on only a sparse subset of the wall measurements. Specifically, we evaluate the reconstruction performance when just $10\%$ of the full-resolution wall data $\bm{\Phi}_{\mathrm{wall}}$ is available. This setup emulates practical sensing constraints in experiments, where sensor coverage is often sparse and nonuniform. The results, summarized in Figure~\ref{fig:sparse10_contour_and_stats_y20}, follow the same structure as in the fully observed case. 
Panel (a) shows an example of the sparse wall measurements, constructed by randomly masking $90\%$ of the wall data. The effective measurement operator is defined as $\mathcal{F}(\mathbf{x}) = \mathcal{J}(\mathcal{F}_{\bm\phi}(\mathbf{x}))$, combining the learned forward operator $\mathcal{F}_{\bm\phi}$ with a binary masking operator $\mathcal{J}$ that selects the observed entries. While the conditioning data contains substantially less information than the full-measurement case, our model still performs well, thanks to the structured prior learned during training.

Figure~\ref{fig:sparse10_contour_and_stats_y20}(b) displays the reconstruction results at $y^+ = 20$. The ensemble mean of the conditionally generated velocity fluctuation samples (middle row) remains in good agreement with the ground truth (top row), particularly for the streamwise component $u'$, where large-scale streaks are reasonably reconstructed. Compared to the full-data case, the spatial error (bottom row) is slightly elevated and more spatially dispersed, especially in $v'$ and $w'$. These differences are due to the increased epistemic uncertainty introduced by reduced wall-flow observability, particularly for velocity components less correlated with wall signals. This is further evidenced by the greater diversity observed in the conditionally generated samples shown in Figure~\ref{fig:sparse10_contour_and_stats_y20}(d). Finally, we assess the statistical fidelity of the generated velocity fields in Figure~\ref{fig:sparse10_contour_and_stats_y20}(c), which compares the pre-multiplied two-dimensional energy spectra of the generated and DNS velocity fluctuations for all $500$ test cases. While minor spectral attenuation is observed, particularly at smaller wavelengths where uncertainty is highest, the generated spectra remain in close agreement with the ground truth. This confirms that, even under sparse sensor deployment, the model maintains accuracy in the dominant energy-carrying modes, and preserves the global structural statistics of the flow.

\subsection{Inference performance and uncertainty quantification analysis}

We now examine how the predictive performance and associated uncertainty of the proposed framework vary across different wall-normal locations and levels of wall measurement sparsity. This analysis provides insight into the model's robustness to decreasing wall-flow coherence and diminishing sensor coverage, two key challenges in real-world turbulence reconstruction. Specifically, we assess the accuracy of the conditional ensemble predictions and quantify the epistemic uncertainty arising from both physical observability limitations and the sparsity of wall data for conditioning.

\begin{figure}[t!]
    \centering
    \includegraphics[width=1.0\textwidth]{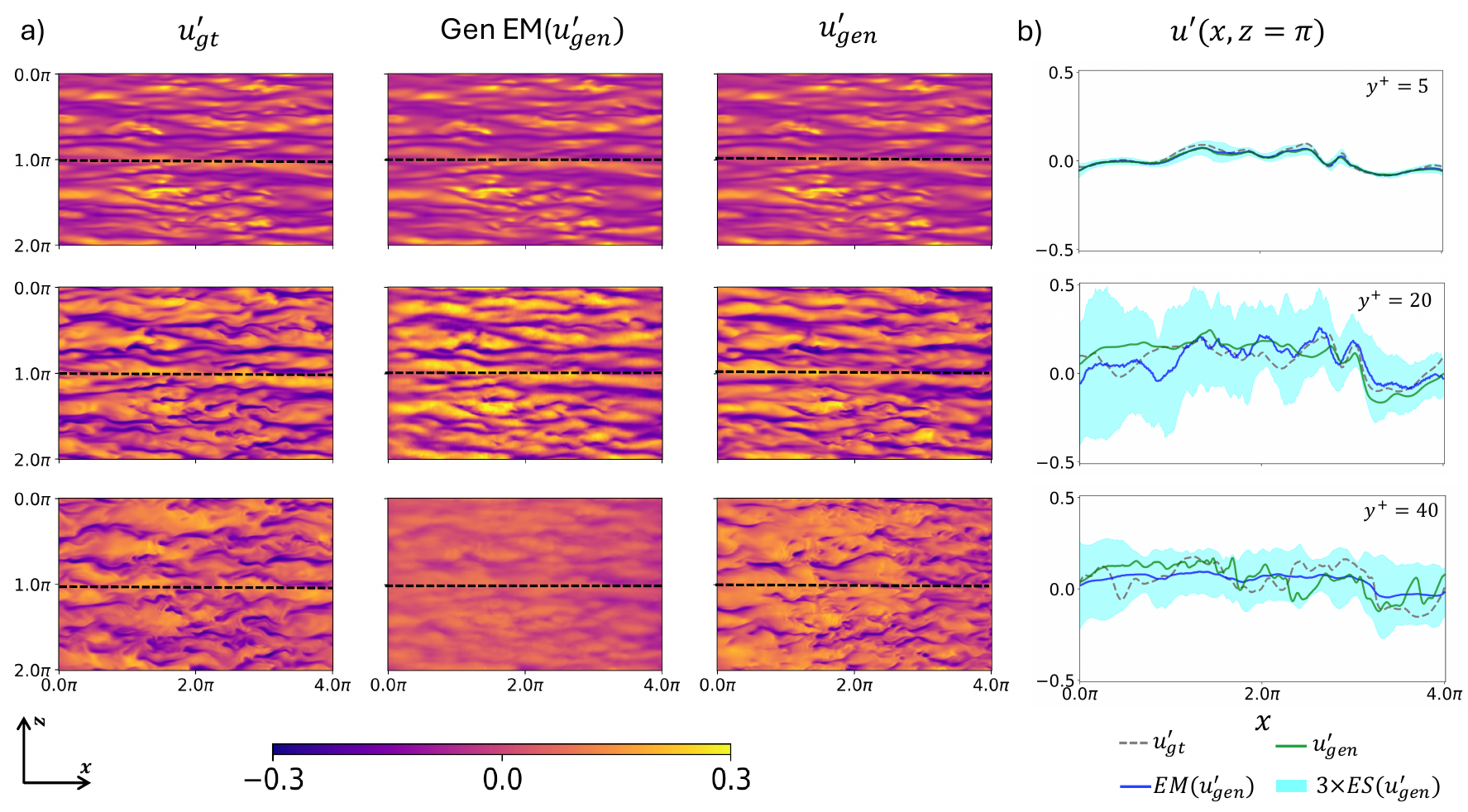}
    \caption{(a) Comparison of streamwise velocity fluctuation $u'$ contours at $y^+=5$  (1st row) $y^+=20$ (2nd row), and $y^+=40$ (3rd row), conditioned on sparse wall measurements ($10\%$ data availability). First column shows the ground truth, the second column is the ensemble mean of conditionally generated $N_{ens} =50$ samples, and the third column displays one representative conditional samples. (b) Uncertainty quantification (\protect\tikz \protect\fill[cyan!40] (0,0) rectangle (0.5,0.25); contour) of $u'$ at $y^+=5, \space 20, \space 40$ along $z=1.0\pi$, with the ground truth (\protect\tikz[baseline=-0.5ex]\protect\draw [gray, thick,dashed] (0,0) -- (0.5,0); line), ensemble mean (\protect\tikz[baseline=-0.5ex]\protect\draw [blue, thick] (0,0) -- (0.5,0); line), and one sample (\protect\tikz[baseline=-0.5ex]\protect\draw [green, thick] (0,0) -- (0.5,0); line).}
    \label{fig:sparse10_yplus_contours_lineplots}
\end{figure}

Figure~\ref{fig:sparse10_yplus_contours_lineplots} addresses the first question by comparing conditional generation results at $y^+ = 5$, $20$, and $40$ using the same $10\%$ sparse sensor configuration previously introduced in Figure~\ref{fig:sparse10_contour_and_stats_y20}(a). To maintain clarity, we focus on the streamwise velocity component $u'$; results for $v'$ and $w'$ follow similar trends and are provided in~\ref{sec:appendix-extra-res}.
Panel (a) shows the ground truth $u'$ (first column), the ensemble mean $\mathrm{EM}(u')$ computed from 50 generated samples (second column), and one representative sample $u'_{\text{gen}}$ from the ensemble (third column). At $y^+=5$, the ensemble mean closely matches the ground truth, with the representative sample exhibiting realistic variability. As $y^+$ increases, the quality of reconstruction gradually degrades. While large-scale streaks are still visible at $y^+=20$, finer structures begin to deviate from the DNS reference. At $y^+=40$, the ensemble mean becomes increasingly smooth and underestimates the true amplitude of fluctuations, an expected result due to the diminishing coherence between wall measurements and flow farther from the wall. However, each individual sample still maintains physically plausible small structures at all $y^+$ thanks to the generative module, even when ensemble mean smooths out.  

To quantify uncertainty and assess the variability of predictions, panel~(b) shows 1D profiles of $u'$ along the streamwise direction at fixed spanwise location $z = \pi$, overlaid with $3\times\mathrm{ES}(u')$ confidence intervals. The shaded bands expand with increasing $y^+$, reflecting greater epistemic uncertainty. Notably, the widest band occurs at $y^+=20$, coinciding with the peak in turbulence intensity $u'_{\text{rms}}$. This suggests that uncertainty is governed not just by wall proximity but also by intrinsic turbulence characteristics, particularly the chaotic amplification at buffer-layer heights. These observations are consistent with physical trends reported in prior works~\cite{guastoni2021convolutional, balasubramanian2023predicting, cuellar2024three, cuellar2024some}, and they validate the model's ability to capture non-monotonic patterns in prediction uncertainty.

\begin{figure}[ht!]
    \centering
    \includegraphics[width=1.\textwidth]{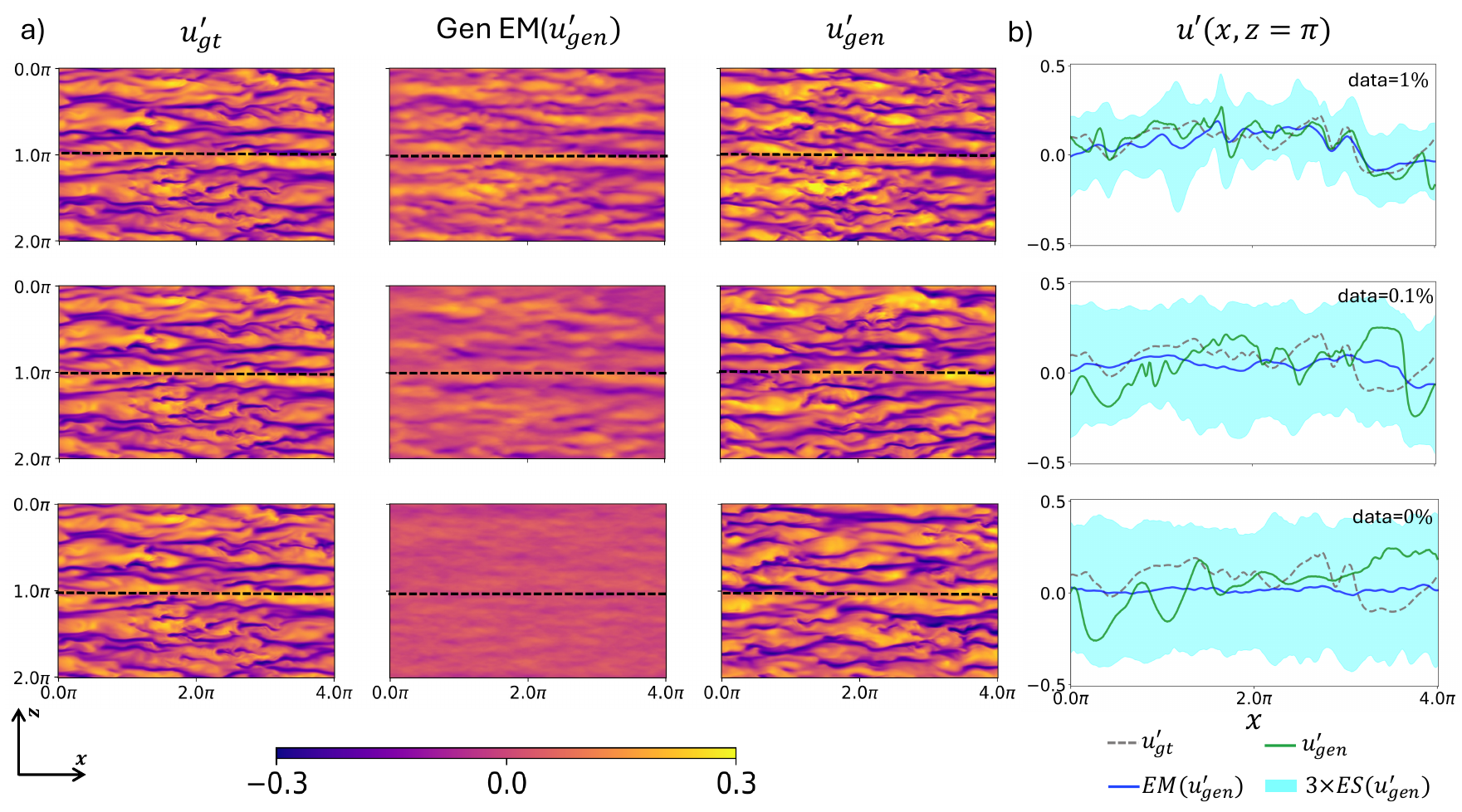}
     \caption{(a) Comparison of streamwise velocity fluctuation $u'$ contours conditioned on $1\%$ (1st row), $0.1\%$ (2nd row), and $0\%$ (3rd row) wall data. First column represents the ground truth DNS, the second column is the ensemble mean of conditionally generated $N_{ens} =50$ samples, and the third column displays one representative conditional samples. (b) Uncertainty quantification (\protect\tikz \protect\fill[cyan!40] (0,0) rectangle (0.5,0.25); contour) of $u'$ at $y^+=5, \space 20, \space 40$ along $z=1.0\pi$, with the ground truth (\protect\tikz[baseline=-0.5ex]\protect\draw [gray, thick,dashed] (0,0) -- (0.5,0); line), ensemble mean (\protect\tikz[baseline=-0.5ex]\protect\draw [blue, thick] (0,0) -- (0.5,0); line), and one sample (\protect\tikz[baseline=-0.5ex]\protect\draw [green, thick] (0,0) -- (0.5,0); line).}
    \label{fig:sparse10_data_contours_lineplots}
\end{figure}

We next investigate how data availability influences prediction fidelity and uncertainty. To this end, we fix the wall-normal location at $y^+ = 20$ and systematically reduce the sensor coverage from $10\%$ down to $0\%$. While the $10\%$ sparse sensor case has already been shown in Figure~\ref{fig:sparse10_contour_and_stats_y20}, we now present additional results for cases with less wall data availability. Figure~\ref{fig:sparse10_data_contours_lineplots}(a) compares the ensemble mean, one representative sample, and the ground truth under three increasingly sparse wall measurement scenarios: $1\%$, $0.1\%$, and $0\%$. 
As sensor coverage decreases, the ensemble mean predictions become progressively smoother and increasingly deviate from the ground truth, particularly in terms of amplitude attenuation. The degradation in predictive accuracy is further illustrated in panel~(b), which presents streamwise profiles of $u'$ along $z = \pi$, overlaid with shaded $3\times\mathrm{ES}(u')$ uncertainty bands. As the amount of available wall data decreases, these uncertainty intervals widen markedly, reflecting the model's decreasing confidence in its predictions. In the extreme case of $0\%$ wall data (i.e., unconditional generation) the ensemble mean collapses to a smooth prior with negligible correlation to the ground truth. Nevertheless, individual samples remain physically plausible due to the inductive bias of the trained generative model, though they no longer reflect the specific realization of the true flow field.

\begin{figure}[t!]
    \centering
    \includegraphics[width=1.\textwidth]{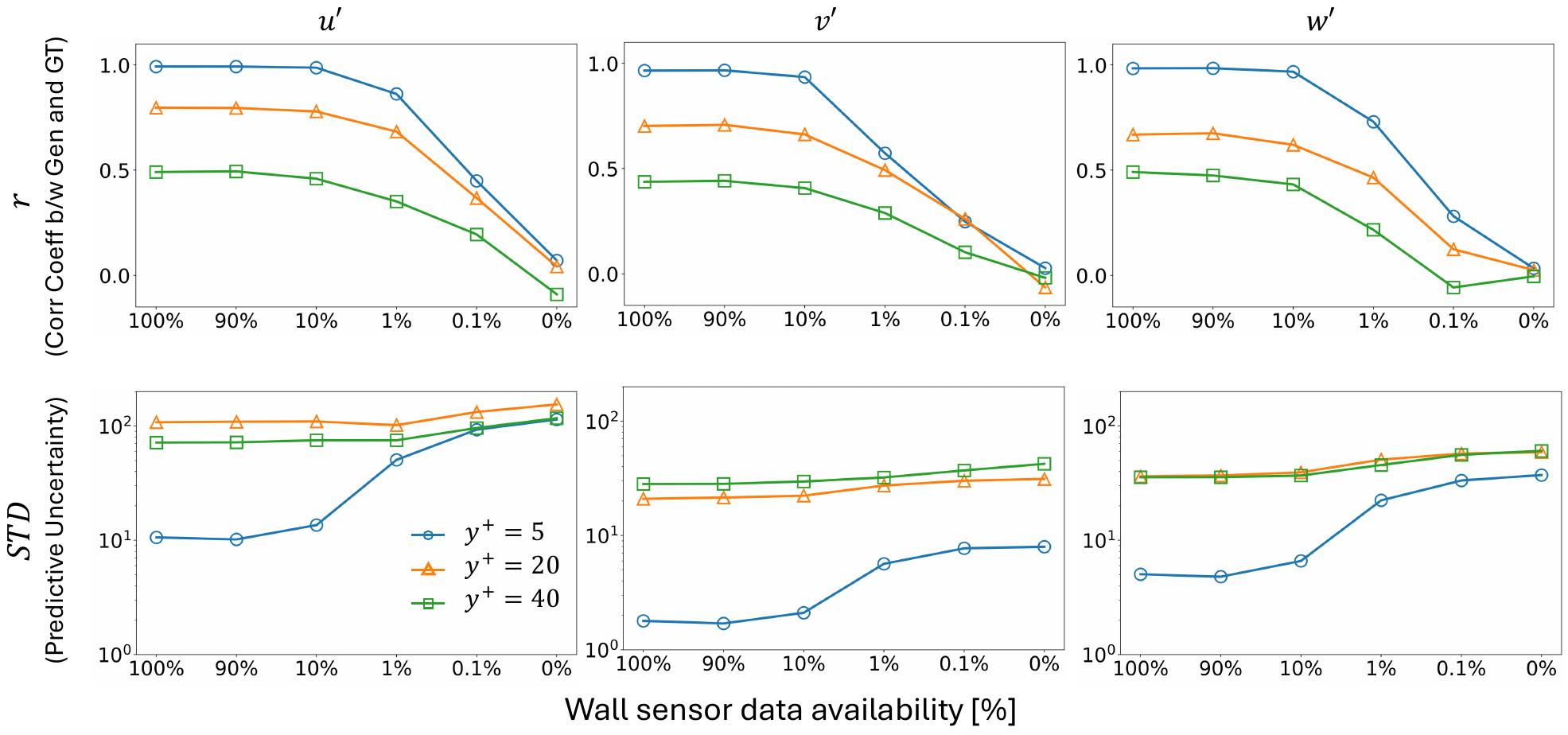}
    \caption{Effect of wall sensor data availability on reconstruction fidelity and predictive uncertainty at different wall-normal locations. Top row: Pearson correlation coefficient $r$ between ensemble-mean predictions and ground truth for streamwise ($u'$), wall-normal ($v'$), and spanwise ($w'$) velocity fluctuations. Bottom row: scalar ensemble standard deviation ($STD$) quantifying predictive uncertainty for each velocity component. Calculations are performed by generating $N_\mathrm{ens} = 50$ samples for all the different measurement scenarios.}
    \label{fig:sparse10_yplus_data_influence_lineplots}
\end{figure}

To quantitatively summarize the trends observed in reconstruction accuracy and predictive uncertainty across wall-normal positions and sensor sparsity levels, Figure~\ref{fig:sparse10_yplus_data_influence_lineplots} presents a set of scalar diagnostics across the full range of tested configurations. The top row reports the Pearson correlation coefficient $r$ between the ensemble mean of the generated samples and the DNS ground truth for $u'$, $v'$, and $w'$ at three wall-normal distances: $y^+ = 5$, $20$, and $40$. The bottom row displays the corresponding global uncertainty level, computed as the scalar standard deviation $\mathrm{STD}$ of the predictive ensemble at each setting,
\begin{equation}
\begin{aligned}
    \mathrm{STD} &= \sqrt{\sum^{N_x}_{i} \sum^{N_z}_{j}{(ES(u'(i,j))}^2}
\end{aligned}
\end{equation}

The correlation plots in the top row clearly demonstrate that reconstruction fidelity degrades with decreasing sensor availability, especially once coverage drops below $10\%$. This trend holds consistently across all velocity components and wall-normal locations, though the impact is most pronounced for the wall-normal ($v'$) and spanwise ($w'$) fluctuations, which are more weakly coupled to wall data. For all three components, predictive accuracy is highest at $y^+ = 5$, where near-wall coherence is strongest, and lowest at $y^+ = 40$, where the flow becomes increasingly decoupled from the wall. These findings are consistent with the qualitative assessments presented earlier and further validate the model's sensitivity to both wall distance and observability.

The bottom row captures how the model's epistemic uncertainty evolves under the same conditions. As expected, the scalar ensemble spread $\mathrm{STD}$ increases as sensor coverage declines, reflecting reduced confidence in conditional predictions. Notably, the uncertainty growth is monotonic for each $y^+$ level and particularly steep between $10\%$ and $1\%$ data availability. Beyond this range, further increasing wall data leads to only marginal reductions in uncertainty, suggesting that the dominant source of epistemic uncertainty arises from the fundamental limitations of wall-flow coupling rather than sensor sparsity. Interestingly, the largest absolute uncertainty is observed around $y^+ = 20$ for $u'$ predictions, consistent with the peak turbulence intensity and the inherently greater variability in the buffer layer. This non-monotonic dependence on wall-normal location reinforces that predictive uncertainty is influenced not only by data sparsity but also by the intrinsic dynamics of wall-bounded turbulence.

\subsection{Off-wall velocity generation with incomplete and low-resolution wall measurements}

We now demonstrate the robustness of the proposed conditional generative framework under three practically relevant wall measurement scenarios: (i) incomplete spatial coverage, where only a portion of the wall is instrumented; (ii) partial observability, where only a subset of wall quantities (e.g., a single shear stress component) is accessible; and (iii) low-resolution sensing, where measurements are available across the full domain but are spatially downsampled due to acquisition constraints. These configurations reflect common limitations in real-world sensor deployments, enabling us to assess the model's performance under degraded and incomplete wall information.

\begin{figure}[htp!]
    \centering
    \includegraphics[width=1.0\textwidth]{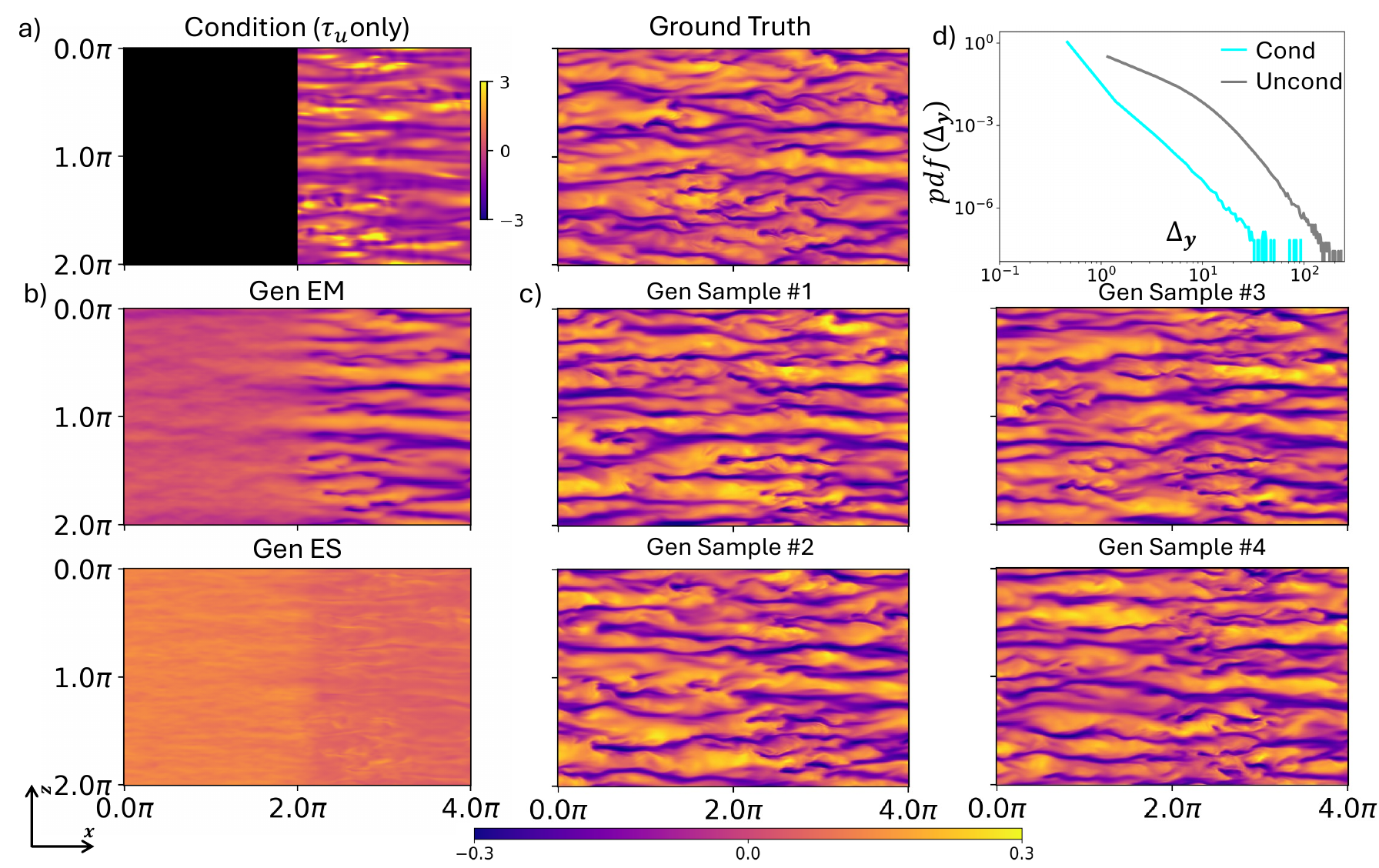}
    \caption{(a) An example of partial wall measurements ($\tau_u$ over half of the domain) (left) and the corresponding ground truth velocity fluctuations ($u'$) at $y^+ =20$ (right). (b) Ensemble mean (top) and standard deviation (bottom) velocity fluctuations for $N_\mathrm{ens} = 50$ generated velocity fluctuations samples (c) Four randomly selected $u'$ velocity fluctuation samples given the wall measurements shown in subfigure (a). (d) Comparison of the distribution of point-wise normalized $L_2$ error ($\Delta_\mathbf{y}$) between measurements ($\mathbf{y}$) of the unconditionally (\protect\tikz[baseline=-0.5ex]\protect\draw [gray, thick] (0,0) -- (0.5,0); line) and conditionally generated (\protect\tikz[baseline=-0.5ex]\protect\draw [cyan, thick] (0,0) -- (0.5,0); line) velocity fluctuations samples corresponding to $500$ different test wall measurements.}
    \label{fig:partial_contour_and_stats_y20}
\end{figure}
Figure~\ref{fig:partial_contour_and_stats_y20} shows the model's performance when only the streamwise wall shear stress $\tau_u$ is measured over the downstream half of the domain ($2\pi \leq x \leq 4\pi$). The wall observations (i.e., $\tau_u$ over the downstream half of the domain) and corresponding DNS ground truth for $u'$ at $y^+=20$ are shown in panel~(a). From this partially available wall information, we generate an ensemble of $N_\mathrm{ens}=50$ velocity fluctuation samples. The ensemble mean (top) and standard deviation (bottom) fields are presented in panel~(b). The mean field captures the prominent large-scale streaks observed in the ground truth, particularly in the region where wall data is available. The ensemble spread decreases markedly in the downstream region where wall measurements are available, confirming the model's ability to quantify epistemic uncertainty. Panel~(c) displays four representative samples from the ensemble, each of which preserves the spatial characteristics and intermittency of the flow, even across the transition between observed and unobserved wall zones ($x=2\pi$). This seamless blending demonstrates the model's ability to infer plausible flow states while avoiding artifacts near sensor discontinuities. 
We also compute the distribution of pointwise normalized $L_2$ errors $\Delta_\mathbf{y}$ over 500 independent test cases to evaluate measurement consistency. As shown in Panel~(d), the conditional samples exhibit a clear reduction in error magnitude compared to their unconditional counterparts. This shift toward lower $\Delta_\mathbf{y}$ values confirms that the model effectively assimilates available wall information, even when spatially limited. 

\begin{figure}[t!]
    \centering
    \includegraphics[width=1.0\textwidth]{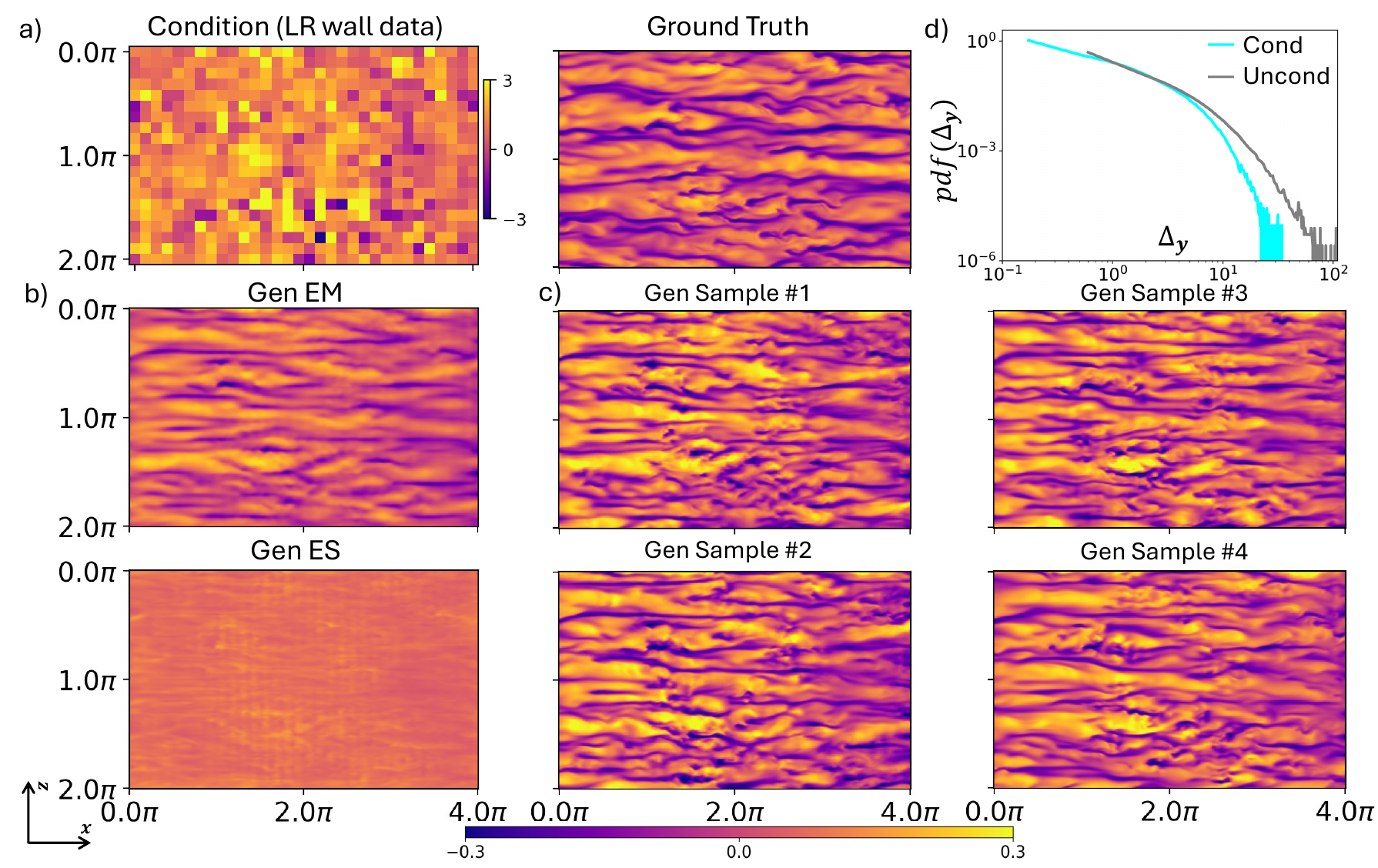}
    \caption{(a) An example of (1/100) low-resolution $\mathbf{y}_{LR}(x, z) = D({\bm\Phi}_{\mathrm{wall}})$, only $D(\tau_u)$ is illustrated (left), where $D$ is the nearest-neighbor interpolation down-sampling operator and the corresponding ground truth velocity fluctuations ($u'$) at $y^+ =20$ (right). (b) Ensemble mean (top) and standard deviation (bottom) velocity fluctuations for $N_\mathrm{ens} = 50$ generated velocity fluctuations samples (c) Four randomly chosen $u'$ velocity fluctuation samples generated for the prescribed wall measurements shown in subfigure (a). (d) Comparison of the distribution of point-wise normalized $L_2$ error ($\Delta_\mathbf{y}$) between measurements ($\mathbf{y}$) of the unconditionally (\protect\tikz[baseline=-0.5ex]\protect\draw [gray, thick] (0,0) -- (0.5,0); line) and conditionally generated (\protect\tikz[baseline=-0.5ex]\protect\draw [cyan, thick] (0,0) -- (0.5,0); line) velocity fluctuations samples corresponding to $500$ different test wall measurements.}
    \label{fig:lowres_contour_and_stats_y20}
\end{figure}

We next examine a low-resolution setting, where wall measurements are uniformly available across the entire domain of interest but lack spatial detail. This setup reflects practical limitations in experimental acquisition systems, such as imaging or pressure-sensitive paint, which often capture wall quantities at coarse spatial resolutions. To mimic this constraint, we apply a $100\times$ downsampling operator $D$, implemented via nearest-neighbor interpolation, on the original wall data $\Phi_{\mathrm{wall}}$, and visualize the resulting low-resolution streamwise shear stress $D(\tau_u)$ in Figure~\ref{fig:lowres_contour_and_stats_y20}(a). Conditioning on this degraded input, we generate an ensemble of $N_\mathrm{ens}=50$ conditional velocity samples. The ensemble mean and pointwise standard deviation fields $ES(u')$ are shown in panel~(b). Despite the severe loss of spatial detail in the wall conditioning, the model reconstructs several dominant large-scale streaks observed in the ground truth DNS field. However, the ensemble spread $ES(u')$ becomes high throughout the domain, indicating amplified epistemic uncertainty in the absence of high-resolution wall data. Panel~(c) shows four representative samples drawn from the ensemble, which contain small-scale turbulent structures thanks to the FM module. These reflect the plausible diversity of flow fields consistent with coarse wall information, highlighting the expressiveness and adaptability of the generative framework.
To quantify how well the generated samples honor the available wall data, we compute the distribution of pointwise normalized $L_2$ error $\Delta_\mathbf{y}$ across 500 randomly selected test cases. Panel~(d) shows that, even under low-resolution conditions, the conditional samples exhibit substantially lower errors than those generated unconditionally. This demonstrates the model's ability to effectively leverage limited information while faithfully capturing the stochasticity and structure of wall-bounded turbulence.

\section{Discussion}
\label{sec:discuss}

\subsection{Comparison with deterministic data-driven baselines}

To evaluate the effectiveness of our proposed stochastic generative model, we benchmark its performance against two widely used baseline methods: a convolutional neural network (CNN) model~\cite{guastoni2021convolutional} and the classical linear stochastic estimation (LSE) method~\cite{adrian1979conditional}. Both methods are representative of state-of-the-art data-driven approaches for wall-based flow reconstruction.
The CNN baseline adopts the fully convolutional neural network architecture introduced by Guastoni et al.~\cite{guastoni2021convolutional}, which learns a nonlinear mapping from wall quantities to velocity fluctuations at a specified wall-normal locations  The LSE method, on the other hand, estimates velocity fluctuations $\bm{u}'(\bm{x}, t)$ via a linear projection of wall measurements $\bm{\Phi}_{\mathrm{wall}}(\bm{x}, t)$ using precomputed correlation kernels:  
\begin{equation}
    \bm{u}' (\bm{x}, t) = \sum h (\bm{x}) {\bm \Phi}_{\mathrm{wall}}(\bm{x}, t) 
\label{eq:LSE}
\end{equation}
where $h$ is obtained by minimizing the least-squares error over the training dataset. While LSE captures linear statistical correlations between wall and off-wall quantities, it cannot model nonlinear interactions or synthesize plausible flow structures in regions of weak observability. Additional implementation details for baselines are provided in~\ref{sec:appendix-baseline}).

\begin{figure}[t!]
    \centering
    \includegraphics[width=0.9\textwidth]{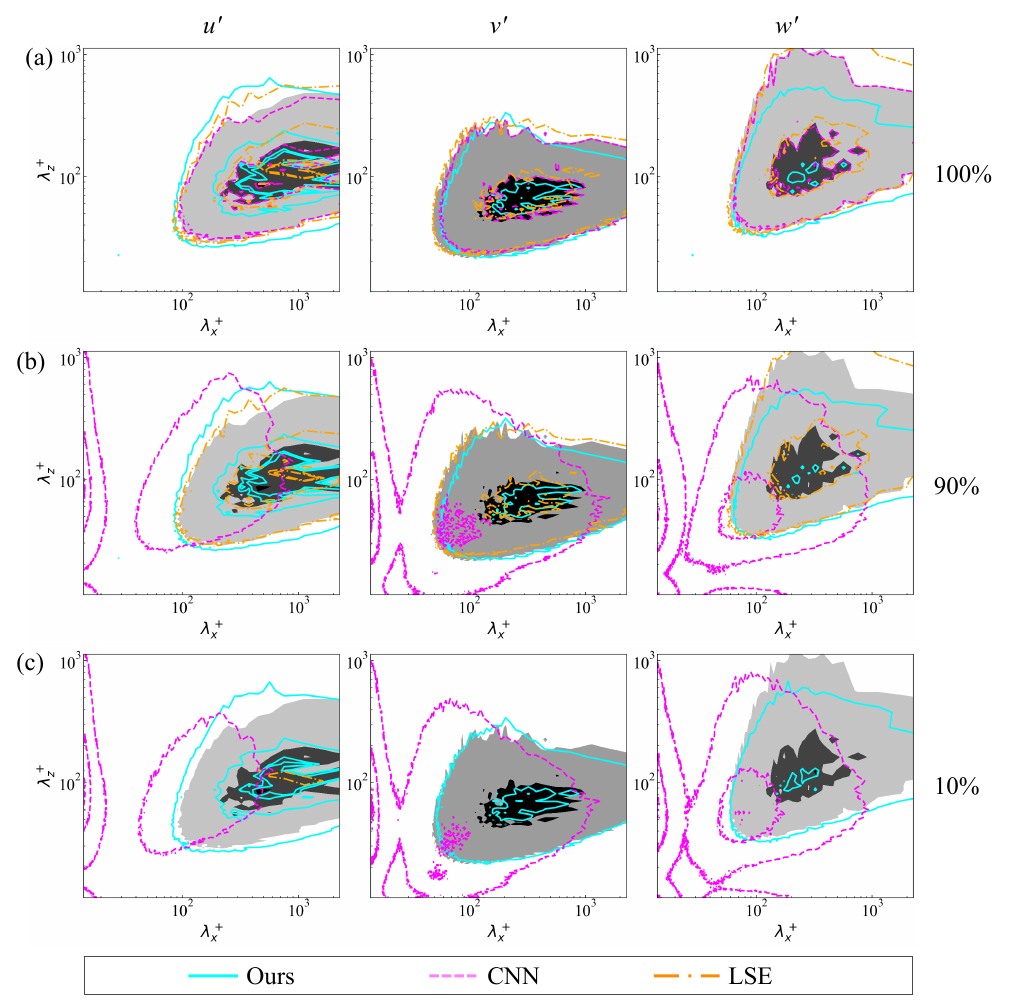}
    \caption{Comparison of pre-multiplied two-dimensional energy spectra between the proposed model and two baseline methods under varying availability of wall quantities at $y^+=20$: (a) $100\%$ wall quantities, (b) $90\%$ wall quantities, and (c) $10\%$ wall quantities. The energy spectra are calculated over 500 test cases. Ground truth DNS is shown as filled contours (\protect\tikz \protect\fill[gray!40] (0,0) rectangle (0.5,0.25); contours), and model predictions are shown as line plots. Streamwise and spanwise wavelengths $\lambda_x^+$ and $\lambda_z^+$ are normalized by wall units. The contour levels contain $10\%$, $50\%$ and $90\%$ of the maximum energy spectra.}
    \label{fig: comparison at 20}
\end{figure}

Figure~\ref{fig: comparison at 20} compares the pre-multiplied two-dimensional energy spectra of velocity fluctuations reconstructed by our model and the two baselines at $y^+=20$ under different levels of wall measurement availability: $100\%$, $90\%$, and $10\%$. When complete wall information is available (panel a), all models recover the dominant energy-containing structures and produce spectra in close agreement with DNS. Under mild sparsity (panel b), however, the CNN baseline begins to lose fidelity, especially at smaller scales, while both our proposed model and LSE maintain strong agreement with DNS. This is consistent with previous observations~\cite{encinar2019logarithmic} indicating that LSE performs reasonably well in recovering wall-attached eddies within the viscous and buffer layers ($y^+ \le 20$) when wall data is only moderately sparse and noise-free.

When the wall information is severely limited to $10\%$ coverage (panel c), both deterministic baseline models fail to reconstruct the turbulent energy distribution accurately. The CNN baseline produces overly smoothed fields, and the LSE spectrum nearly vanishes due to underprediction of fluctuation amplitudes. In stark contrast, our model remains stable and physically realistic, preserving the large-scale streaks and the correct energy content across wavelengths. This robustness stems from two key features: (i) the use of a generative prior that synthesizes statistically consistent flow fields in underdetermined regions and (ii) explicit uncertainty quantification that allows the model to balance observed data with prior structure during sampling.

\begin{figure}[!ht]
    \centering
    \includegraphics[width=0.9\textwidth]{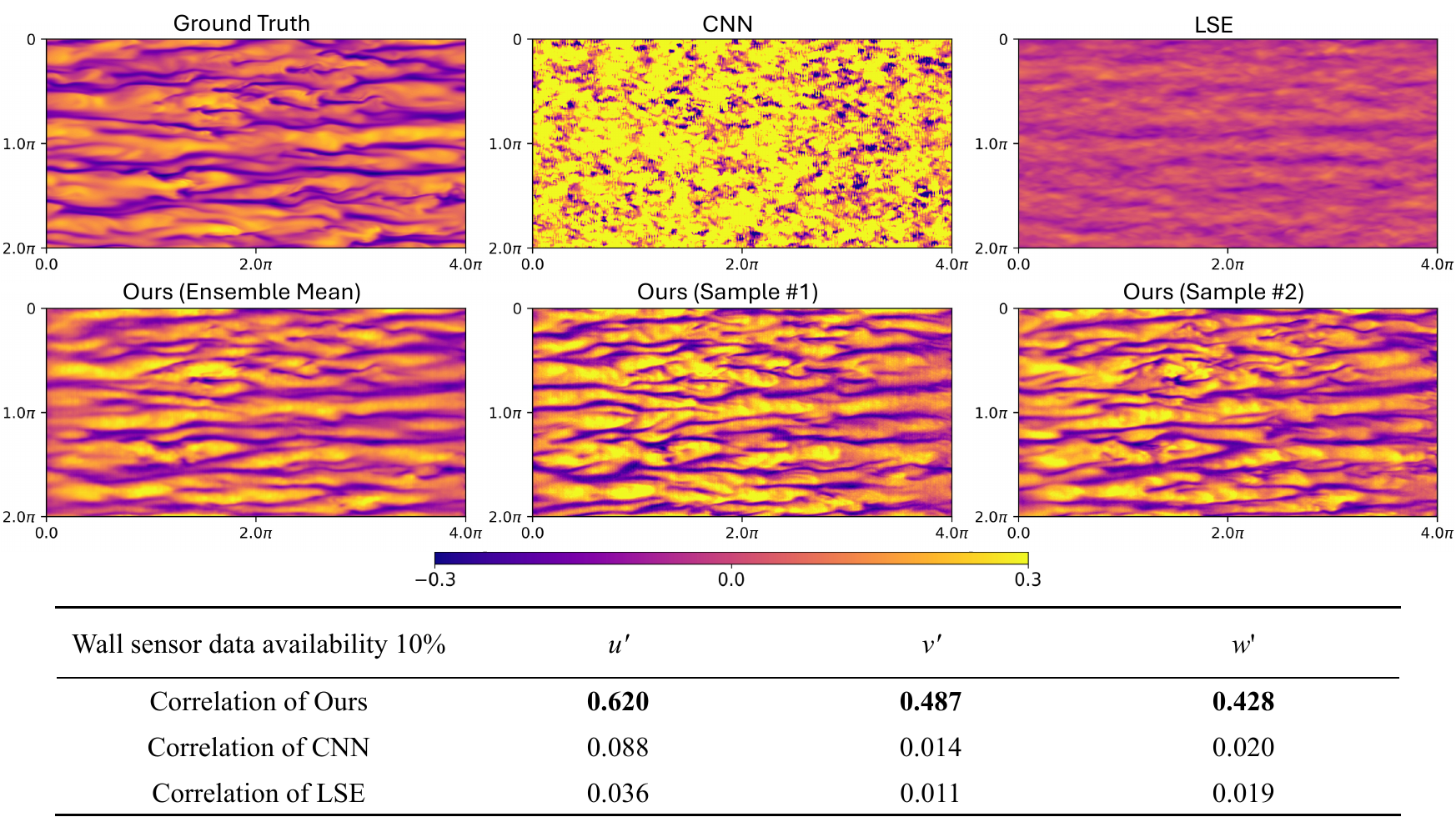}
    \caption{Comparison of instantaneous velocity fluctuations $u'$ at $y^+=20$ reconstructed from $10\%$ wall measurements using the proposed method, CNN, and LSE. Correlation values between model predictions and DNS ground truth are also reported for each model.}
    \label{fig: comparison_contour at 20}
\end{figure}

To provide further qualitative insight, Figure~\ref{fig: comparison_contour at 20} shows instantaneous contours of $u'$ reconstructed by all models at $y^+=20$ with only $10\%$ wall data. 

Our method clearly produces sharper, more structured flow fields that closely resemble the DNS ground truth, whereas both baselines produce heavily smoothed or spatially decorrelated predictions. Correlation metrics listed in the accompanying table confirm the superiority of our method across all velocity components.

In summary, both baselines produce deterministic point estimates and offer no measure of confidence, making them less reliable in underdetermined or weakly observable regimes. In contrast, our framework not only delivers superior accuracy under sparse or degraded sensing, but also quantifies epistemic uncertainty—a critical capability for decision-making in wall modeling, sensor placement, and active flow control applications.

\subsection{The SWAG-based forward operator and implications for wall modeling}

The forward operator $\mathcal{F}_{\bm \phi}$ serves as measurement model in our conditional generation framework by mapping predicted velocity fluctuations to corresponding wall quantities. As it is used during the inference to guide conditional sampling, it must be both differentiable and capable of quantifying uncertainty, especially in regions where wall-flow coupling is weak. To this end, we adopt a Bayesian training strategy, enabling the operator to capture epistemic uncertainty in its predictions. 

\begin{figure}[htp!]
    \centering
    \includegraphics[width=1.\textwidth]{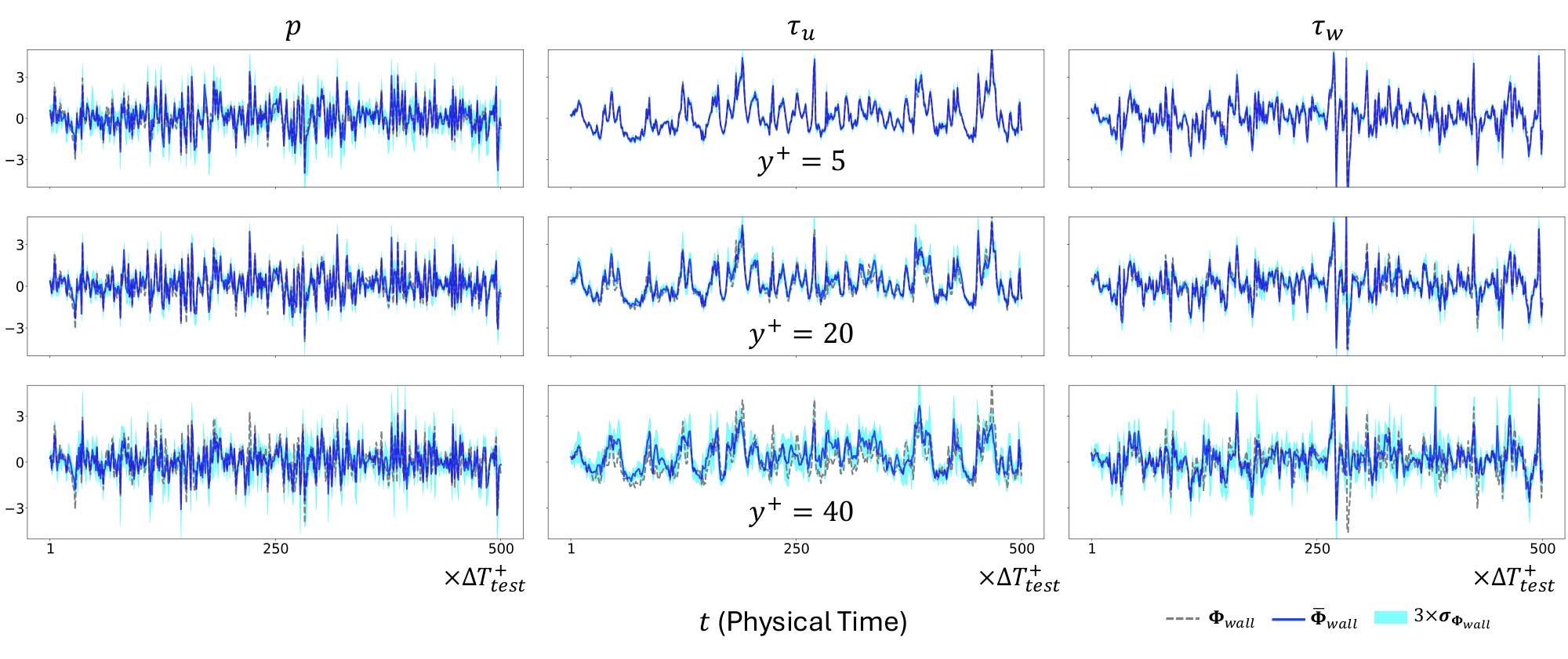}
    \caption{SWAG-based forward operator predictions of wall quantities $\bm{\Phi}_{\mathrm{wall}}$ for velocity inputs at $y^+ = 5$ (top), $20$ (middle), and $40$ (bottom), evaluated along $x = 2\pi$, $z = \pi$ over 500 time steps. Shaded bands show $\pm3\sigma$ uncertainty intervals (\protect\tikz \protect\fill[cyan!40] (0,0) rectangle (0.5,0.25); contour); 
    ensemble mean (\protect\tikz[baseline=-0.5ex]\protect\draw [blue, thick] (0,0) -- (0.5,0); line) and ground truth (\protect\tikz[baseline=-0.5ex]\protect\draw [gray, thick,dashed] (0,0) -- (0.5,0); line).}
    \label{fig:discussion_forward_uncertain}
\end{figure}

To evaluate the uncertainty quantification capability of the trained forward model, we apply it to $500$ temporally ordered test samples at three wall-normal distances ($y^+ = 5$, $20$, and $40$) and extract the reconstructed wall quantities $\bm{\Phi}_{\mathrm{wall}}$ at the probe location $(x=2\pi, z=\pi)$. Figure~\ref{fig:discussion_forward_uncertain} shows the resulting time series of ensemble mean predictions and $3\sigma$ uncertainty intervals computed from $m = 50$ SWAG posterior weight samples. As $y^+$ increases, two trends are notable: (1) the ensemble mean becomes slightly less aligned with the ground truth, and (2) the uncertainty band of the prediction expands accordingly. This behavior reflects the diminishing influence of velocity fluctuations on wall quantities with increasing distance from the wall, and confirms that the forward operator correctly captures epistemic uncertainty induced by weakening wall-flow coherence.  

\begin{figure}[t!]
    \centering
    \includegraphics[width=0.85\textwidth]{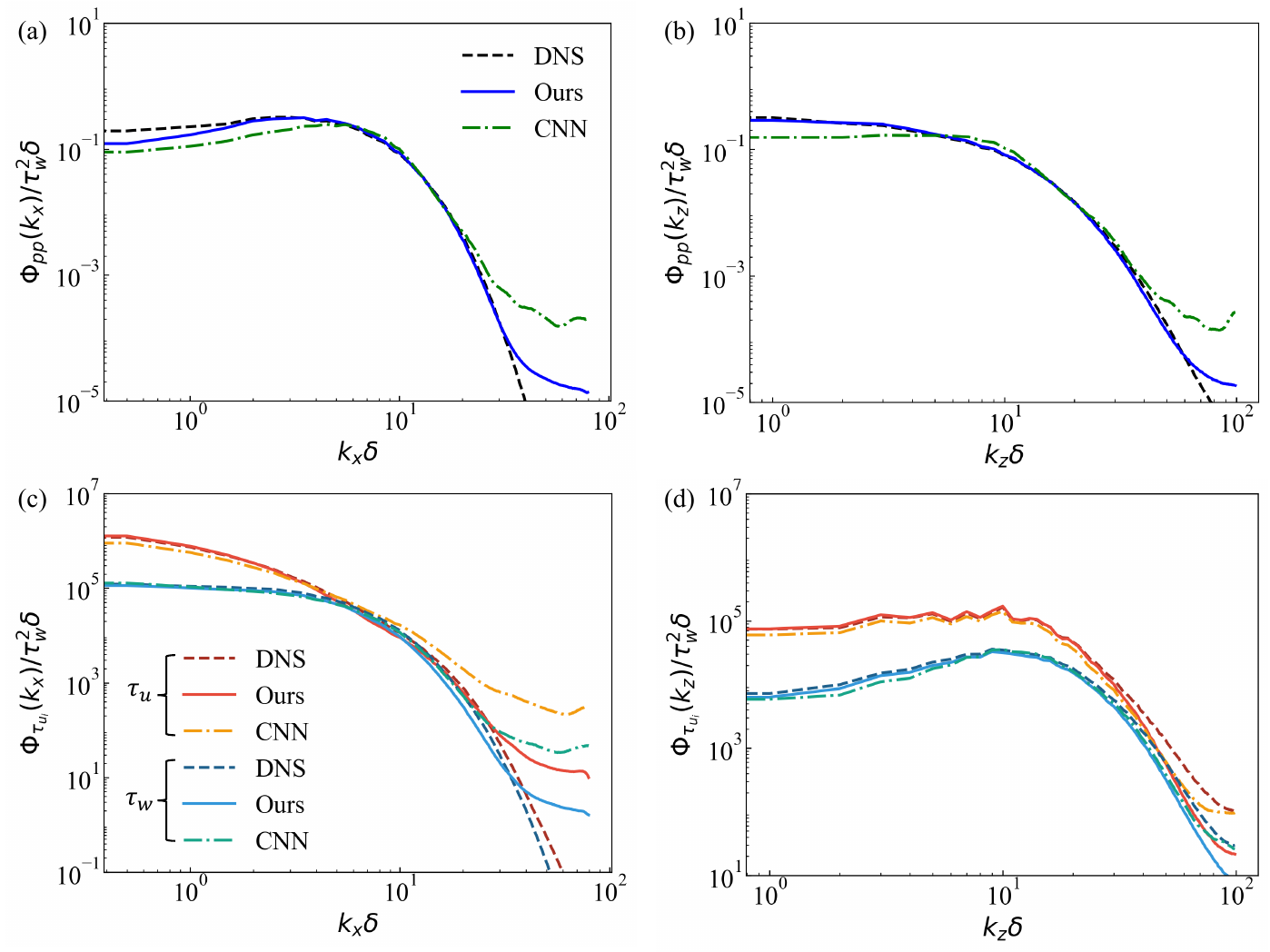}
    \caption{ Comparison of statistics from different forward models for mapping velocities at $y^+=20$ to wall: (a) streamwise and (b) spanwise spectra of pressure fluctuations; (b) spanwise spectrum of pressure fluctuations; (c) streamwise and (d) spanwise spectra of wall shear stresses $\tau_u$ and $\tau_w$.}
    \label{fig:discussion_forward_stat_y20}
\end{figure}

We further assess our SWAG-based probabilistic neural operator $\mathcal{F}_{\bm \phi}$ by comparing its performance against a baseline convolutional neural network (CNN) model adapted from~\cite{guastoni2021convolutional}. Although originally developed to infer velocity fields from wall data, we reverse its input–output direction for a fair comparison, leveraging its purely data-driven nature. Figure~\ref{fig:discussion_forward_stat_y20} shows the one-dimensional pre-multiplied spectra of predicted wall quantities at $y^+ = 20$ for both models.
Across all wall quantities, our SWAG-based model consistently outperforms the CNN baseline, particularly for pressure spectra in both streamwise and spanwise directions (panels a, b). The CNN significantly overestimates energy at high wavenumbers, suggesting the presence of spurious high-frequency artifacts, likely a result of overfitting and the absence of uncertainty regularization. In contrast, our model produces smoother spectra that align closely with DNS, demonstrating improved multiscale fidelity. While both models show reduced accuracy at low wavenumbers, the gap is smaller in our model and likely reflects the intrinsic difficulty of recovering large-scale pressure structures rather than model capacity. For wall shear stress spectra (panels c, d), both models recover the dominant energy content, though the SWAG-based operator again achieves better agreement with DNS across a broader range of scales. It worth noting that the CNN baseline is trained on four times more data (36K samples vs 9K samples for $\mathcal{F}_{\bm \phi}$), consistent with prior study~\cite{guastoni2021convolutional, guemes2021coarse, balasubramanian2023predicting}.

Beyond conditioning guidance, the forward operator also has potential use as a data-driven wall model in turbulence simulations. To evaluate this possibility, we conduct an \textit{a priori} test, in comparison with a traditional algebraic wall model based on the Spalding function~\cite{spalding1961single}, using input velocity fields at $y^+ = 20$.
\begin{figure}[t!]
    \centering
    \includegraphics[width=1.0\textwidth]{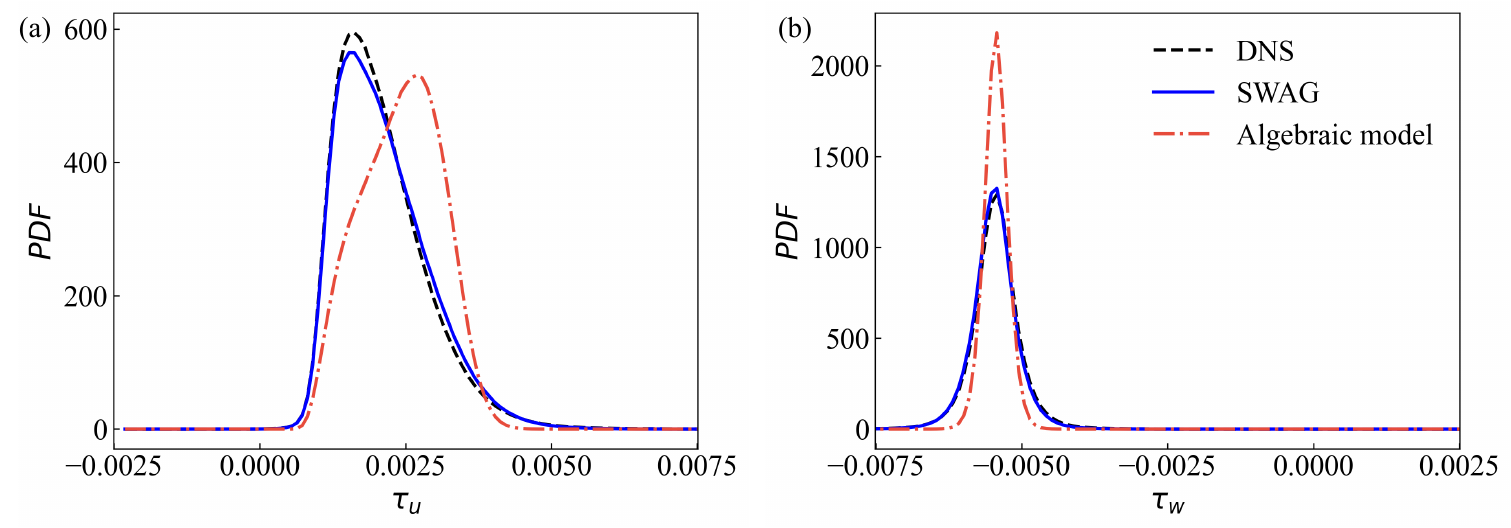}
    \caption{Comparison of predicted instantaneous wall shear stress: (a) PDF of streamwise wall shear stress $\tau_u$ ; (b) PDF of spanwise wall shear stress $\tau_w$. The input velocity field from $y^+=20$.}
    \label{fig:forward_wall_model}
\end{figure}
Figure~\ref{fig:forward_wall_model} shows that our operator produces predictions of wall shear stress that closely match DNS ground truth, while the Spalding-based model overpredicts $\tau_u$ by approximately $15\%$, consistent with the well-documented log-layer mismatch observed in algebraic wall models~\cite{yang2017log}. These results highlight the potential of $\mathcal{F}_{\bm \phi}$ as a differentiable, uncertainty-aware wall model. Its ability to learn from data while maintaining physical consistency makes it a promising candidate for hybrid RANS–LES or wall-modeled LES frameworks. More extensive \textit{a posteriori} validation will be explored in future work.

\subsection{Training-Free conditional generation via flow matching}
\label{sec:train-free-method-inpainting}

A key strength of the proposed framework lies in its ability to perform test-time conditional generation without retraining or modifying the learned generative model. This property is enabled by the proposed training-free conditional inference strategy for FM. In contrast to traditional supervised approaches, which require explicit re-training for each new measurement type or sensor configuration, our approach flexibly assimilates conditioning data by incorporating it as a guidance term in the FM sampling dynamics. This decoupling of model training and inference enables a unified generative framework for diverse inverse problems in turbulence modeling.

\begin{figure}[t!]
    \centering
    \includegraphics[width=1.0\textwidth]{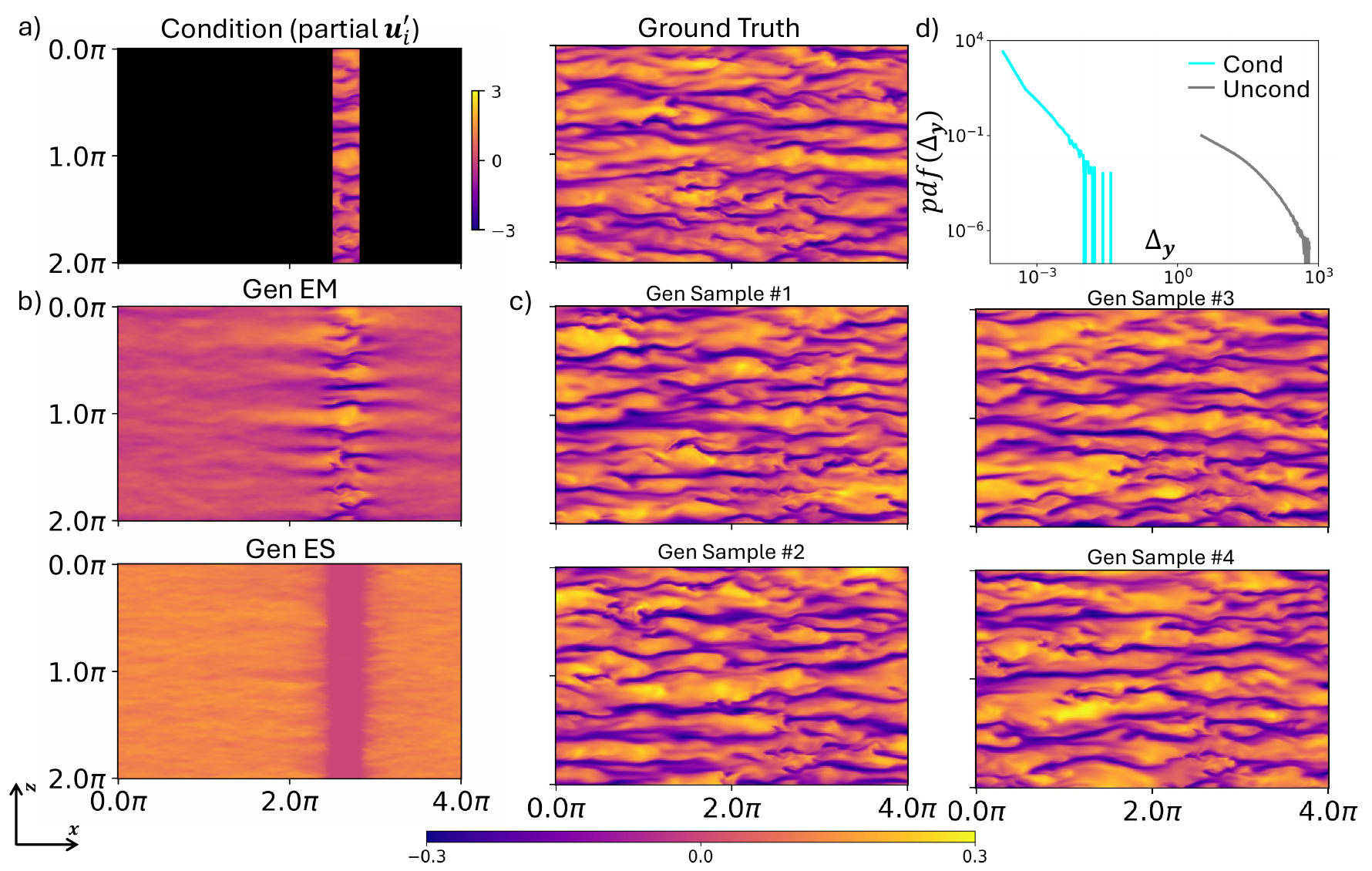}
    \caption{(a) An example of partial velocity fluctuation measurements ($\mathbf{y}(x,z) = \mathcal{J}(x, z) \odot \bm{u}'_i(x, z)$) (left) and the corresponding ground truth velocity fluctuations ($u'$) at $y^+ =20$ (right). (b) Ensemble mean  (top) and standard deviation (bottom) of $N_\mathrm{ens} = 50$ generated velocity fluctuations samples (c) Four randomly chosen $u'$ velocity fluctuation samples generated for the prescribed wall measurements shown in subfigure (a). (d) Comparison of the distribution of point-wise normalized $L_2$ error ($\Delta_\mathbf{y}$) between measurements ($\mathbf{y}$) of the unconditionally (\protect\tikz[baseline=-0.5ex]\protect\draw [gray, thick] (0,0) -- (0.5,0); line) and conditionally generated (\protect\tikz[baseline=-0.5ex]\protect\draw [cyan, thick] (0,0) -- (0.5,0); line) velocity fluctuations samples corresponding to $500$ different wall measurements.}
    \label{fig:inpainting_discussion}
\end{figure}

To illustrate this capability in a simplified setting, we consider an internal field recovery task in which the generative model is conditioned directly on partial measurements of the velocity fluctuation field $\bm{u}'_i$. This setting serves as a conceptual analogue to classical inpainting tasks in computer vision and highlights the model's behavior under idealized but informative conditions. Here, the measurement operator $\mathcal{F}(\bm{u}'_i)$ is defined as a binary masking operator $\mathcal{J}(x,z)$ that zeroes out the velocity field at unobserved locations: 
\begin{equation}
    \mathbf{y}(x, z) = \mathcal{J}(x, z) \odot \bm{u}'_i(x, z),
\end{equation}
where $\odot$ denotes element-wise multiplication and $\mathcal{J}(x,z) \in {0,1}$ indicates the sensor layout. We apply this masking operator to approximately $90\%$ of the streamwise velocity field at $y^+=20$, leaving only sparse values available for conditioning (Figure~\ref{fig:inpainting_discussion}a).

To generate posterior-consistent samples, we apply the flow-matching ODE starting from Gaussian noise $\mathbf{x}_0 \sim \mathcal{N}(0, I)$ and compute a corrected trajectory using the training-free guidance method introduced in Section~\ref{sub-sec:infer}. At each time step, an approximate one-step prediction $\hat{\mathbf{x}}_1$ is obtained using the learned velocity field $\bm{\nu}_\theta(t, \mathbf{x})$, from which an estimated measurement $\hat{\mathbf{y}} = \mathcal{J} \odot \hat{\mathbf{x}}_1$ is constructed. The mismatch between $\hat{\mathbf{y}}$ and the true observation $\mathbf{y}$ defines the conditioning error, whose gradient is used to form the correction term $\bm{\nu}'(t, \mathbf{x}, \mathbf{y})$ as detailed in Equation~\ref{eqn:condition-term}.

Figure~\ref{fig:inpainting_discussion}b shows the ensemble mean and standard deviation computed from $N_{\text{ems}} = 50$ conditionally generated samples. The ensemble mean accurately reconstructs both the observed and nearby unobserved regions, reflecting spatial coherence in the generated fields. Importantly, the ensemble standard deviation adapts to the conditioning mask, exhibiting higher uncertainty in unobserved regions and lower variance near conditioned points. These results confirm that the model captures uncertainty in a physically consistent manner.
Panel~(c) visualizes four representative conditional samples, each of which exhibits realistic turbulence structures with fine-scale variability and flow intermittency. The visual plausibility of the samples, despite strong underdetermination, demonstrates the inductive bias encoded by the generative model and the effectiveness of flow-based guidance in posterior exploration.

To quantify measurement consistency, Figure~\ref{fig:inpainting_discussion}d compares the distribution of the pointwise normalized $L_2$ error $\Delta_{\mathbf{y}}$ for 500 test-time realizations. Samples generated with conditioning exhibit significantly lower errors compared to unconditional samples, confirming that the guided flow-matching trajectory conforms to the measurement constraints. Together, these results highlight the effectiveness and robustness of the training-free inference strategy, enabling high-fidelity conditional generation under arbitrary sensor configurations without retraining the generative model.

\section{Conclusion}
\label{sec:conclude}

In this work, we have proposed a data-driven generative framework for reconstructing near-wall turbulent velocity fluctuations from wall-based measurements, grounded in the theory of conditional flow matching and Bayesian neural operators. The method effectively decouples training from inference through a training-free conditional generation strategy, allowing flexible assimilation of wall data in various forms without the need for retraining across different sensor modalities. Our integration of flow matching with a SWAG-trained forward operator enables consistent propagation of both observations and their epistemic uncertainty into the generative sampling process.

We have demonstrated that the model performs robustly across a range of wall measurement configurations, including fully observed, sparse, partial, and low-resolution cases, and provides physically realistic reconstructions with quantified uncertainty. The results highlight how generative modeling can bridge data-driven inference and physically structured variability in turbulent flows. Compared to existing deterministic data-driven approaches such as CNNs and LSE, our proposed framework offers significantly improved robustness and fidelity, particularly under weak observability conditions. It accurately reconstructs fine-scale turbulent structures even at large wall-normal distances or with severely limited wall data, and crucially, it provides principled uncertainty quantification to capture the confidence of each prediction. This enables consistent, high-quality reconstruction of buffer-layer and logarithmic-layer turbulence where traditional methods often fail.

Looking forward, this framework opens several directions for future exploration. First, coupling with physics-based constraints (e.g., via differentiable solvers or spectral regularization) may further enhance physical consistency. Second, integrating temporal dynamics via conditional generative trajectories could extend this work toward real-time state estimation and data assimilation. Finally, integrating the model with experimental measurements opens the door to practical applications in data-driven wall modeling, real-time flow control, and the development of high-fidelity digital twin systems. Through its flexibility, uncertainty-awareness, and sampling efficiency, this framework represents a significant step toward robust data-driven modeling of wall-bounded turbulence.


\begin{thebibliography}{10}
\expandafter\ifx\csname url\endcsname\relax
  \def\url#1{\texttt{#1}}\fi
\expandafter\ifx\csname urlprefix\endcsname\relax\def\urlprefix{URL }\fi
\expandafter\ifx\csname href\endcsname\relax
  \def\href#1#2{#2} \def\path#1{#1}\fi

\bibitem{townsend1961equilibrium}
A.~Townsend, Equilibrium layers and wall turbulence, Journal of Fluid Mechanics 11~(1) (1961) 97--120.

\bibitem{pope2001turbulent}
S.~B. Pope, Turbulent flows, Measurement Science and Technology 12~(11) (2001) 2020--2021.

\bibitem{adrian2007hairpin}
R.~J. Adrian, Hairpin vortex organization in wall turbulence, Physics of fluids 19~(4) (2007).

\bibitem{hwang2016inner}
J.~Hwang, J.~Lee, H.~J. Sung, T.~A. Zaki, Inner--outer interactions of large-scale structures in turbulent channel flow, Journal of Fluid Mechanics 790 (2016) 128--157.

\bibitem{kim2007active}
J.~Kim, Active control of turbulent boundary layers for drag reduction, in: Industrial and Environmental Applications of Direct and Large-Eddy Simulation: Proceedings of a Workshop Held in Istanbul, Turkey, 5--7 August 1998, Springer, 2007, pp. 142--152.

\bibitem{lofdahl1999mems}
L.~L{\"o}fdahl, M.~Gad-el Hak, Mems-based pressure and shear stress sensors for turbulent flows, Measurement Science and Technology 10~(8) (1999) 665.

\bibitem{choi1994active}
H.~Choi, P.~Moin, J.~Kim, Active turbulence control for drag reduction in wall-bounded flows, Journal of Fluid Mechanics 262 (1994) 75--110.

\bibitem{zaki2024turbulence}
T.~A. Zaki, Turbulence from an observer perspective, Annual Review of Fluid Mechanics 57 (2024).

\bibitem{abe2004very}
H.~Abe, H.~Kawamura, H.~Choi, Very large-scale structures and their effects on the wall shear-stress fluctuations in a turbulent channel flow up to re $\tau$= 640, J. Fluids Eng. 126~(5) (2004) 835--843.

\bibitem{mathis2009large}
R.~Mathis, N.~Hutchins, I.~Marusic, Large-scale amplitude modulation of the small-scale structures in turbulent boundary layers, Journal of Fluid Mechanics 628 (2009) 311--337.

\bibitem{lozano2014time}
A.~Lozano-Dur{\'a}n, J.~Jim{\'e}nez, Time-resolved evolution of coherent structures in turbulent channels: characterization of eddies and cascades, Journal of fluid mechanics 759 (2014) 432--471.

\bibitem{colburn2011state}
C.~Colburn, J.~Cessna, T.~Bewley, State estimation in wall-bounded flow systems. part 3. the ensemble kalman filter, Journal of Fluid Mechanics 682 (2011) 289--303.

\bibitem{wang2025variational}
M.~Wang, T.~A. Zaki, Variational data assimilation in wall turbulence: from outer observations to wall stress and pressure, Journal of Fluid Mechanics 1008 (2025) A26.

\bibitem{suzuki2017estimation}
T.~Suzuki, Y.~Hasegawa, Estimation of turbulent channel flow at based on the wall measurement using a simple sequential approach, Journal of Fluid Mechanics 830 (2017) 760--796.

\bibitem{amaral2021resolvent}
F.~R. Amaral, A.~V. Cavalieri, E.~Martini, P.~Jordan, A.~Towne, Resolvent-based estimation of turbulent channel flow using wall measurements, Journal of Fluid Mechanics 927 (2021) A17.

\bibitem{adrian1988stochastic}
R.~J. Adrian, P.~Moin, Stochastic estimation of organized turbulent structure: homogeneous shear flow, Journal of Fluid Mechanics 190 (1988) 531--559.

\bibitem{marusic2010predictive}
I.~Marusic, R.~Mathis, N.~Hutchins, Predictive model for wall-bounded turbulent flow, Science 329~(5988) (2010) 193--196.

\bibitem{baars2016spectral}
W.~J. Baars, N.~Hutchins, I.~Marusic, Spectral stochastic estimation of high-reynolds-number wall-bounded turbulence for a refined inner-outer interaction model, Physical Review Fluids 1~(5) (2016) 054406.

\bibitem{encinar2019logarithmic}
M.~P. Encinar, J.~Jim{\'e}nez, Logarithmic-layer turbulence: a view from the wall, Physical Review Fluids 4~(11) (2019) 114603.

\bibitem{towne2018spectral}
A.~Towne, O.~T. Schmidt, T.~Colonius, Spectral proper orthogonal decomposition and its relationship to dynamic mode decomposition and resolvent analysis, Journal of Fluid Mechanics 847 (2018) 821--867.

\bibitem{guemes2019sensing}
A.~G{\"u}emes, S.~Discetti, A.~Ianiro, Sensing the turbulent large-scale motions with their wall signature, Physics of Fluids 31~(12) (2019).

\bibitem{guastoni2021convolutional}
L.~Guastoni, A.~G{\"u}emes, A.~Ianiro, S.~Discetti, P.~Schlatter, H.~Azizpour, R.~Vinuesa, Convolutional-network models to predict wall-bounded turbulence from wall quantities, Journal of Fluid Mechanics 928 (2021) A27.

\bibitem{balasubramanian2023predicting}
A.~G. Balasubramanian, L.~Guastoni, P.~Schlatter, H.~Azizpour, R.~Vinuesa, Predicting the wall-shear stress and wall pressure through convolutional neural networks, International Journal of Heat and Fluid Flow 103 (2023) 109200.

\bibitem{cuellar2024three}
A.~Cu{\'e}llar, A.~G{\"u}emes, A.~Ianiro, {\'O}.~Flores, R.~Vinuesa, S.~Discetti, Three-dimensional generative adversarial networks for turbulent flow estimation from wall measurements, Journal of Fluid Mechanics 991 (2024) A1.

\bibitem{hora2024physics}
G.~S. Hora, P.~Gentine, M.~Momen, M.~G. Giometto, Physics-informed data-driven reconstruction of turbulent wall-bounded flows from planar measurements, Physics of Fluids 36~(11) (2024).

\bibitem{yousif2023deep}
M.~Z. Yousif, L.~Yu, S.~Hoyas, R.~Vinuesa, H.~Lim, A deep-learning approach for reconstructing 3d turbulent flows from 2d observation data, Scientific Reports 13~(1) (2023) 2529.

\bibitem{wang2022observable}
Q.~Wang, M.~Wang, T.~A. Zaki, What is observable from wall data in turbulent channel flow?, Journal of Fluid Mechanics 941 (2022) A48.

\bibitem{arranz2024informative}
G.~Arranz, A.~Lozano-Dur{\'a}n, Informative and non-informative decomposition of turbulent flow fields, Journal of Fluid Mechanics 1000 (2024) A95.

\bibitem{guemes2021coarse}
A.~G{\"u}emes, S.~Discetti, A.~Ianiro, B.~Sirmacek, H.~Azizpour, R.~Vinuesa, From coarse wall measurements to turbulent velocity fields through deep learning, Physics of fluids 33~(7) (2021).

\bibitem{cuellar2024some}
A.~Cu{\'e}llar, A.~Ianiro, S.~Discetti, Some effects of limited wall-sensor availability on flow estimation with 3d-gans, Theoretical and Computational Fluid Dynamics 38~(5) (2024) 729--745.

\bibitem{ruhling2023dyffusion}
S.~R{\"u}hling~Cachay, B.~Zhao, H.~Joren, R.~Yu, Dyffusion: A dynamics-informed diffusion model for spatiotemporal forecasting, Advances in neural information processing systems 36 (2023) 45259--45287.

\bibitem{shu2023physics}
D.~Shu, Z.~Li, A.~B. Farimani, A physics-informed diffusion model for high-fidelity flow field reconstruction, Journal of Computational Physics 478 (2023) 111972.

\bibitem{kohl2023benchmarking}
G.~Kohl, L.-W. Chen, N.~Thuerey, Benchmarking autoregressive conditional diffusion models for turbulent flow simulation, arXiv preprint arXiv:2309.01745 (2023).

\bibitem{li2023multi}
T.~Li, M.~Buzzicotti, L.~Biferale, F.~Bonaccorso, S.~Chen, M.~Wan, Multi-scale reconstruction of turbulent rotating flows with proper orthogonal decomposition and generative adversarial networks, Journal of Fluid Mechanics 971 (2023) A3.

\bibitem{gao2024bayesian}
H.~Gao, X.~Han, X.~Fan, L.~Sun, L.-P. Liu, L.~Duan, J.-X. Wang, Bayesian conditional diffusion models for versatile spatiotemporal turbulence generation, Computer Methods in Applied Mechanics and Engineering 427 (2024) 117023.

\bibitem{gao2024generative}
H.~Gao, S.~Kaltenbach, P.~Koumoutsakos, Generative learning for forecasting the dynamics of high-dimensional complex systems, Nature Communications 15~(1) (2024) 8904.

\bibitem{shehataimproved}
Y.~Shehata, B.~Holzschuh, N.~Thuerey, Improved sampling of diffusion models in fluid dynamics with tweedie's formula, in: The Thirteenth International Conference on Learning Representations, 2025.

\bibitem{gao2025generative}
H.~Gao, S.~Kaltenbach, P.~Koumoutsakos, Generative learning of the solution of parametric partial differential equations using guided diffusion models and virtual observations, Computer Methods in Applied Mechanics and Engineering 435 (2025) 117654.

\bibitem{dong2024data}
X.~Dong, C.~Chen, J.-L. Wu, Data-driven stochastic closure modeling via conditional diffusion model and neural operator, arXiv preprint arXiv:2408.02965 (2024).

\bibitem{molinaro2024generative}
R.~Molinaro, S.~Lanthaler, B.~Raoni{\'c}, T.~Rohner, V.~Armegioiu, S.~Simonis, D.~Grund, Y.~Ramic, Z.~Y. Wan, F.~Sha, et~al., Generative ai for fast and accurate statistical computation of fluids, arXiv preprint arXiv:2409.18359 (2024).

\bibitem{zhuang2025spatially}
Y.~Zhuang, S.~Cheng, K.~Duraisamy, Spatially-aware diffusion models with cross-attention for global field reconstruction with sparse observations, Computer Methods in Applied Mechanics and Engineering 435 (2025) 117623.

\bibitem{du2024conditional}
P.~Du, M.~H. Parikh, X.~Fan, X.-Y. Liu, J.-X. Wang, Conditional neural field latent diffusion model for generating spatiotemporal turbulence, Nature Communications 15~(1) (2024) 10416.

\bibitem{fan2025neural}
X.~Fan, D.~Akhare, J.-X. Wang, Neural differentiable modeling with diffusion-based super-resolution for two-dimensional spatiotemporal turbulence, Computer Methods in Applied Mechanics and Engineering 433 (2025) 117478.

\bibitem{liu2024confild}
X.-Y. Liu, M.~H. Parikh, X.~Fan, P.~Du, Q.~Wang, Y.-F. Chen, J.-X. Wang, Confild-inlet: Synthetic turbulence inflow using generative latent diffusion models with neural fields, arXiv preprint arXiv:2411.14378 (2024).

\bibitem{lipman2022flow}
Y.~Lipman, R.~T. Chen, H.~Ben-Hamu, M.~Nickel, M.~Le, Flow matching for generative modeling, arXiv preprint arXiv:2210.02747 (2022).

\bibitem{maddox2019simple}
W.~J. Maddox, P.~Izmailov, T.~Garipov, D.~P. Vetrov, A.~G. Wilson, A simple baseline for bayesian uncertainty in deep learning, Advances in neural information processing systems 32 (2019).

\bibitem{bae2021life}
H.~J. Bae, M.~Lee, Life cycle of streaks in the buffer layer of wall-bounded turbulence, Physical Review Fluids 6~(6) (2021) 064603.

\bibitem{dhariwal2021diffusion}
P.~Dhariwal, A.~Nichol, Diffusion models beat gans on image synthesis, Advances in neural information processing systems 34 (2021) 8780--8794.

\bibitem{jacobsen2025cocogen}
C.~Jacobsen, Y.~Zhuang, K.~Duraisamy, Cocogen: Physically consistent and conditioned score-based generative models for forward and inverse problems, SIAM Journal on Scientific Computing 47~(2) (2025) C399--C425.

\bibitem{fu2024unveil}
H.~Fu, Z.~Yang, M.~Wang, M.~Chen, Unveil conditional diffusion models with classifier-free guidance: A sharp statistical theory, arXiv preprint arXiv:2403.11968 (2024).

\bibitem{chung2022diffusion}
H.~Chung, J.~Kim, M.~T. Mccann, M.~L. Klasky, J.~C. Ye, Diffusion posterior sampling for general noisy inverse problems, arXiv preprint arXiv:2209.14687 (2022).

\bibitem{adrian1979conditional}
R.~J. Adrian, Conditional eddies in isotropic turbulence, The Physics of Fluids 22~(11) (1979) 2065--2070.

\bibitem{spalding1961single}
D.~B. Spalding, et~al., A single formula for the law of the wall, Journal of Applied Mechanics 28~(3) (1961) 455--458.

\bibitem{yang2017log}
X.~I. Yang, G.~I. Park, P.~Moin, Log-layer mismatch and modeling of the fluctuating wall stress in wall-modeled large-eddy simulations, Physical review fluids 2~(10) (2017) 104601.

\bibitem{perez2018film}
E.~Perez, F.~Strub, H.~de~Vries, V.~Dumoulin, A.~C. Courville, Film: Visual reasoning with a general conditioning layer, in: AAAI, 2018.

\end{thebibliography}

\section*{Acknowledgements}
 The authors would like to acknowledge the funds from Office of Naval Research under award numbers N00014-23-1-2071 and National Science Foundation under award numbers OAC-2047127.

\section*{Compliance with Ethical Standards}
Conflict of Interest: The authors declare that they have no conflict of interest.

\appendix 

\section{Network architectures and training details}
\label{sec:appendix-training}

This appendix outlines the architecture and training specifications of the neural networks employed in our framework. Particularly, two key components are built: the flow matching model that learns the optimal flow transport ${\bm \nu}_{\bm \theta}$ between the standard normal distribution $\mathcal{N}(\bm{0}, \mathbf{I})$ and that of the velocity fluctuations ${\bm u}'_i$, and the measurement operator $\mathcal{F}_{\bm \phi}$, which maps velocity fluctuations ${\bm u}’_i$ to wall quantities ${\bm \Phi}_{\mathrm{wall}}$ with quantified epistemic uncertainty via SWAG.

\begin{figure}[H]
    \centering
    \includegraphics[width=0.9\linewidth]{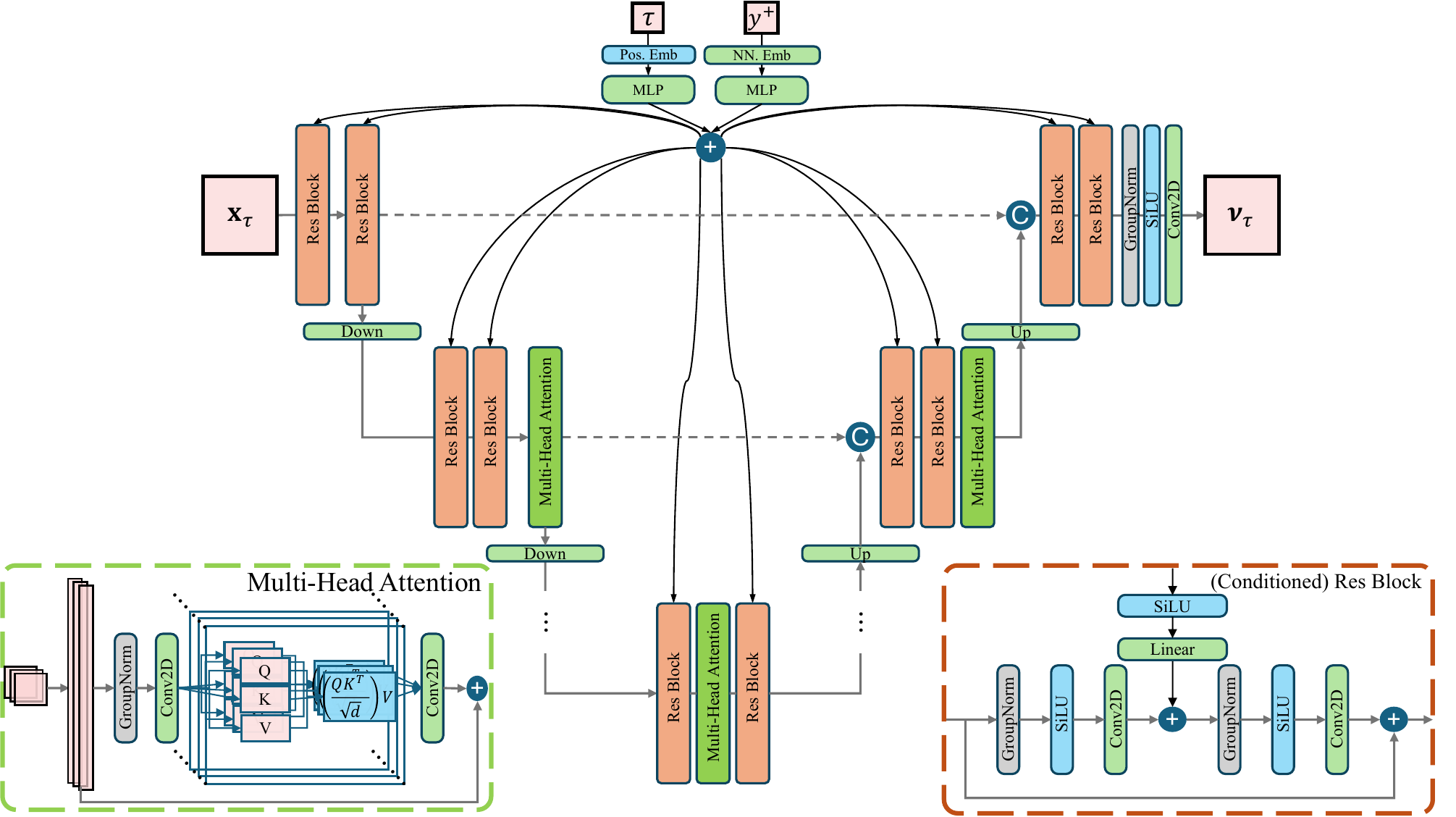}    
    \caption{Network architecture of the U-Net used for estimating the neural velocity ${\bm\nu}_{\bm\theta}$ within the flow matching framework.}
    \label{fig:net-arch-ours}
\end{figure}

\begin{figure}[H]
    \centering
    \includegraphics[width=.9\linewidth]{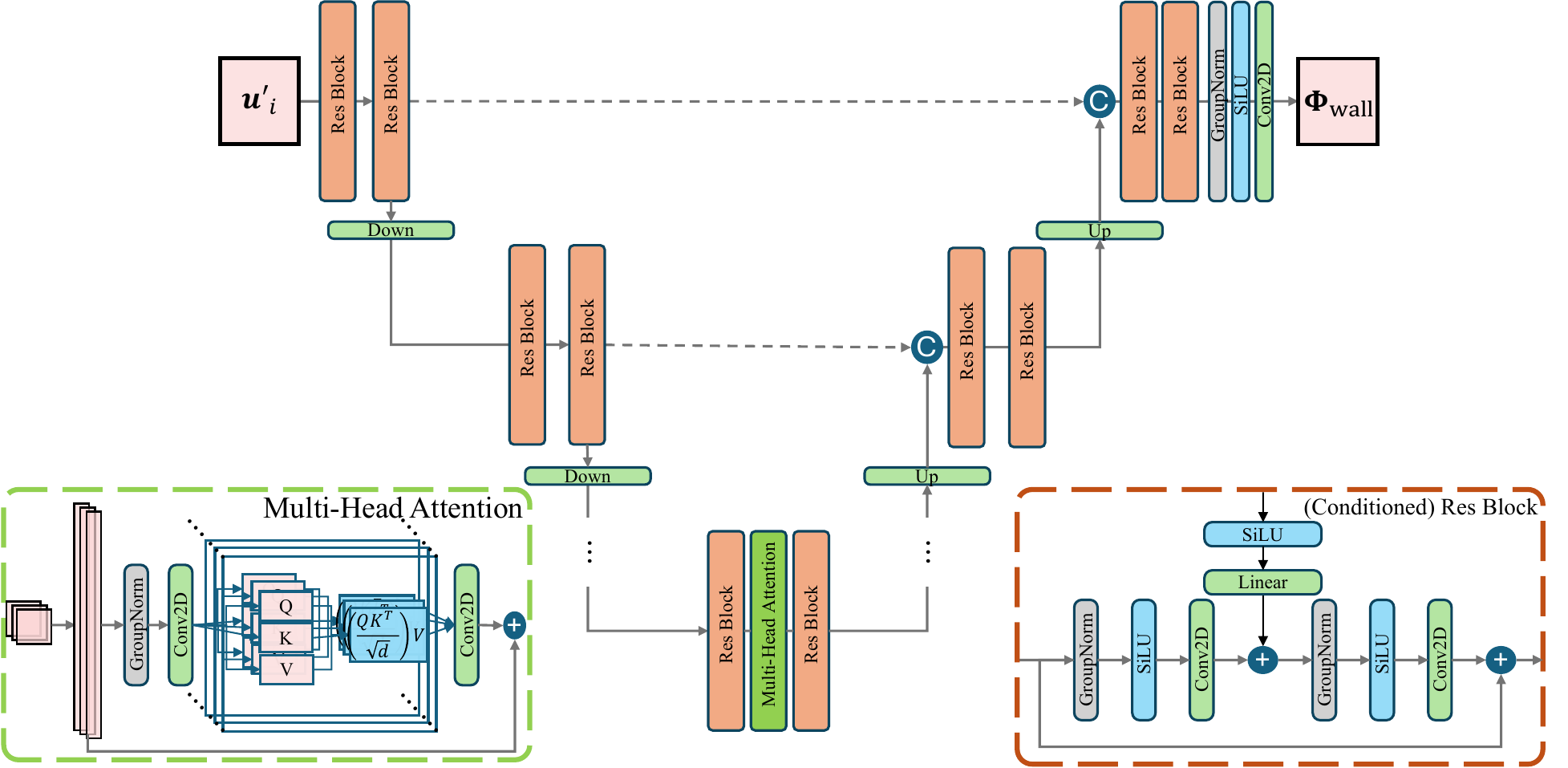}
    \caption{Network architecture of the U-Net used for estimating for estimating wall quantities ${\bm \Phi}_{\mathrm{wall}}$ by using the velocity fluctuations ${\bm u}'_i$ as input.}
    \label{fig:net-arch-ours-for}
\end{figure}

Both networks adopt a modernized U-Net architecture with residual connections and hierarchical resolution structure, although their configurations differ in size and input domain. Schematics of the network architecture for the flow transport model and the measurement operator are shown in Figures~\ref{fig:net-arch-ours} and~\ref{fig:net-arch-ours-for}, respectively. The U-Net consists of downsampling and upsampling residual blocks at each resolution level. Each block incorporates SiLU activations, convolutional layers, optional self-attention mechanisms, and trainable downsampling or upsampling layers. A central bottleneck module contains a self-attention layer sandwiched between two residual blocks. For the flow transport model, the network is further conditioned on the fictitious transport time $\tau$ using Feature-wise Linear Modulation (FiLM) layers~\cite{perez2018film}. Class embeddings are used to encode the wall-normal position $y^+ \in {5, 20, 40}$ during training.
 
The detailed architectural configurations and training parameters are summarized in Table~\ref{tab:train-params}. Notably, the flow-matching network operates on full-domain fields ($3 \times 320 \times 200$), whereas the measurement operator is trained patch-wise on subdomains of size $3 \times 32 \times 32$ to enhance statistical diversity and enable efficient uncertainty quantification.

\begin{table}[htb]
\begin{tabular}{|l|l|l|}
\hline
Hyperparameters/Networks   & Flow Operator (${\bm \nu}_{\bm \theta}$)   & Measurement Operator ($\mathcal{F}_{\bm \phi}$)   \\ \hline
Trainable parameters                         & $105,260,163$ & $7,896,643$ \\ \hline
Input and Output Data Dimensions                   & $ 3\times 320 \times 200$                & $3 \times 32 \times 32$                 \\ \hline
Number of levels                & $4$         & $4$       \\ \hline
Number of Channels at each resolution & $[128,\space 256,\space 512,\space 512]$ & $[64,\space 128,\space 128,\space 128]$ \\ \hline
Number of Residual Blocks at each resolution & $2$         & $2$       \\ \hline
Attention  Resolution                        & $40 \times 25$        & -         \\ \hline
Number of Classes                            & $3$         & -         \\ \hline
\end{tabular}
\small \caption{Architectural and training details of flow-matching models and SWAG-based forward operator.}
\label{tab:train-params}
\end{table}

\section{Deterministic CNN baseline}
\label{sec:appendix-baseline}

\begin{figure}[t!]
    \centering
    \includegraphics[width=.9\linewidth]{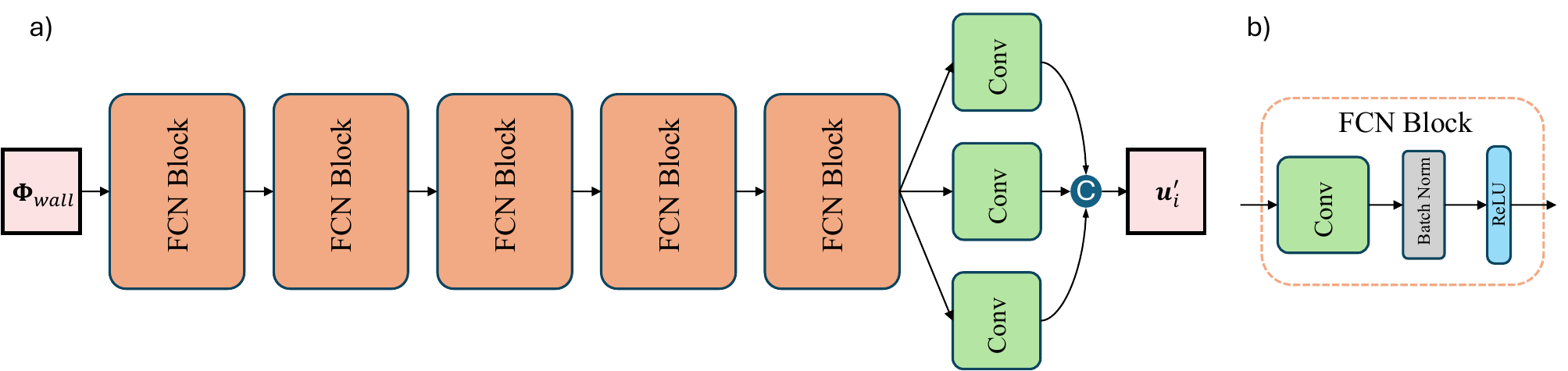}
    \caption{a) Network architecture of baseline CNN forward model, b) Components of the FCN block}
    \label{fig:net-arch-baseline}
\end{figure}

To benchmark the performance of our proposed probabilistic framework, we compare against a representative deterministic data-driven model introduced by Guastoni et al.\cite{guastoni2021convolutional}. Their architecture is a fully convolutional neural network (FCN) designed to predict velocity fluctuations from wall-based measurements, specifically the streamwise and spanwise wall shear stresses ($\tau_u$, $\tau_w$) and wall pressure ($p$). The model is trained using a mean squared error loss on the velocity fluctuations, and the network structure is illustrated in Figure~\ref{fig:net-arch-baseline}. The FCN consists of a series of convolutional residual blocks, where each block includes batch normalization and ReLU activation (except for the final layer). A key architectural feature of the original design is its explicit enforcement of periodic boundary conditions, achieved by aligning the convolutional kernel structure with the periodicity of the input domain. This ensures that the model strictly respects the physical periodicity of the channel flow and leverages the locality of convolutional filters effectively.


\section{Extra Results}
\label{sec:appendix-extra-res}
This section compiles supplementary visualizations and quantitative results that were omitted from the main text for clarity and conciseness. 

\subsection{Conditional generation}
\label{sec:appendix-cond-gen}

\begin{figure}[H]
    \centering
    \includegraphics[width=1.0\textwidth]{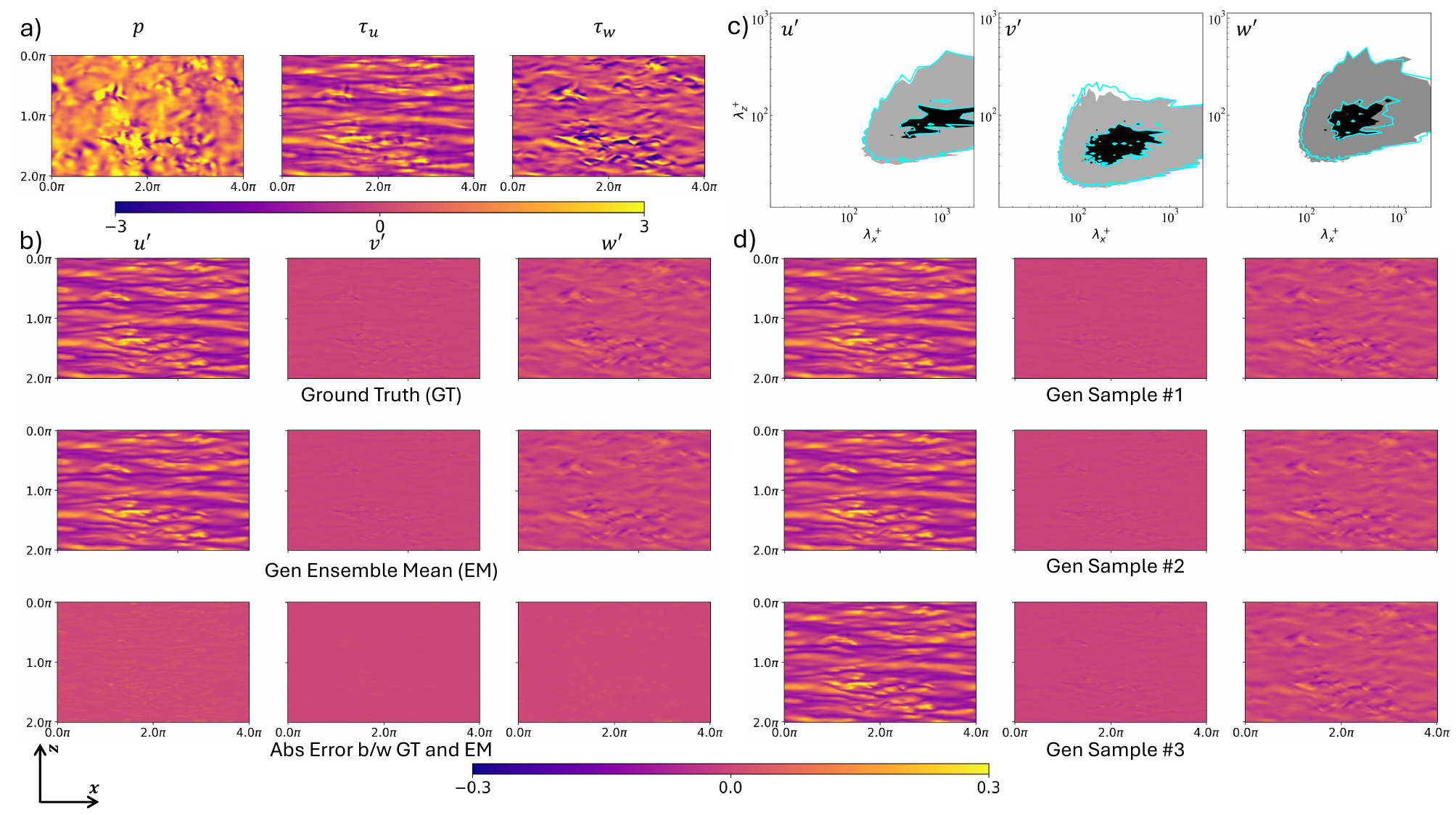}
    \caption{(a) An example of fully observed wall measurements $\bm{\Phi}_{\mathrm{wall}} = [p, \tau_u, \tau_w]$ used as the condition for generating corresponding velocity fluctuations $\bm{u}'_i$ for $y^+=5$. (b) Comparison between the ground truth velocity fluctuations (top row), the ensemble mean of $50$ conditionally generated samples (middle row), and the absolute error between the ensemble mean and ground truth (bottom row), for all three velocity components. (c) Pre-multiplied two-dimensional energy spectra of the generated samples (\protect\tikz[baseline=-0.5ex]\protect\draw [cyan, thick] (0,0) -- (0.5,0); lines) versus ground truth (\protect\tikz \protect\fill[gray!40] (0,0) rectangle (0.5,0.25); contours), computed from $500$ different test cases. Streamwise and spanwise wavelengths $\lambda_x^+$ and $\lambda_z^+$ are normalized by wall units; contours indicate $10\%$, $50\%$, and $90\%$ of the maximum ground-truth energy. (d) Three representative samples from the ensemble, illustrating the diversity of generated flow realizations consistent with the wall measurements in (a).}
    \label{fig:full_contour_and_stats_y5}
\end{figure}

\begin{figure}[H]
    \centering
    \includegraphics[width=1.0\textwidth]{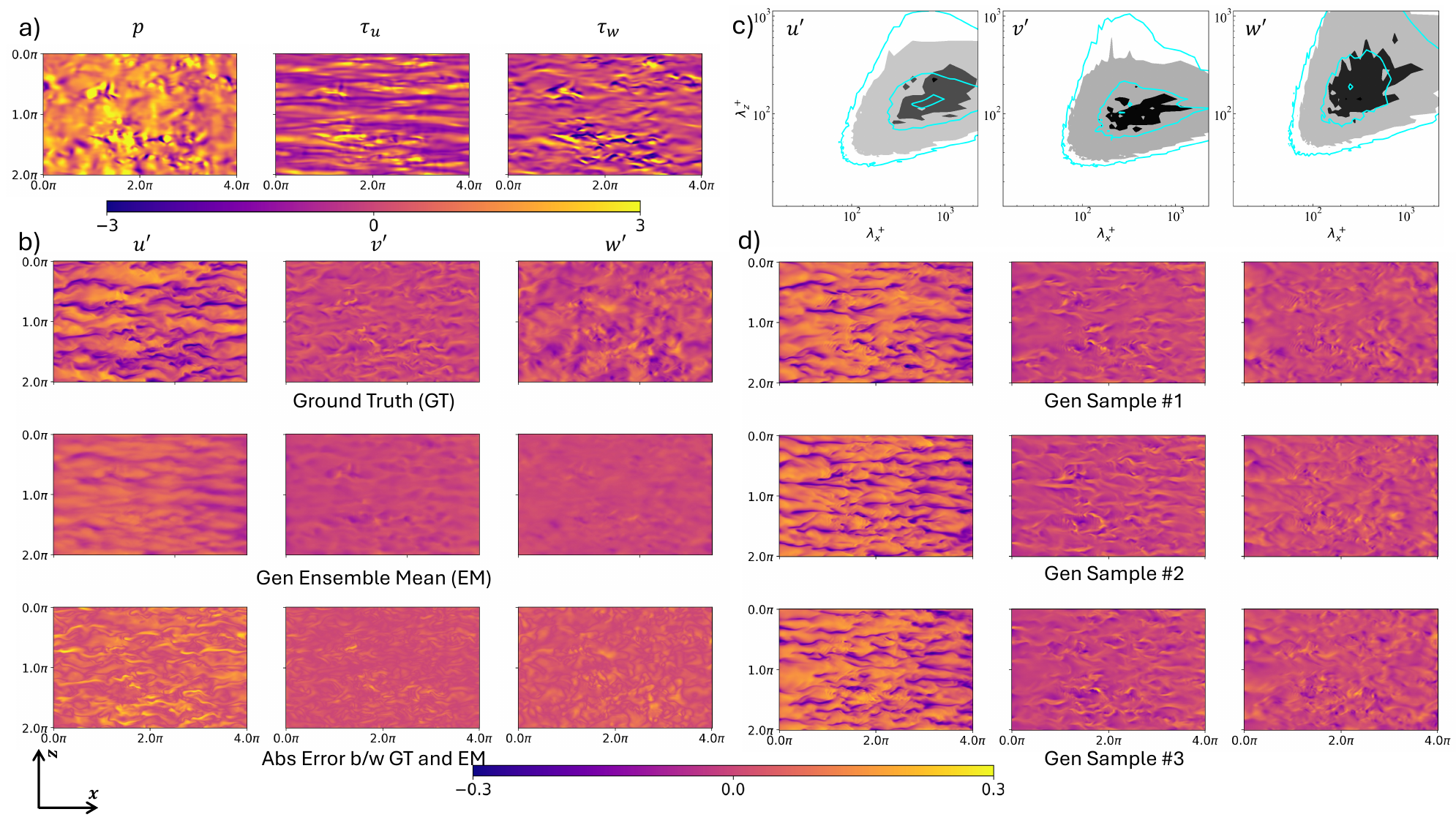}
    \caption{(a) An example of fully observed wall measurements $\bm{\Phi}_{\mathrm{wall}} = [p, \tau_u, \tau_w]$ used as the condition for generating corresponding velocity fluctuations $\bm{u}'_i$ for $y^+=40$. (b) Comparison between the ground truth velocity fluctuations (top row), the ensemble mean of $50$ conditionally generated samples (middle row), and the absolute error between the ensemble mean and ground truth (bottom row), for all three velocity components. (c) Pre-multiplied two-dimensional energy spectra of the generated samples (\protect\tikz[baseline=-0.5ex]\protect\draw [cyan, thick] (0,0) -- (0.5,0); lines) versus ground truth (\protect\tikz \protect\fill[gray!40] (0,0) rectangle (0.5,0.25); contours), computed from $500$ different test cases. Streamwise and spanwise wavelengths $\lambda_x^+$ and $\lambda_z^+$ are normalized by wall units; contours indicate $10\%$, $50\%$, and $90\%$ of the maximum ground-truth energy. (d) Three representative samples from the ensemble, illustrating the diversity of generated flow realizations consistent with the wall measurements in (a).}
    \label{fig:full_contour_and_stats_y40}
\end{figure}

\begin{figure}[H]
    \centering
    \includegraphics[width=1.0\textwidth]{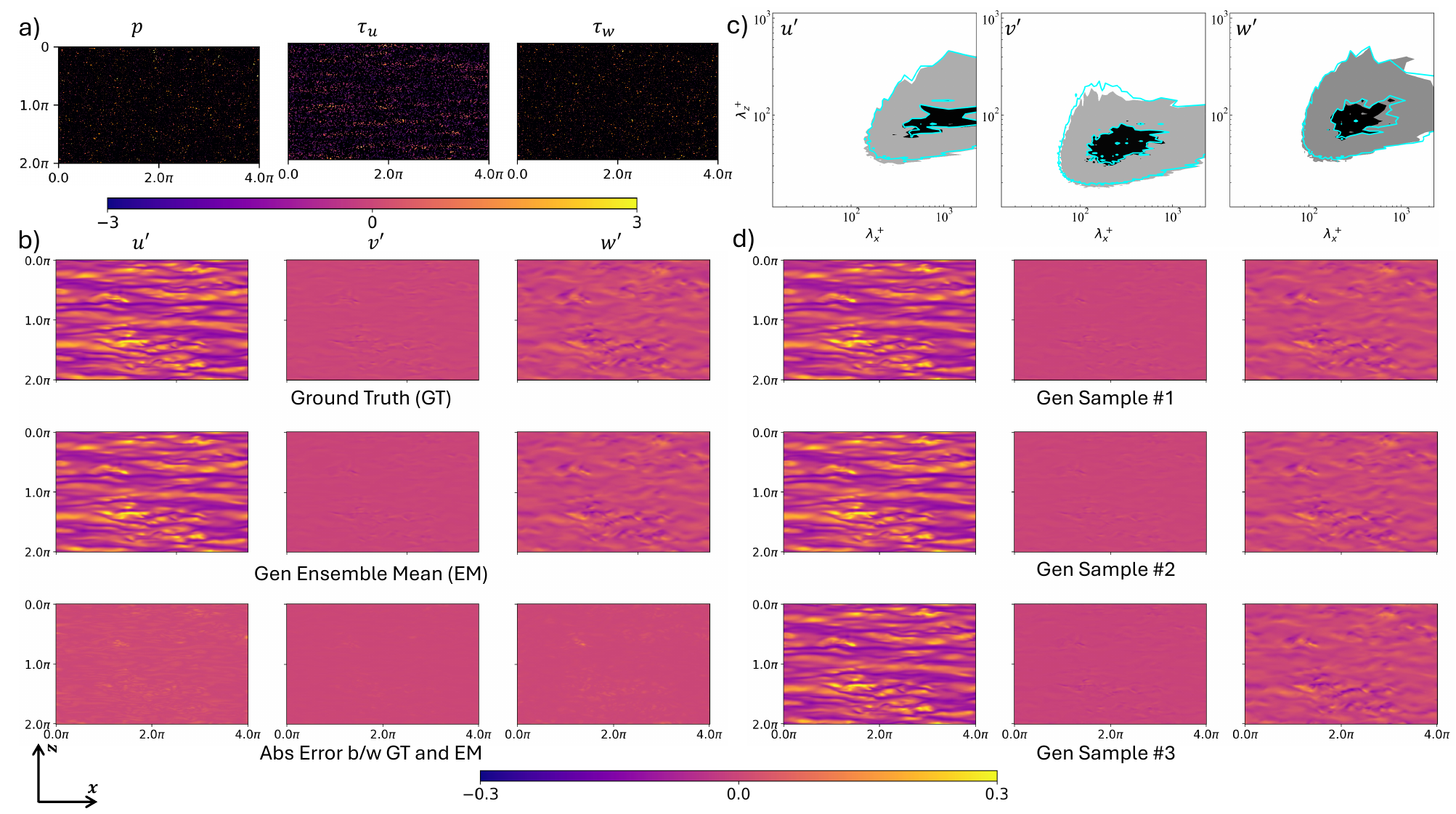}
    \caption{(a) An example of sparse wall measurements ($10\%$ data availability) ${\bm \Phi}_\mathrm{wall}$ for which the corresponding velocity fluctuations $\bm{u}'_i$ at $y^+ =5$ are generated. (b) Comparison between the ground truth velocity fluctuations (top row), the ensemble mean of $50$ conditionally generated samples (middle row), and the absolute error between the ensemble mean and ground truth (bottom row), for all three velocity components. (c) Pre-multiplied two-dimensional energy spectra of the generated samples (\protect\tikz[baseline=-0.5ex]\protect\draw [cyan, thick] (0,0) -- (0.5,0); lines) versus ground truth (\protect\tikz \protect\fill[gray!40] (0,0) rectangle (0.5,0.25); contours), computed from $500$ different test cases. Streamwise and spanwise wavelengths $\lambda_x^+$ and $\lambda_z^+$ are normalized by wall units; contours indicate $10\%$, $50\%$, and $90\%$ of the maximum ground-truth energy. (d) Three representative samples from the ensemble, illustrating the diversity of generated flow realizations consistent with the wall measurements in (a).}
    \label{sparse10_contour_and_stats_y5}
\end{figure}

\begin{figure}[H]
    \centering
    \includegraphics[width=1.0\textwidth]{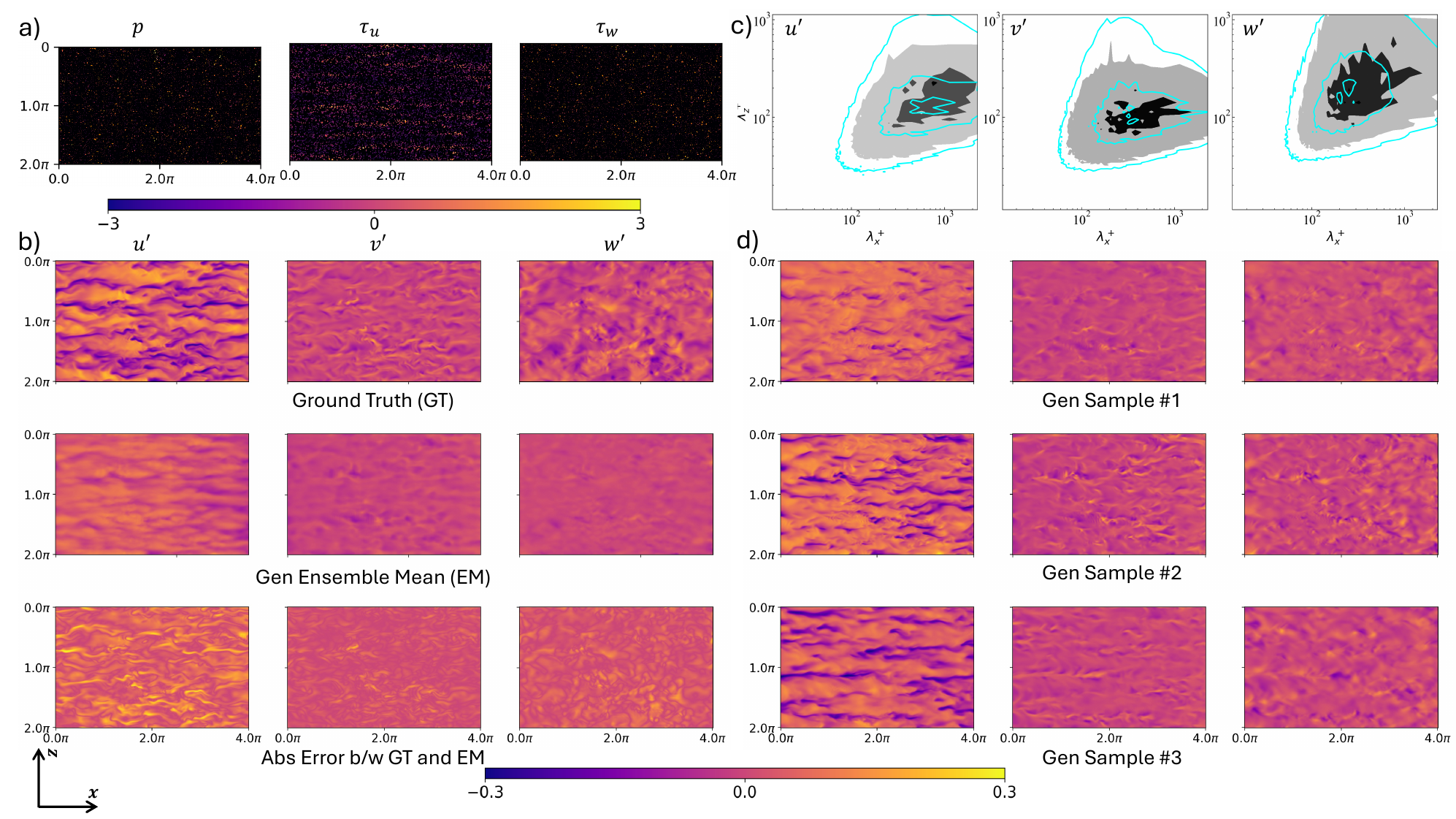}
    \caption{(a) An example of sparse wall measurements ($10\%$ data availability) ${\bm \Phi}_\mathrm{wall}$ for which the corresponding velocity fluctuations $\bm{u}'_i$ at $y^+ =40$ are generated.(b) Comparison between the ground truth velocity fluctuations (top row), the ensemble mean of $50$ conditionally generated samples (middle row), and the absolute error between the ensemble mean and ground truth (bottom row), for all three velocity components. (c) Pre-multiplied two-dimensional energy spectra of the generated samples (\protect\tikz[baseline=-0.5ex]\protect\draw [cyan, thick] (0,0) -- (0.5,0); lines) versus ground truth (\protect\tikz \protect\fill[gray!40] (0,0) rectangle (0.5,0.25); contours), computed from $500$ different test cases. Streamwise and spanwise wavelengths $\lambda_x^+$ and $\lambda_z^+$ are normalized by wall units; contours indicate $10\%$, $50\%$, and $90\%$ of the maximum ground-truth energy. (d) Three representative samples from the ensemble, illustrating the diversity of generated flow realizations consistent with the wall measurements in (a).}
    \label{sparse10_contour_and_stats_y40}
\end{figure}

\begin{figure}[H]
    \centering
    \includegraphics[width=1.0\textwidth]{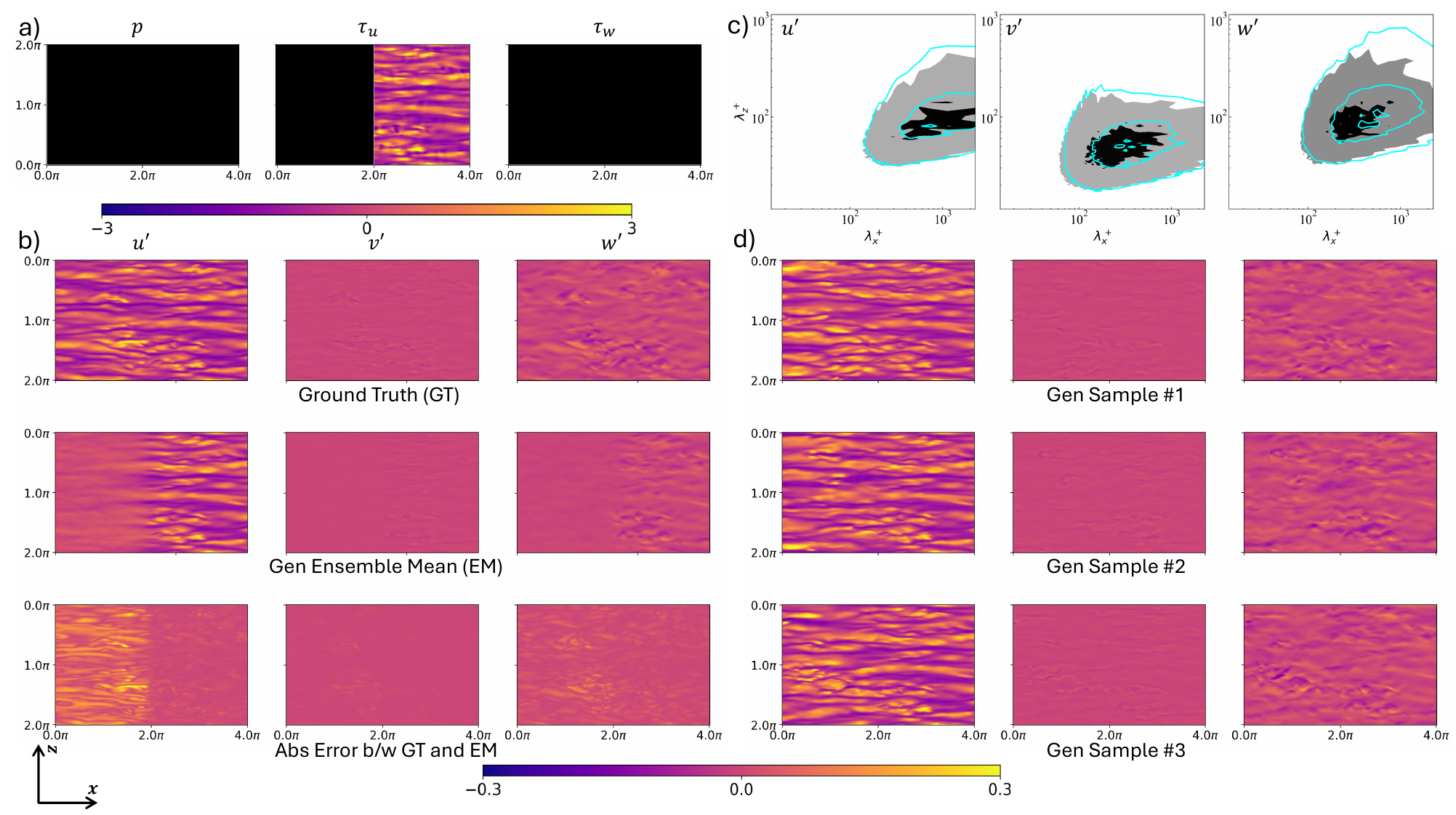}
    \caption{(a) An example of partial ($\mathbf{y} = \left(0, \tau_u, 0\right) \cdot \mathbf{1}_{2\pi \leq x \leq 4\pi \land 0 \leq z \leq 2\pi}$) wall measurements ${\bm \Phi}_\mathrm{wall}$ for which the corresponding velocity fluctuations $\bm{u}'_i$ at $y^+ =5$ are generated. (b) Comparison between the ground truth velocity fluctuations (top row), the ensemble mean of $50$ conditionally generated samples (middle row), and the absolute error between the ensemble mean and ground truth (bottom row), for all three velocity components. (c) Pre-multiplied two-dimensional energy spectra of the generated samples (\protect\tikz[baseline=-0.5ex]\protect\draw [cyan, thick] (0,0) -- (0.5,0); lines) versus ground truth (\protect\tikz \protect\fill[gray!40] (0,0) rectangle (0.5,0.25); contours), computed from $500$ different test cases. Streamwise and spanwise wavelengths $\lambda_x^+$ and $\lambda_z^+$ are normalized by wall units; contours indicate $10\%$, $50\%$, and $90\%$ of the maximum ground-truth energy. (d) Three representative samples from the ensemble, illustrating the diversity of generated flow realizations consistent with the wall measurements in (a).}
    \label{partial_contour_and_stats_y5}
\end{figure}

\begin{figure}[H]
    \centering
    \includegraphics[width=1.0\textwidth]{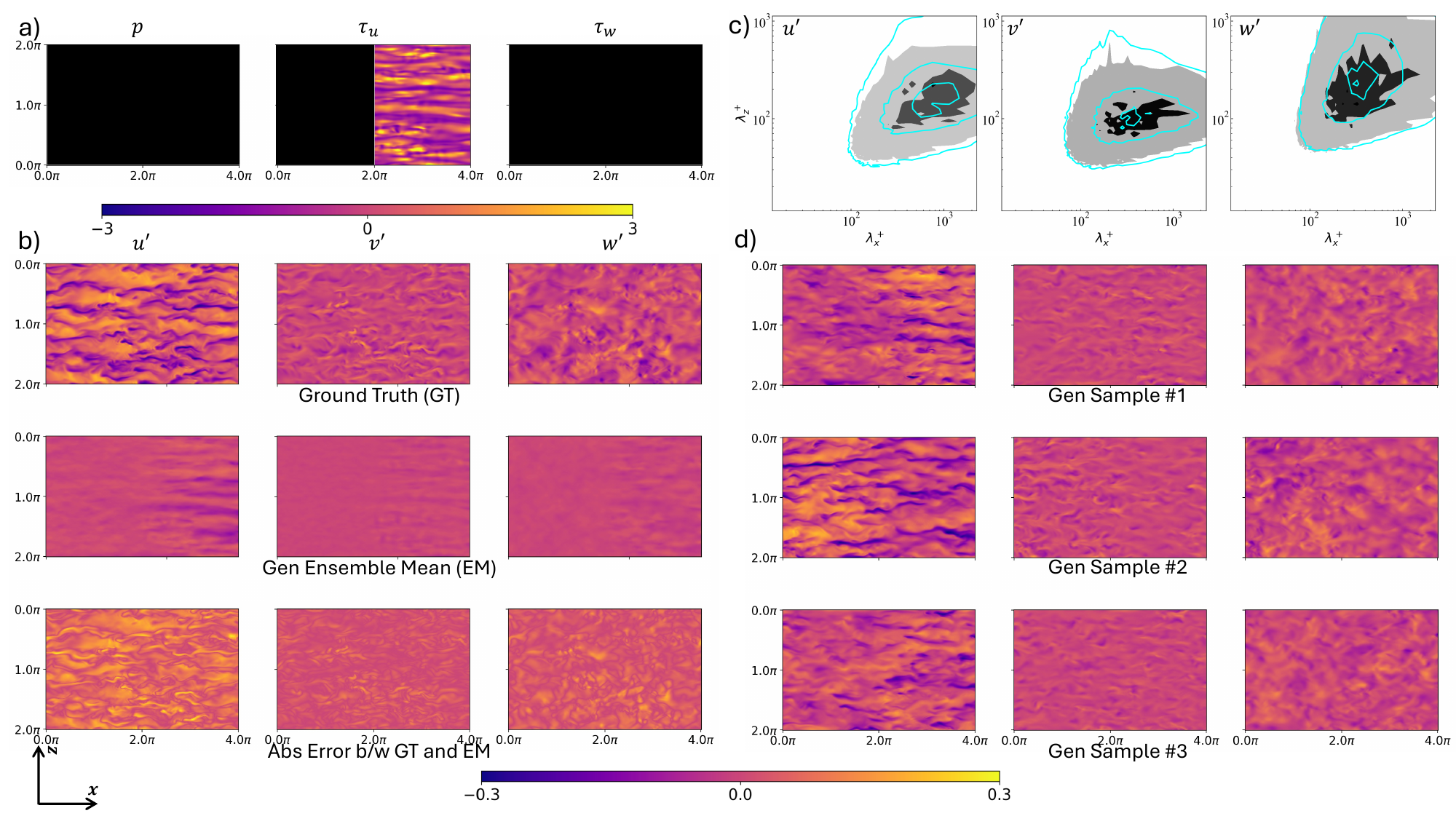}
    \caption{(a) An example of partial ($\mathbf{y}(x, z) = \left(0, \tau_u(x, z), 0\right) \cdot \mathbf{1}_{2\pi \leq x \leq 4\pi \land 0 \leq z \leq 2\pi}$) wall measurements ${\bm \Phi}_\mathrm{wall}$ for which the corresponding velocity fluctuations $\bm{u}'_i$ at $y^+ =40$ are generated. (b) The ground truth velocity fluctuations (top row), ensemble mean of $50$ velocity fluctuations samples generated using proposed method (middle row), and the absolute error between the ground truth and ensemble mean velocity fluctuations (bottom row). (c) The comparison of the pre-multiplied two-dimensional energy spectra calculated between ground truth (\protect\tikz \protect\fill[gray!40] (0,0) rectangle (0.5,0.25); contours) and generated velocity fluctuations (\protect\tikz[baseline=-0.5ex]\protect\draw [cyan, thick] (0,0) -- (0.5,0); lines) corresponding to $500$ different wall measurements, where $\lambda_x^+$ and $\lambda_z^+$ are streamwise and spanwise wavelength normalized by wall unit. The contour levels contain $10\%$, $50\%$ and $90\%$ of the maximum ground-truth energy spectra. (d) $3$ randomly chosen velocity fluctuation samples generated for the prescribed wall measurements shown in subfigure (a).}
    \label{partial_contour_and_stats_y40}
\end{figure}

\begin{figure}[H]
    \centering
    \includegraphics[width=1.0\textwidth]{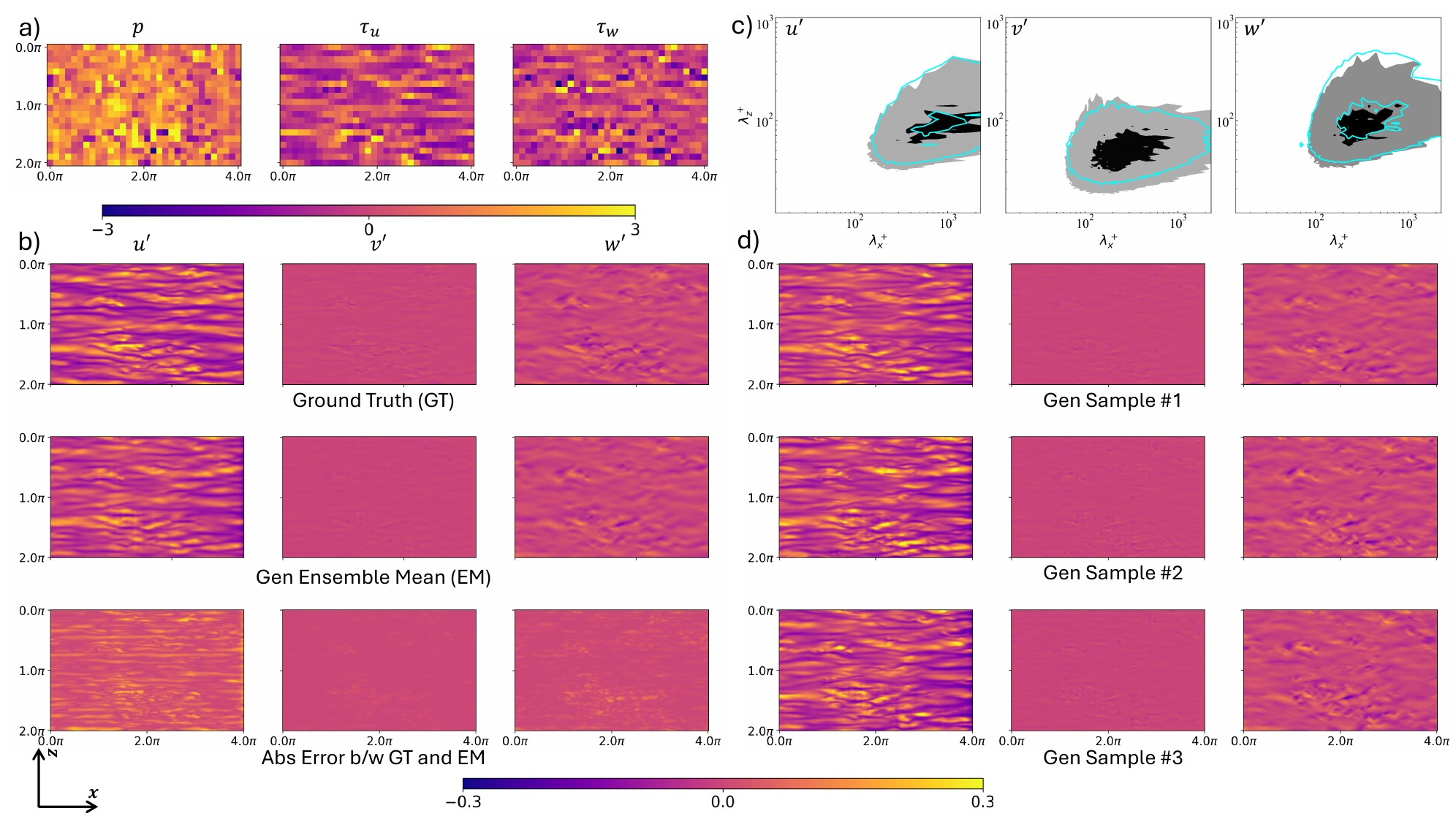}
    \caption{(a) An example of low-resolution ($\frac{1}{100}\times$) $\mathbf{y}_{LR}(x, z) = D({\bm \Phi}_{\mathrm{wall}})$ wall measurements ${\bm \Phi}_\mathrm{wall}$ for which the corresponding velocity fluctuations $\bm{u}'_i$ at $y^+ =5$ are generated. (b) Comparison between the ground truth velocity fluctuations (top row), the ensemble mean of $50$ conditionally generated samples (middle row), and the absolute error between the ensemble mean and ground truth (bottom row), for all three velocity components. (c) Pre-multiplied two-dimensional energy spectra of the generated samples (\protect\tikz[baseline=-0.5ex]\protect\draw [cyan, thick] (0,0) -- (0.5,0); lines) versus ground truth (\protect\tikz \protect\fill[gray!40] (0,0) rectangle (0.5,0.25); contours), computed from $500$ different test cases. Streamwise and spanwise wavelengths $\lambda_x^+$ and $\lambda_z^+$ are normalized by wall units; contours indicate $10\%$, $50\%$, and $90\%$ of the maximum ground-truth energy. (d) Three representative samples from the ensemble, illustrating the diversity of generated flow realizations consistent with the wall measurements in (a).}
    \label{lowres_contour_and_stats_y5}
\end{figure}

\begin{figure}[H]
    \centering
    \includegraphics[width=1.0\textwidth]{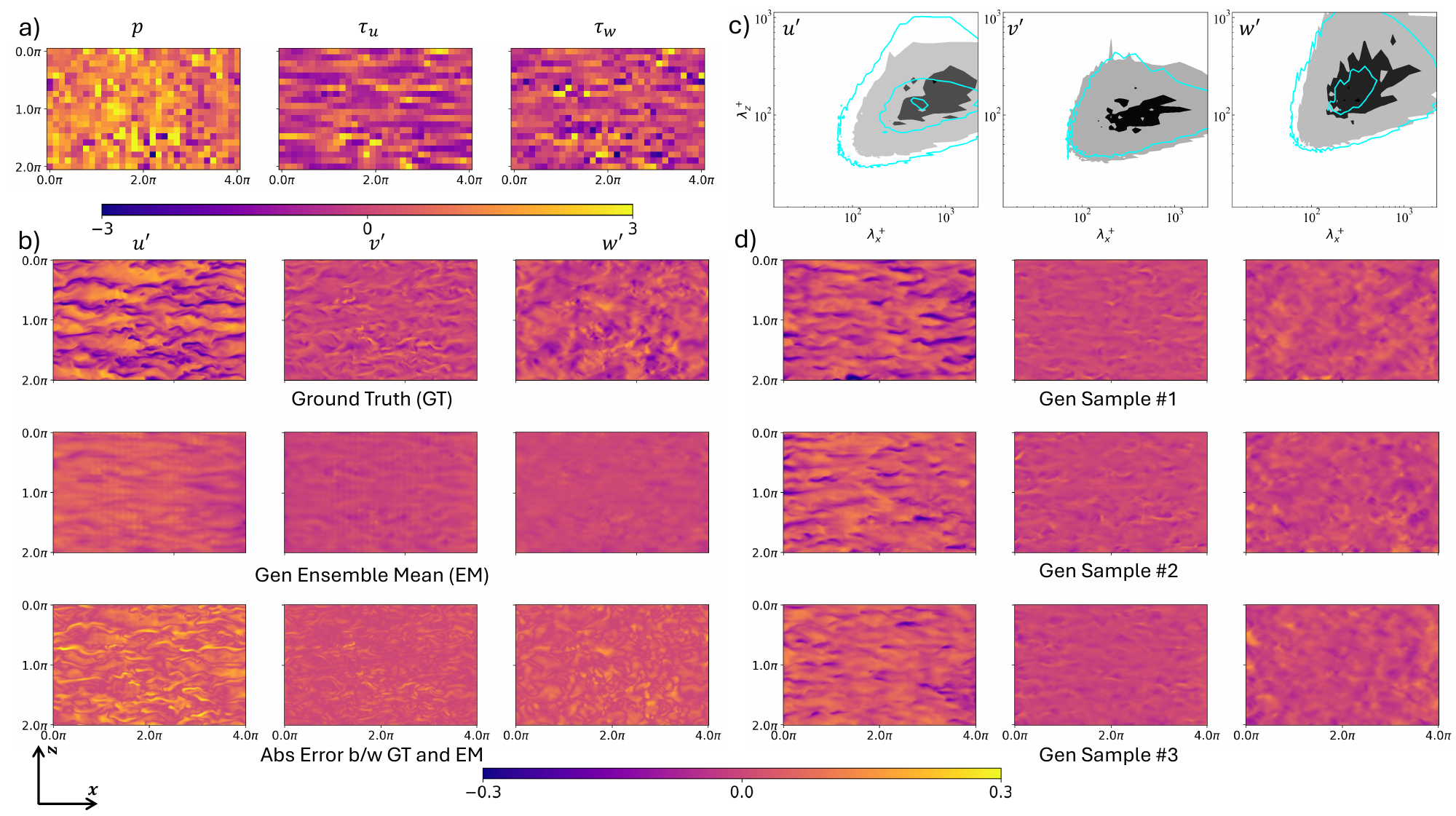}
    \caption{(a) An example of low-resolution ($\frac{1}{100}\times$) $\mathbf{y}_{LR}(x, z) = D({\bm \Phi}_{\mathrm{wall}})$ wall measurements ${\bm \Phi}_\mathrm{wall}$ for which the corresponding velocity fluctuations $\bm{u}'_i$ at $y^+ =40$ are generated. (b) Comparison between the ground truth velocity fluctuations (top row), the ensemble mean of $50$ conditionally generated samples (middle row), and the absolute error between the ensemble mean and ground truth (bottom row), for all three velocity components. (c) Pre-multiplied two-dimensional energy spectra of the generated samples (\protect\tikz[baseline=-0.5ex]\protect\draw [cyan, thick] (0,0) -- (0.5,0); lines) versus ground truth (\protect\tikz \protect\fill[gray!40] (0,0) rectangle (0.5,0.25); contours), computed from $500$ different test cases. Streamwise and spanwise wavelengths $\lambda_x^+$ and $\lambda_z^+$ are normalized by wall units; contours indicate $10\%$, $50\%$, and $90\%$ of the maximum ground-truth energy. (d) Three representative samples from the ensemble, illustrating the diversity of generated flow realizations consistent with the wall measurements in (a).}
    \label{lowres_contour_and_stats_y40}
\end{figure}

\subsection{Forward flow-to-wall model}
\label{sec:appendix-forward}

\begin{figure}[H]
    \centering
    \includegraphics[width=0.85\textwidth]{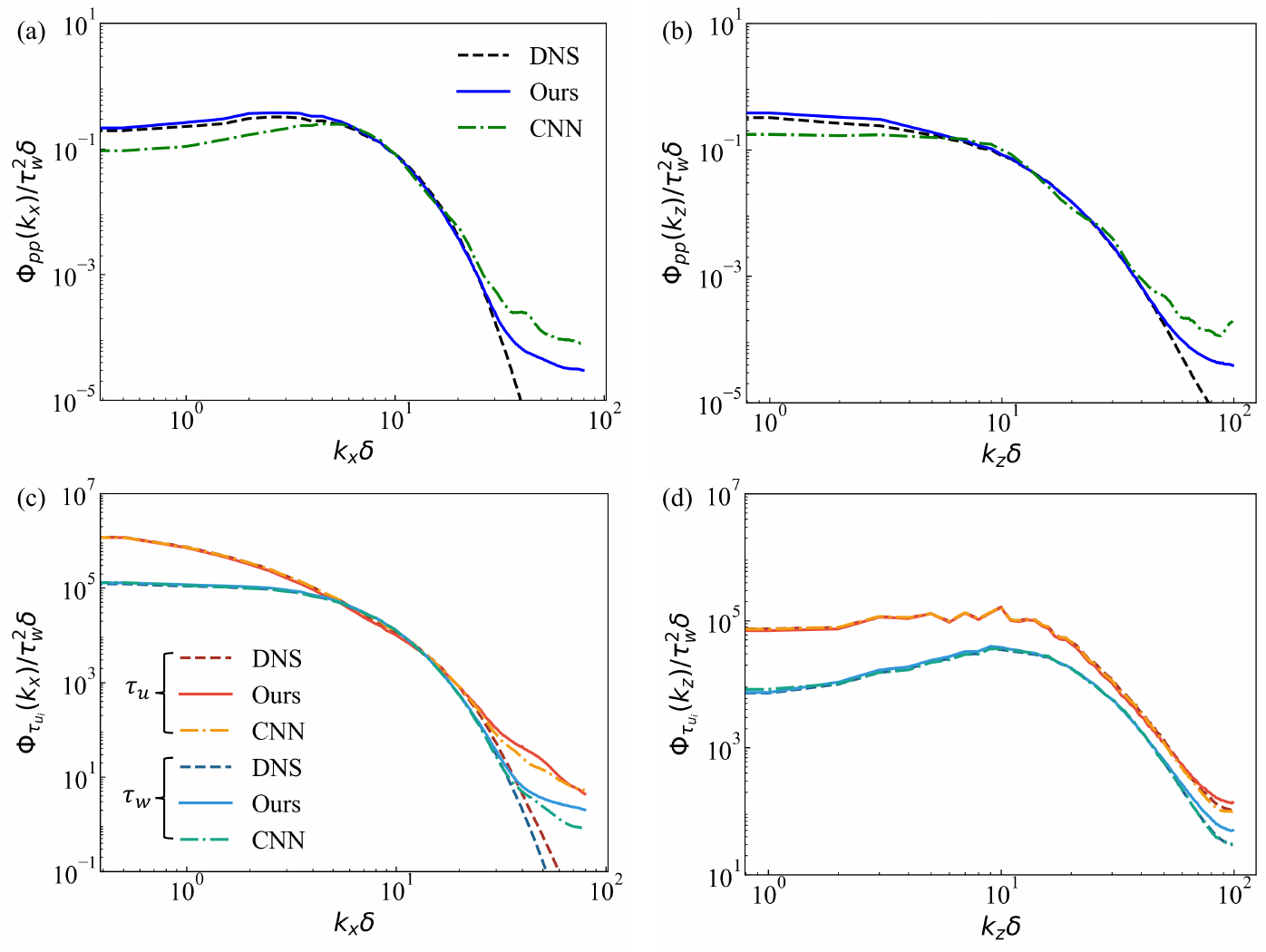}
    \caption{ Comparison of statistics from different forward models for mapping velocities at $y^+=5$ to wall: (a) streamwise and (b) spanwise spectra of pressure fluctuations; (b) spanwise spectrum of pressure fluctuations; (c) streamwise and (d) spanwise spectra of wall shear stresses $\tau_u$ and $\tau_w$.}
    \label{fig:forward_stat_y5}
\end{figure}

\begin{figure}[H]
    \centering
    \includegraphics[width=0.85\textwidth]{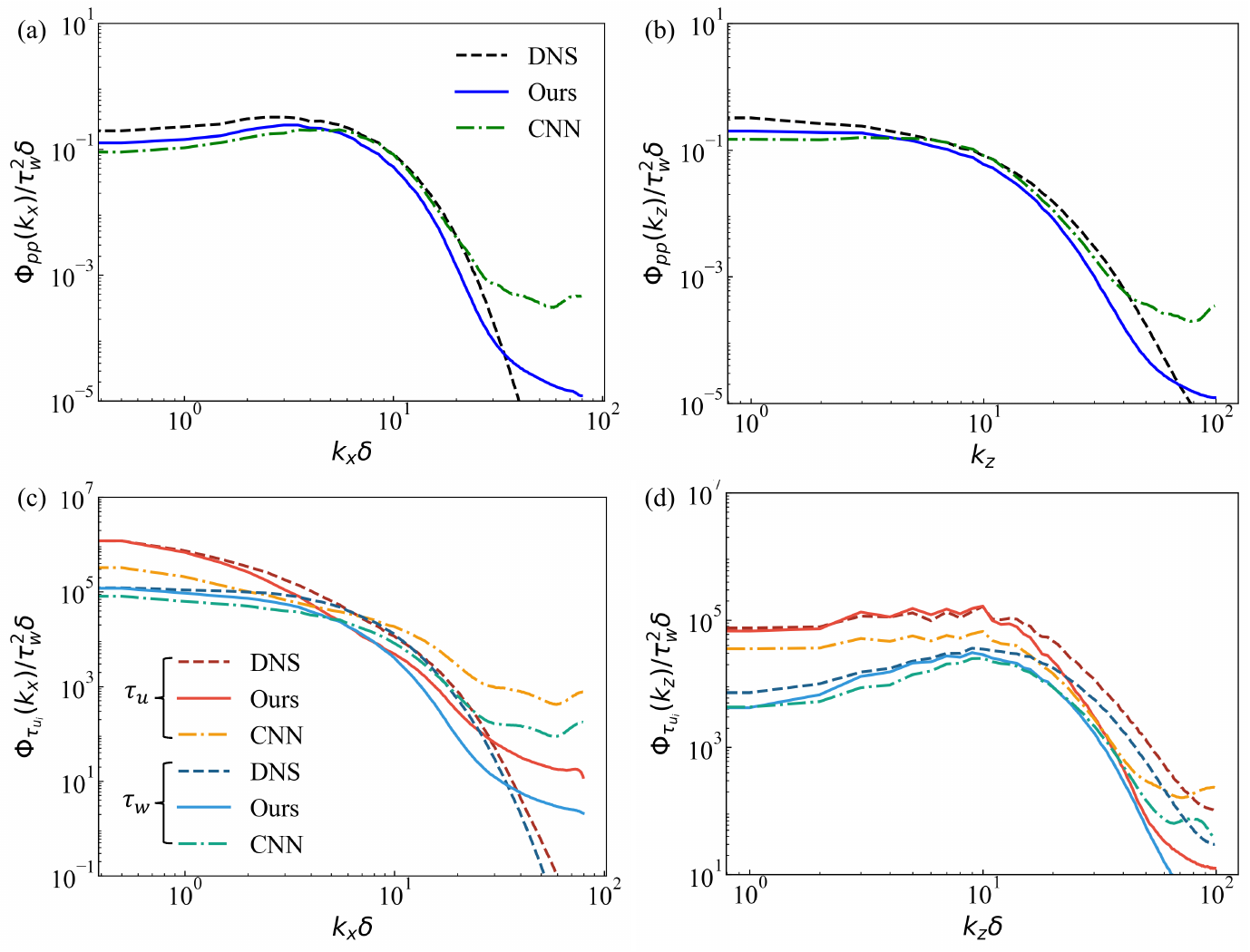}
    \caption{ Comparison of statistics from different forward models for mapping velocities at $y^+=40$ to wall: (a) streamwise and (b) spanwise spectra of pressure fluctuations; (b) spanwise spectrum of pressure fluctuations; (c) streamwise and (d) spanwise spectra of wall shear stresses $\tau_u$ and $\tau_w$.}
    \label{fig:forward_stat_y40}
\end{figure}


\end{document}